\documentclass[superscriptaddress,twocolumn,showpacs,prb]{revtex4-1}
\usepackage[utf8]{inputenc} 
\usepackage{mathtools}
\usepackage{amsmath}
\usepackage{braket}
\usepackage{comment}
\usepackage{booktabs} 
\usepackage{siunitx} 
\usepackage{booktabs}
\usepackage{graphicx}
\usepackage{multirow}
\usepackage{graphicx}
\usepackage{pgfplots}
\pgfplotsset{compat=1.14}
\usepackage{csquotes}
\usepackage{hhline}
\usepackage{tabularx}
\usepackage{subcaption}
\usepackage{amssymb}
\usepackage{amsmath}
\usepackage{braket}
\usepackage{graphicx}
\usepackage{tikz}
\usepackage{geometry}
\usepackage{quantikz}
\usepackage[normalem]{ulem} 
\usepackage{textcomp} 
\usepackage{dcolumn}
\usepackage{tabularx}
\usepackage[colorlinks=true,linkcolor=blue,citecolor=blue,urlcolor=blue]{hyperref}

\usepackage{braket}
\usepackage{float}
\usepackage{listings}
\usepackage{color}
\usepackage{subcaption}
\usepackage{placeins}
\definecolor{mygreen}{rgb}{0,0.6,0}
\definecolor{mygray}{rgb}{0.5,0.5,0.5}
\definecolor{mymauve}{rgb}{0.58,0,0.82}

\newcolumntype{C}{>{\centering\arraybackslash}X}

\begin{document}

\title{QMIMO: Circuit Based Quantum MIMO Design with Variational Receiver }

\author{Sayeda Bipanchi Ahmed}
\email{sayedabianchiahmed@gmail.com}
\affiliation{Department of Physics and Astronomy, National Institute of Technology, Rourkela, 769008, Odisha, India}
\author{Bikash K. Behera}
\email{bikas.riki@gmail.com}
\affiliation{Bikash's Quantum (OPC) Private Limited, Mohanpur, 741246, West Bengal, India, and Università degli Studi di Cagliari, Via Is Mirrions, Cagliari, 09123, Italy}
\author{Mandar Thatte}
\email{mandarthatte1042@gmail.com}
\affiliation{Department of Computer Science, Université de Sherbrooke, Quebec, J1N3C6, Canada  and CQST,  Siksha O Anusandhan, Khandagiri, Bhubaneshwar, 751030, Odisha, India}
\author{Prasanta K. Panigrahi}
\email{pprasanta@iiserkol.ac.in}
\affiliation{Department of Physical Sciences, Indian Institute of Science Education and Research Kolkata, Mohanpur 741246, West Bengal, India}
\affiliation{CQST, Siksha O Anusandhan, Khandagiri, Bhubaneswar, 751030, Odisha, India}

\begin{abstract}

This paper investigates a quantum extension of classical Multiple-Input Multiple-Output (MIMO) communication in which the conventional linear channel model is replaced by a parameterized multi-qubit unitary transformation. Within this framework, interference is represented through coherent quantum interactions rather than additive signal coupling. To recover transmitted information, a Variational Quantum Circuit (VQC) receiver is introduced that learns an approximate inverse channel transformation through supervised variational optimization. The proposed system is evaluated under realistic noisy intermediate-scale quantum (NISQ) conditions incorporating depolarizing noise, thermal relaxation, and measurement imperfections, and its performance is compared with that of standard classical detection methods. The results reveal a trade-off between the two approaches: classical detectors achieve substantially lower bit-error rates across much of the investigated parameter range but exhibit pronounced performance degradation for specific channel configurations, whereas the VQC receiver maintains a more uniform error profile as channel complexity increases, albeit at a higher average BER. These findings suggest that variational quantum receivers are not a direct replacement for classical detection methods, but rather a complementary approach that may offer increased performance stability in communication scenarios characterized by strong coupling and complex interference patterns.

\end{abstract}
\maketitle
\section{Introduction}

The rapid growth of wireless data traffic and the emergence of beyond-5G and sixth-generation (6G) communication systems continue to place increasing demands on spectral efficiency, receiver scalability, and computational complexity. Modern wireless networks must simultaneously support massive numbers of users across increasingly congested radio-frequency bands while operating under strict constraints on power consumption, delay, and hardware resources. Multiple-input multiple-output (MIMO) communication has emerged as one of the most effective solutions to these challenges by exploiting spatial multiplexing and diversity to transmit multiple independent data streams over the same frequency band without requiring additional spectral resources~\cite{tse2005fundamentals,telatar1999capacity}. By employing multiple transmitting and receiving antennas, MIMO systems significantly improve channel capacity and transmission reliability, and under favorable propagation conditions the capacity of a multi-antenna communication link scales with the minimum of the numbers of transmit and receive antennas~\cite{telatar1999capacity}.

Despite these advantages, the performance of classical MIMO systems remains fundamentally constrained by interference, additive noise, imperfect channel estimation, and increasing signal-processing complexity. Linear detection methods such as Zero-Forcing (ZF) and Minimum Mean-Square Error (MMSE) equalization have been extensively studied within estimation and detection theory~\cite{poor1994introduction,kay1993}. Among these approaches, MMSE detection provides a practical compromise between interference suppression and noise amplification, making it one of the most widely used receiver architectures in modern wireless systems~\cite{tse2005fundamentals}. However, MMSE equalization requires inversion of large channel matrices, and the computational cost associated with matrix inversion grows rapidly as the system dimension increases. Furthermore, the effectiveness of such receivers depends strongly on accurate channel-state estimation, which itself introduces additional pilot overhead and computational burden.

Over the past three decades, significant progress has also been made in the development of quantum information processing theory~\cite{nielsen2010quantum,wilde2013quantum}.Quantum channels obey capacity limits that differ in important respects from those of the classical framework~\cite{shannon1948mathematical,holevo1998capacity,schumacher1997sending}, enabling communication and signal-processing strategies that exploit genuine quantum resources. The intersection of classical multi-antenna signal processing and quantum information theory has consequently begun to attract increasing research attention~\cite{mikki2019quantum,koudia2025crosstalk,oleynik2025diversity,rehman2025mimo}.

Beyond providing a quantum representation of channel evolution, the entangling operations employed in quantum MIMO systems introduce a fundamentally new communication resource. In classical MIMO channels, information is often interpreted as being transmitted through spatial modes associated with the channel matrix and its singular-value structure~\cite{tse2005fundamentals,telatar1999capacity}. In contrast, the coherent interactions generated by entangling quantum operations allow information to be distributed across collective multi-qubit degrees of freedom that cannot be decomposed into independent transmission paths~\cite{nielsen2010quantum,wilde2013quantum}. From this perspective, entanglement acts as a communication resource that enables information processing through correlated quantum modes rather than solely through classical spatial channels~\cite{wilde2013quantum,schumacher1997sending,holevo1998capacity}. The proposed framework exploits this principle by modeling interference through entangling dynamics and training a variational receiver to recover information from the resulting correlated quantum states. The parameterized entangling quantum channel performs the role of multiplexing by coherently coupling multiple transmitted information streams into a correlated multi-qubit quantum state through superposition and entanglement~\cite{tse2005fundamentals,telatar1999capacity,nielsen2010quantum,wilde2013quantum}. The variational quantum circuit (VQC) at the receiver subsequently performs the complementary demultiplexing operation by learning an inverse transformation that reconstructs the transmitted information from the entangled quantum state through hybrid quantum-classical optimization~\cite{cerezo2021variational,mitarai2018quantum,kandala2017hardware}. Although these operations are implemented through quantum evolution rather than linear signal processing, they retain the same conceptual roles as the multiplexing and equalization stages in classical MIMO communication~\cite{tse2005fundamentals,telatar1999capacity}.

The advent of noisy intermediate-scale quantum (NISQ) hardware~\cite{preskill2018nisq} has shifted attention toward near-term algorithms that can tolerate gate imperfections and limited qubit counts. Variational quantum circuits (VQCs), which encode trainable parameters directly into gate rotation angles, have emerged as the leading paradigm for extracting useful computation from NISQ devices~\cite{cerezo2021variational,mcclean2018barren}. Analytically exact gradient evaluation through the parameter-shift rule~\cite{mitarai2018quantum}, hardware-efficient ansatz compilation~\cite{kandala2017hardware}, and a growing suite of error mitigation techniques~\cite{temme2017error,li2017efficient} have collectively made VQC-based methods viable candidates for realistic signal-processing applications.

\subsection{Related Works}

The literature on MIMO detection and equalization is extensive. Capacity scaling laws and MMSE receiver design are treated comprehensively in the foundational theoretical framework~\cite{tse2005fundamentals,telatar1999capacity}, and robustness under channel estimation error is well characterized within classical estimation theory~\cite{poor1994introduction,kay1993}. The mutual information between transmitted and received signal vectors serves as a natural figure of merit that links receiver design directly to channel capacity~\cite{shannon1948mathematical,cover2006elements}. 

The application of quantum computing to wireless communication problems is still in its early stages, but several important research directions have begun to emerge. The reformulation of electromagnetic propagation within a quantum-information framework has been explored as a conceptual bridge between antenna theory and quantum channel models~\cite{mikki2019quantum}, while recent studies on quantum MIMO channels under realistic decoherence conditions have established information-theoretic benchmarks for practical quantum receivers and have investigated diversity-multiplexing tradeoffs and correlated quantum communication channels~\cite{koudia2025crosstalk,oleynik2025diversity,rehman2025mimo}. Parallel developments in quantum channel learning, variational decoding, and quantum communication architectures have investigated tomography-inspired process reconstruction, parameterized quantum circuits for state recovery, and entanglement-assisted communication strategies~\cite{schuld2020circuit,cerezo2021variational,wilde2013quantum}. These efforts build on quantum detection theory and quantum channel coding theorems~\cite{helstrom1976quantum,schumacher1997sending,holevo1998capacity}, and together they highlight the potential for quantum-mechanical correlations to provide a fundamentally different framework for interference mitigation and signal detection compared with conventional classical communication systems.

On the algorithmic side, variational quantum algorithms have attracted substantial interest as trainable signal-processing primitives~\cite{cerezo2021variational}. The barren-plateau phenomenon places important constraints on the circuit architectures and initialization strategies that are viable in practice~\cite{mcclean2018barren}. Error mitigation schemes including zero-noise extrapolation~\cite{temme2017error} and probabilistic error cancellation~\cite{li2017efficient} have been proposed to extend the reach of variational methods on near-term hardware. The theoretical foundations of VQCs as machine-learning models, including the circuit-centric classifier framework and the role of data encoding, have been worked out in a series of contributions~\cite{schuld2019quantum,schuld2020circuit,mitarai2018quantum}.

Several optimizer families are relevant to VQC training. The simultaneous perturbation stochastic approximation (SPSA) method~\cite{spall1998implementation,spall1998overview} estimates gradients by random simultaneous perturbation of all parameters, making it well-suited to noisy objective landscapes. Adaptive moment estimation (Adam)~\cite{kingma2014adam} and its convergence properties under non-convex stochastic objectives~\cite{bottou2018optimization} provide a second optimizer class. The constrained optimization by linear approximation (COBYLA) method~\cite{powell1994direct} offers a derivative-free alternative that can be effective when gradient evaluations are expensive. Open-source quantum computing frameworks~\cite{qiskit} have made it straightforward to benchmark all three approaches on noise-model-augmented simulators.

\subsection{Research Gap and Motivation}

Although these studies have substantially advanced quantum communication theory and variational quantum algorithms, they have largely focused on individual aspects of the communication pipeline rather than an integrated communication framework. Existing quantum communication research has primarily emphasized channel characterization, information-theoretic limits, and quantum channel capacity~\cite{koudia2025crosstalk,holevo1998capacity,schumacher1997sending}, whereas variational quantum algorithms have been developed mainly as general-purpose optimization and learning techniques without being specifically tailored to wireless MIMO equalization~\cite{cerezo2021variational,mcclean2018barren,kandala2017hardware}. To the best of our knowledge, an end-to-end hybrid quantum-classical framework that jointly incorporates parameterized quantum channel modeling, realistic noisy quantum evolution, and variational receiver optimization within a unified quantum MIMO communication architecture has not yet been reported.

These limitation motivates the present work and highlights a clear research gap at the intersection of quantum variational computing and MIMO equalization. Although classical MMSE receivers are well established for mitigating linear interference in conventional communication systems~\cite{poor1994introduction,kay1993}, they are not designed to exploit the non-classical correlations generated by entangled quantum channels. Consequently, the use of parameterized quantum circuits as trainable MMSE-inspired quantum equalizers, capable of learning an approximate inverse transformation of a noisy entangling channel through variational optimization, has received limited attention in the literature.

This gap is particularly relevant in the NISQ regime, where small-scale multi-qubit systems can already be implemented with moderate fidelity on near-term quantum hardware~\cite{preskill2018nisq}. At the same time, mutual information provides a natural connection between communication-theoretic receiver performance and quantum channel capacity~\cite{cover2006elements,holevo1998capacity}. A hybrid quantum-classical variational framework, in which quantum circuits perform channel evolution and decoding while classical optimization updates the trainable receiver parameters, therefore offers a promising route toward practical quantum MIMO communication architectures.

Motivation for the present work also arises from computational scalability. Classical MMSE equalization requires inversion of an $n \times n$ channel matrix, an operation whose computational cost scales approximately as $\mathcal{O}(n^3)$ for dense systems. Quantum circuits evolve in exponentially large Hilbert spaces while being described through sequences of gate operations whose number typically scales polynomially with the number of qubits. Although no claim of asymptotic quantum advantage is made here, the possibility that a variational circuit may learn an effective inverse-channel representation motivates a systematic empirical investigation of VQC-based MMSE approximation as a function of circuit depth, qubit count, noise strength, and optimizer choice.

\subsection{Novelty and Contributions}

The present paper makes several contributions toward the study of quantum-assisted MIMO MMSE equalization.

\begin{itemize}

\item Unlike conventional MIMO formulations in which the propagation medium is described by a linear channel matrix, the proposed framework represents the communication channel as a parameterized many-body quantum unitary acting on an $n$-qubit register. The channel coefficients are encoded through a sequence of entangling $R_{YY}$ rotations whose angles are derived directly from the entries of the channel matrix. This construction provides a quantum-native description of interference coupling and provides a direct bridge between classical MIMO channel modeling and variational quantum circuit design~\cite{kandala2017hardware,mitarai2018quantum}.

\item The MMSE equalization problem is reformulated within a hybrid quantum-classical learning framework, where the equalizer is realized as a parameterized quantum circuit and its trainable parameters are optimized using classical optimization algorithms. Building on the principles of variational quantum circuits and hybrid quantum-classical optimization~\cite{cerezo2021variational}, the circuit parameters are optimized through minimization of an average bit-error-rate objective evaluated over a representative ensemble of training states, enabling the receiver to adaptively reconstruct the transmitted information. This approach replaces the conventional analytical MMSE inversion procedure with a data-driven variational optimization framework implemented within a quantum computational architecture.

\item The complete framework is evaluated under realistic noisy quantum hardware conditions through completely positive trace-preserving (CPTP) noise modeling that includes single- and two-qubit depolarizing errors, thermal relaxation processes, and asymmetric readout errors~\cite{preskill2018nisq,temme2017error,li2017efficient}. Performance is assessed using both bit-error rate (BER) and mutual information between transmitted and recovered bit strings, enabling simultaneous characterization of decoding accuracy and information preservation under realistic operating conditions~\cite{cover2006elements,holevo1998capacity,schumacher1997sending}.

\end{itemize}

Collectively, these contributions present an end-to-end framework for variational quantum MMSE equalization that bridges classical MIMO signal processing~\cite{tse2005fundamentals,telatar1999capacity,poor1994introduction,kay1993}, quantum information theory~\cite{nielsen2010quantum,wilde2013quantum,shannon1948mathematical,holevo1998capacity,schumacher1997sending}, and near-term quantum algorithm design~\cite{cerezo2021variational,preskill2018nisq,mitarai2018quantum,kandala2017hardware}. 

The structure of the paper is organized as follows:
Section~I introduces the background and briefly reviews existing work related to quantum communication and classical MIMO systems.
Section~II introduces the basic concepts and mathematical preliminaries required for the study of the proposed system.
Section~III then describes the methodology in detail, which includes the problem formulation, mathematical modeling, circuit design, and the overall framework of the system.
Section~IV discusses and analyzes the results, focusing on the interpretation of the observations and their implications in terms of system performance.
Finally, Section~V concludes the paper by summarizing the main outcomes and contributions while outlining several possible directions for future research.

\section{Background}\label{SecII}

In recent years, communication models have been increasingly explored in the quantum domain, where information is encoded into quantum states. Unlike classical systems, quantum communication exploits superposition and entanglement~\cite{nielsen2010quantum,preskill2018nisq}, allowing qubits to evolve coherently. As a result, interference is not represented as a simple additive effect, but is inherently embedded in multi-qubit transformations governed by unitary operations~\cite{nielsen2010quantum}.

Quantum channels are commonly modeled using quantum circuits, where system evolution is described through sequences of quantum gates~\cite{nielsen2010quantum}. In multi-qubit systems, entangling operations introduce correlations between qubits, naturally capturing interactions between input streams. This provides a suitable framework for Quantum MIMO systems, where interference can be represented through parameterized multi-qubit operations~\cite{schuld2020circuit}. However, recovering transmitted information remains challenging due to the high-dimensional and nonlinear nature of these transformations.

Practical quantum communication systems are also affected by noise, including depolarization, dephasing, relaxation, and measurement imperfections~\cite{preskill2018nisq}. These effects distort quantum states and accumulate throughout circuit execution, making realistic noise modeling essential for accurate system analysis.

Parameterized quantum circuits offer a scalable framework for representing unknown quantum transformations, with trainable parameters controlling qubit interactions~\cite{schuld2020circuit}. In particular, Variational Quantum Circuits (VQCs) provide an effective approach for approximating complex channel inverses through hybrid quantum-classical optimization~\cite{cerezo2021variational}, without requiring explicit channel knowledge.

Training such circuits involves optimizing measurement-based cost functions over noisy and non-convex landscapes. Optimization methods such as ADAM, SPSA, and COBYLA are commonly employed~\cite{kingma2014adam,spall1998overview,powell1994direct}, offering different trade-offs in convergence, robustness, and computational efficiency.

System performance is evaluated using Bit Error Rate (BER) and mutual information, which respectively measure decoding accuracy and information transmission efficiency~\cite{poor1994introduction,cover2006elements}. Together, these metrics provide a comprehensive assessment of quantum MIMO receiver performance under different noise conditions and optimization strategies.

\section{Methodology}\label{SecIII}

\subsection{Quantum MIMO System Model}

\begin{figure*}[t]
\centering
\includegraphics[width=0.9\textwidth]{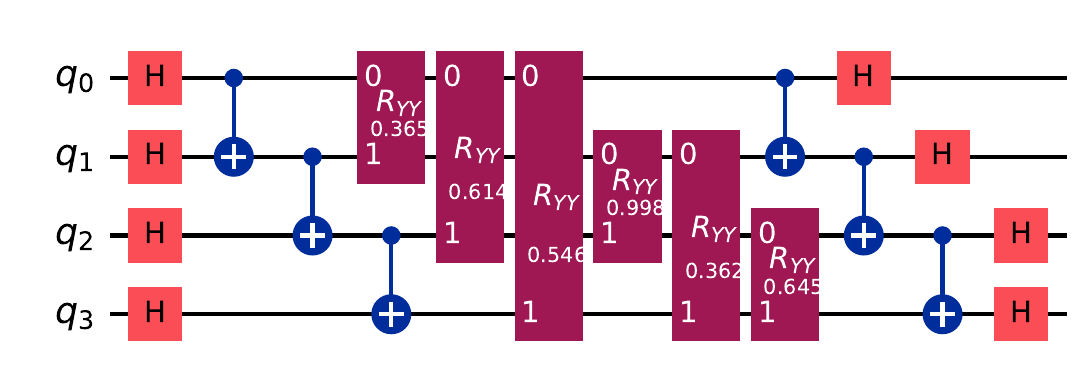}
\caption{Circuit representation of the proposed quantum MIMO channel for an $n=4$-qubit system. The channel consists of Hadamard layers, nearest-neighbor entangling operations, and pairwise $R_{YY}(\theta_{ij})$ interactions.}
\label{fig:channel}
\end{figure*}

The proposed quantum Multiple-Input Multiple-Output (MIMO) system is developed using a circuit-based approach. In this architecture, both the communication channel and the receiver operations are modeled as operations within an $n$-qubit Hilbert space $\mathcal{H} = \mathbb{C}^{2^n}$ \cite{nielsen2010quantum}. 

Classical information is mapped onto a set of $2^n$ orthogonal quantum states; we use the computational basis \cite{nielsen2010quantum}, which corresponds to the set of all possible binary strings of length $n$. Each possible transmitted symbol $x$ is mapped to a corresponding quantum state:

\begin{equation}
|x\rangle = |b_1 b_2 \dots b_n\rangle, \quad b_i \in \{0, 1\}
\end{equation}

The encoding process originates from the initial state, $|0\rangle^{\otimes n}$, which serves as a reference state from which the desired input state can be constructed.

The encoding mechanism utilizes the Pauli-$X$ gate due to its bit-flip property in the computational basis \cite{nielsen2010quantum}. The mapping is defined as:

\begin{equation}
b_i = 0 \rightarrow |0\rangle, \quad b_i = 1 \rightarrow X|0\rangle = |1\rangle
\end{equation}

The Pauli-$X$ operator is applied only when the corresponding classical bit value is $1$, which ensures that the classical information is preserved with high fidelity within the quantum state representation.

Extending this to the $n$-qubit system, the encoded state is modeled as the tensor product of individual qubit operations:

\begin{equation}
|x\rangle = X_1^{b_1} \otimes X_2^{b_2} \otimes \dots \otimes X_n^{b_n} \ket{0}^{\otimes n}
\end{equation}

The selection of this encoding system provides a direct, low-complexity mapping from classical binary streams to quantum states while preserving orthogonality between transmitted symbols.

The quantum MIMO channel is modeled as a structured multi-qubit unitary transformation that captures interference through coherent interactions within the quantum system \cite{mikki2019quantum, koudia2025crosstalk}. For an encoded input state $|x\rangle$, the channel evolution is represented as

\begin{equation}
|\psi_{\text{out}}\rangle = U_{\text{ch}}(\Theta)|x\rangle,
\end{equation}

where $U_{\text{ch}}(\Theta)$ is a parameterized unitary operator that serves as the quantum analogue of the channel matrix in classical MIMO systems \cite{tse2005fundamentals, telatar1999capacity}. 

In conventional MIMO communication, the received signal is typically expressed as

\begin{equation}
y = Hx + n,
\end{equation}

where $H$ denotes the channel matrix and $n$ represents additive noise. In this formulation, interference arises through linear mixing of transmitted signal amplitudes. In contrast, the proposed quantum framework models signal propagation through coherent unitary evolution in Hilbert space, where information is redistributed through correlated quantum interactions rather than classical linear superposition. Consequently, the resulting interference process incorporates superposition and entanglement effects that are absent in conventional linear channel models.

The unitary operator is constructed in a way that each stage contributes differently to the overall transformation. This channel model combines superposition, entanglement, and controlled interactions to demonstrate how information propagates through the system, which can be defined as

\begin{equation}
U_{\text{ch}}(\Theta) = H^{\otimes n} U_{\text{ent}} U_{\text{mix}}(\Theta) U_{\text{ent}} H^{\otimes n}
\end{equation}

The initial Hadamard layer projects the input state across a superposition of all computational basis states \cite{nielsen2010quantum}, allowing the encoded information to evolve simultaneously along multiple computational paths. The final Hadamard layer performs the inverse role by bringing the resulting interference pattern back into the computational basis so that it can be measured.

Entanglement is introduced by the operator $U_{\text{ent}}$, which consists of a sequence of nearest-neighbor CNOT gates \cite{nielsen2010quantum}:

\begin{equation}
U_{\text{ent}} = \prod_{i=1}^{n-1} \text{CNOT}(i \rightarrow i+1),
\end{equation}

which creates correlations between adjacent qubits, allowing information to propagate across the system. The generated correlations imply that the quantum state can no longer be represented as an independent product of individual qubit states, thereby introducing collective system dynamics that are absent in classical channel models.

To model global interference between all transmission paths, we introduce a parameterized mixing operator:

\begin{equation}
U_{\text{mix}}(\boldsymbol{\Theta}) = \prod_{i<j} R_{YY}^{(i,j)}(\theta_{ij}),
\end{equation}

where each two-qubit interaction is defined as:

\begin{equation}
R_{YY}^{(i,j)}(\theta_{ij}) = \exp\left(-i \frac{\theta_{ij}}{2} Y_i \otimes Y_j \right).
\end{equation}

The interaction strength is governed by the parameter $\theta_{ij}$, where $\theta_{ij} = 0$ reduces the operator to the identity. As $\theta_{ij}$ increases, the resulting rotations generate non-separable many-body correlations and effective mixing among the components of the quantum state.

From a communication-theoretic perspective, this mechanism diverges fundamentally from classical MIMO systems. While classical models represent interference as linear amplitude mixing via matrix multiplication, this framework utilizes the coherent redistribution of probability amplitudes across an exponentially large Hilbert space. By modeling the channel as a unitary evolution $|\psi_{\text{out}}\rangle = U_{\text{ch}}(\boldsymbol{\Theta})|x\rangle$, the decoding process is transformed into a correlated inverse estimation problem. In this high-dimensional space, information recovery depends on the collective correlation structure established during the evolution rather than independent signal components. Consequently, the entangling quantum channel performs the multiplexing function by coherently coupling the transmitted information into correlated quantum states, while the variational quantum receiver performs the complementary demultiplexing function by learning an approximate inverse transformation that reconstructs the transmitted information from the resulting correlated quantum state~\cite{tse2005fundamentals,telatar1999capacity,cerezo2021variational,nielsen2010quantum,wilde2013quantum}.

The complete set of parameters $\boldsymbol{\Theta} = \{\theta_{ij}\}$ characterizes the channel behavior with a scaling of $O(n^2)$, mirroring the structural complexity of a classical MIMO channel matrix. Unlike classical systems, however, this scaling drives a significant increase in circuit depth and many-body entanglement, which redistributes probability amplitudes across  computational basis states. This dispersion introduces measurement ambiguity, making the recovery of the original input state more challenging as interaction strength grows. Furthermore, the expanding parameter set creates a high-dimensional, non-convex optimization landscape marked by flatter gradients and stronger parameter correlations, providing an analytical basis for the observed decline in decoding stability as the system scales.

The primary novelty of this framework lies in its departure from both classical communication theory and existing Variational Quantum Algorithms (VQAs). Unlike conventional VQAs or channel learning methods designed for state preparation or process approximation, this formulation explicitly treats the parameterized quantum circuit as a structured communication channel. By incorporating entanglement, coherent amplitude redistribution, and non-classical correlations between transmission paths, the framework interprets MIMO interference as a many-body quantum evolution process. This approach establishes a physically grounded extension of communication theory, treating the interference properties as emergent features of multi-qubit interactions within the quantum domain.

\begin{figure*}[t]
\centering
\renewcommand{\arraystretch}{1.2}
\setlength{\fboxsep}{6pt}

\tikzset{
block/.style={
    draw=black!80,
    rounded corners=5pt,
    align=center,
    minimum height=1.35cm,
    minimum width=2.6cm,
    font=\small,
    fill=gray!4
},
arrow/.style={-latex, thick},
feedback/.style={-latex, thick, dashed, gray!70}
}

\begin{tikzpicture}


\node[block] (input) {
\textbf{Classical Source}\\
$\mathbf{x}\in\{0,1\}^n$
};

\node[block, right=1.4cm of input] (encode) {
\textbf{Quantum Encoding}\\
$|x\rangle = X^{\mathbf{b}}|0\rangle^{\otimes n}$
};

\node[block, right=1.4cm of encode] (channel) {
\textbf{Quantum MIMO Channel}\\
$U_{\mathrm{ch}}(\Theta)$\\
Entangling Dynamics\\
Pairwise $R_{YY}(\Theta)$ Couplings
};

\node[block, right=1.4cm of channel] (noise) {
\textbf{NISQ Noise Model}\\
CPTP Map $\mathcal{E}(\rho)$\\
Thermal Relaxation\\
+ Depolarizing Noise\\
+ Readout Errors
};


\node[block, below=1.8cm of noise] (vqc) {
\textbf{Variational Decoder}\\
$U_{\mathrm{VQC}}(\phi)$\\
Trainable Recovery Circuit};

\node[block, left=1.4cm of vqc] (measure) {
\textbf{Measurement}\\
Computational Basis\\
$P_\phi(y|x)$
};

\node[block, left=1.4cm of measure] (decode) {
\textbf{Reconstruction}\\
$\hat{\mathbf{x}}$
};

\node[block, left=1.4cm of decode] (ber) {
\textbf{BER Metric}\\
$\mathrm{BER}=\frac{1}{n}d(x,\hat{x})$
};

\node[block, below=0.9 cm of decode] (mi) {
\textbf{Mutual Information}\\
$I(X;Y)$
};


\node[block, below=1.8cm of measure, fill=blue!3] (opt) {
\textbf{Classical Optimizer}\\
ADAM / SPSA / COBYLA\\
$\phi \rightarrow \phi^{\ast}$
};


\draw[arrow] (input) -- (encode);
\draw[arrow] (encode) -- (channel);
\draw[arrow] (channel) -- (noise);

\draw[arrow] (noise) -- (vqc);

\draw[arrow] (vqc) -- (measure);
\draw[arrow] (measure) -- (decode);
\draw[arrow] (decode) -- (ber);
\draw[arrow] (decode.south) -- (mi.north);

\draw[feedback] (ber.south) |- (opt.west);
\draw[feedback] (opt.east) -| node[pos=0.25,right, font=\small] {parameter update} (vqc.south);


\node[font=\bfseries\small, above=0.4cm of channel] {
Forward Quantum MIMO Transmission
};

\node[font=\bfseries\small, below=0.5cm of opt] {
Hybrid Quantum--Classical Training Loop
};

\end{tikzpicture}
\caption{
Hybrid quantum--classical architecture for the proposed
$n\times n$ quantum MIMO communication system.
Classical information is encoded into multi-qubit basis states
and transmitted through a parameterized entangling quantum channel
with pairwise $R_{YY}(\Theta)$ interactions.
A variational quantum receiver performs adaptive quantum decoding
under a realistic NISQ noise environment modeled using a CPTP map
incorporating thermal relaxation, depolarizing gate errors,
and readout noise.The reconstructed bitstrings are used to evaluate BER and mutual information, while variational parameters are iteratively optimized using ADAM, SPSA, or COBYLA.
}
\label{fig:hybrid_qmimo_architecture}
\end{figure*}
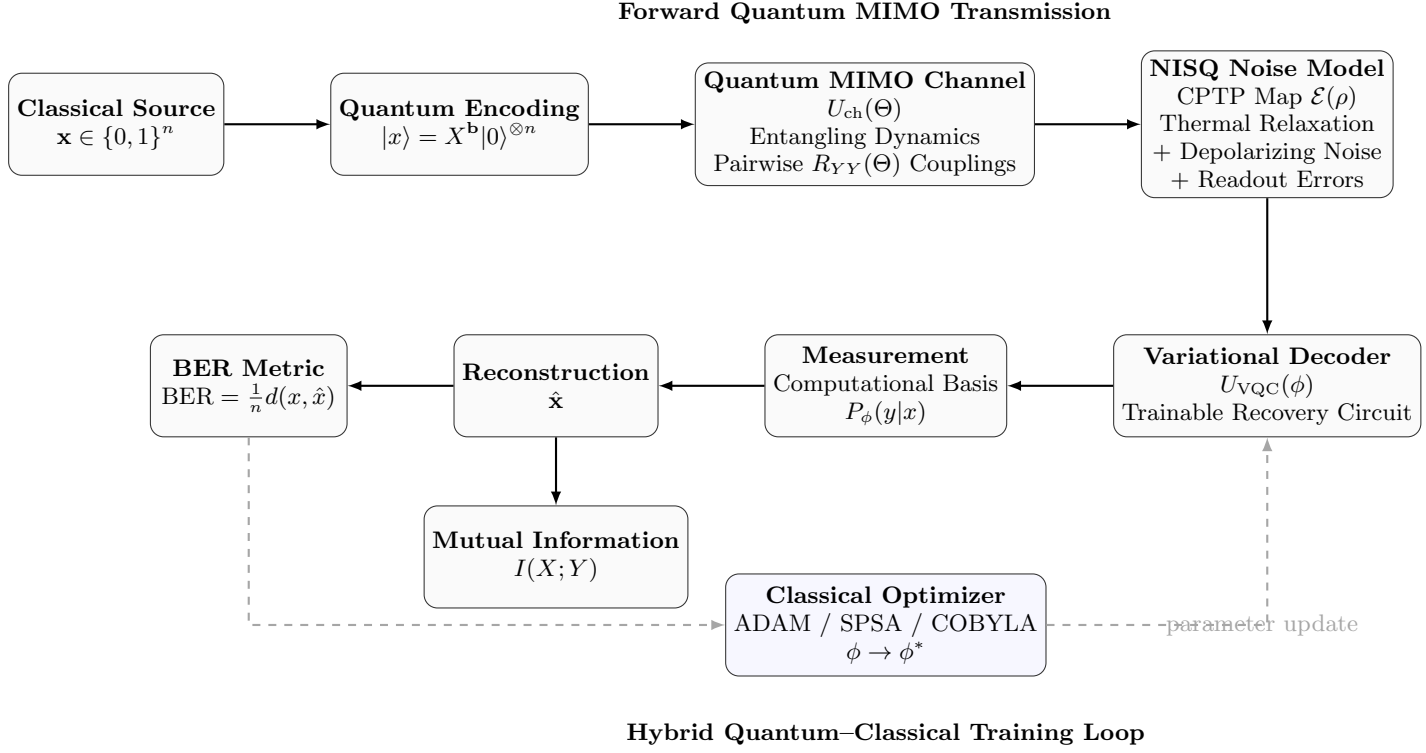

\subsection{Noise Modeling and CPTP Formalism}

In practical quantum systems, the evolution of a quantum state is not strictly unitary due to unavoidable interactions with the surrounding environment and imperfections in quantum hardware \cite{preskill2018nisq}. Consequently, the evolution of the transmitted quantum state cannot be described solely through the unitary channel operator $U_{\text{ch}}(\Theta)$. Instead, the overall system dynamics must account for decoherence, dissipation, gate imperfections, and measurement disturbances that collectively perturb the ideal quantum evolution.

For an input density operator $\rho_{\text{in}}$, the noisy channel evolution is modeled as
\begin{equation}
\rho_{\text{out}} = \mathcal{E}
\Big(
U_{\text{ch}}(\Theta)\rho_{\text{in}}U_{\text{ch}}^\dagger(\Theta)
\Big),
\end{equation}
where $\mathcal{E}(\cdot)$ denotes the overall noise process acting on the system.

This formulation reflects the fact that the transmitted quantum information undergoes both coherent evolution through the channel unitary and incoherent evolution due to environmental interactions. Unlike classical additive noise models, quantum noise affects both the probability amplitudes and phase coherences of the quantum state, thereby altering the interference and entanglement structure established during the channel evolution \cite{nielsen2010quantum,wilde2013quantum}.

The noisy dynamics are implemented using Qiskit’s quantum noise modeling framework \cite{qiskit}, where noise channels are inserted after specific quantum operations to emulate realistic device behavior. The resulting model incorporates gate-level noise, decoherence effects, and readout imperfections that are characteristic of noisy intermediate-scale quantum (NISQ) hardware \cite{preskill2018nisq}.

From a mathematical perspective, the noise process $\mathcal{E}$ is described within the framework of Completely Positive Trace Preserving (CPTP) maps \cite{wilde2013quantum}. CPTP maps provide a general mathematical framework for describing the evolution of open quantum systems while ensuring that the resulting density operator remains physically valid throughout the transformation. Accordingly, the noisy channel evolution admits a Kraus operator representation:
\begin{equation}
\rho_{\text{out}}
=
\sum_k
K_k
\Big(
U_{\text{ch}}\rho_{\text{in}}U_{\text{ch}}^\dagger
\Big)
K_k^\dagger,
\end{equation}
where the Kraus operators $\{K_k\}$ satisfy the completeness relation
\begin{equation}
\sum_k K_k^\dagger K_k = I.
\end{equation}
This representation provides a unified formalism capable of modeling depolarization, thermal relaxation, decoherence, correlated gate errors, and measurement imperfections within a single mathematical framework \cite{nielsen2010quantum,wilde2013quantum}.

In the proposed quantum MIMO system, noise is introduced at the level of individual quantum operations. For single-qubit gates such as Pauli-$X$, Hadamard, and identity operations, the noise model combines depolarizing noise with thermal relaxation effects:
\begin{equation}
\mathcal{E}_1
=
\mathcal{E}_{\text{thermal}}
\circ
\mathcal{E}_{\text{dep}}^{(1)}.
\end{equation}
The single-qubit depolarizing channel is defined as \cite{nielsen2010quantum}
\begin{equation}
\mathcal{E}_{\text{dep}}^{(1)}(\rho)
=
(1-p)\rho
+
\frac{p}{3}
(X\rho X + Y\rho Y + Z\rho Z),
\end{equation}
where $p$ denotes the depolarizing noise strength.

Physically, this channel models stochastic Pauli errors that randomly perturb the quantum state during gate execution. As the noise strength increases, the quantum state progressively loses coherence and approaches a maximally mixed state, thereby reducing the distinguishability between transmitted quantum symbols \cite{nielsen2010quantum}.

The thermal relaxation component models incoherent energy exchange between the quantum processor and its surrounding environment. This process is characterized by the relaxation times $T_1$ and $T_2$, corresponding to amplitude damping and phase decoherence, respectively. The parameter $T_1$ governs the decay of excited-state populations, while $T_2$ determines the loss of phase coherence between basis states. In this implementation, the relaxation parameters are selected as $T_1 = 1/p$ and $T_2 = 0.7T_1$, thereby ensuring that larger noise strengths correspond to faster coherence degradation \cite{preskill2018nisq}.

For multi-qubit operations such as controlled-NOT and $R_{YY}$ gates, a two-qubit depolarizing channel is employed:
\begin{equation}
\mathcal{E}_2(\rho)
=
(1-p)\rho
+
\frac{p}{15}
\sum_{P_i \neq I} P_i \rho P_i,
\end{equation}
where $\{P_i\}$ denotes the set of all non-identity two-qubit Pauli operators.

This model captures correlated gate-level disturbances arising during entangling operations. Since two-qubit gates require substantially greater control precision than single-qubit operations, they are generally more sensitive to decoherence and hardware imperfections \cite{temme2017error}. Consequently, entangling operations become one of the dominant contributors to cumulative error accumulation within the proposed quantum MIMO architecture.

In addition to gate-level noise, the measurement process itself is imperfect. Readout errors are incorporated using a classical confusion matrix of the form
\begin{equation}
M =
\begin{bmatrix}
1-\alpha & \alpha \\
\beta & 1-\beta
\end{bmatrix},
\end{equation}
where $\alpha$ and $\beta$ denote the probabilities of incorrectly identifying the states $|0\rangle$ and $|1\rangle$, respectively.

Under ideal measurement conditions,

\begin{equation}
P(m|x) = \delta_{m,x},
\end{equation}

whereas measurement noise modifies the observed probability distribution such that
\begin{equation}
P(m|x) \neq \delta_{m,x}.
\end{equation}
Consequently, the measured bit string may differ from the actual computational basis state generated during the quantum evolution, thereby introducing additional uncertainty into the decoding process \cite{qiskit}.

The impact of two-qubit noise becomes particularly significant due to the structure of the mixing operator
\begin{equation}
U_{\text{mix}}(\Theta)
=
\prod_{i<j}
R_{YY}^{(i,j)}(\theta_{ij}),
\end{equation}
which introduces interactions between all pairs of qubits. For an $n$-qubit system, the total number of pairwise interactions becomes
\begin{equation}
N_{YY}
=
\sum_{i=1}^{n-1}(n-i).
\end{equation}
Evaluating the summation gives
\begin{equation}
N_{YY}
=
(n-1)+(n-2)+\cdots+1
=
\frac{n(n-1)}{2}.
\end{equation}
Therefore, the number of entangling operations scales quadratically with the system size:
\begin{equation}
N_{YY} = \mathcal{O}(n^2).
\end{equation}
This quadratic scaling implies that larger quantum MIMO systems require a significantly greater number of entangling operations, increasing the overall circuit complexity and cumulative exposure to noise processes during the channel evolution.

The complete noisy channel evolution can therefore be expressed as a sequence of alternating unitary transformations and CPTP noise channels:
\begin{equation}
\rho_{\text{out}} = \prod_{k=1}^{N_{\text{gates}}} (\mathcal{E}_k \circ U_k)(\rho_{\text{in}}),
\end{equation}
where each unitary operation $U_k$ is followed by its corresponding noise channel $\mathcal{E}_k$.

Based on this operational structure, the cumulative degradation of the system can be characterized by introducing $p_g$ as the average error rate associated with an individual quantum operation. Under the assumption that successive noise processes are approximately independent and identically distributed, the probability of an ideal execution for a single gate is $(1-p_g)$. For a circuit comprising $N_{\text{gates}}$ stochastic operations, the global fidelity representing the probability that the entire circuit evolves without a single error event is given by:
\begin{equation}
P_{\text{ideal}} = (1-p_g)^{N_{\text{gates}}}.
\end{equation}
Consequently, the aggregate probability of at least one noise event occurring during the circuit evolution defines the effective noise parameter $p_{\text{eff}}$:
\begin{equation}
p_{\text{eff}}
=
1-(1-p_g)^{N_{\text{gates}}}.
\end{equation}
This parameter $p_{\text{eff}}$ serves as a macroscopic metric for the total decoherence incurred, explicitly incorporating the combined influence of single-qubit depolarizing noise, thermal relaxation, correlated two-qubit gate errors, measurement imperfections, and cumulative gate-level decoherence.

In the high-dimensional regime, the circuit complexity is dominated by the layer of entangling operations, such that
\begin{equation}
N_{\text{gates}} \sim \mathcal{O}(n^2).
\end{equation}
Substituting this structural constraint into the expression for $p_{\text{eff}}$ yields the scaling relation:
\begin{equation}
p_{\text{eff}}
=
1-(1-p_g)^{\mathcal{O}(n^2)}.
\end{equation}
In the weak-noise regime characteristic of NISQ hardware $(p_g \ll 1)$, a first-order Taylor expansion allows for a linear approximation of the cumulative error:
\begin{equation}
(1-p_g)^{N_{\text{gates}}}
\approx
1-N_{\text{gates}}p_g,
\end{equation}
leading to the simplified analytical form:
\begin{equation}
p_{\text{eff}}
\approx
N_{\text{gates}}p_g.
\end{equation}
By applying the quadratic scaling of the gate count, we establish that the effective noise strength increases quadratically with the number of qubits:
\begin{equation}
p_{\text{eff}}
\propto
\mathcal{O}(n^2)p_g.
\end{equation}
This derivation establishes a direct analytical connection between circuit scaling and cumulative noise accumulation. As the number of qubits increases, the growth in entangling operations requires deeper quantum circuits and longer coherence times, thereby increasing the overall exposure to decoherence processes. Consequently, the effective noise strength grows progressively with system size, causing larger deviations from the ideal noiseless evolution and making accurate state recovery increasingly difficult.

For analytical interpretation, the cumulative noise accumulation is modeled under the assumption that individual noise contributions act approximately independently across successive quantum operations. Under this approximation, higher-order correlations between different noise processes are neglected, allowing the overall system degradation to be characterized through the aggregate parameter $p_{\text{eff}}$. This assumption is most accurate in the weak-noise regime, where correlated many-body error propagation remains relatively limited \cite{preskill2018nisq,temme2017error}.

\subsection{Variational Receiver and Optimization Framework}
\label{sec:receiver}

The recovery of quantum information is performed using a Variational Quantum Circuit (VQC) within a hybrid quantum-classical receiver architecture~\cite{cerezo2021variational,kandala2017hardware}. The quantum processor executes the parameterized decoding circuit and performs quantum measurements, while a classical optimizer iteratively updates the circuit parameters based on the measured decoding performance through a classical optimization loop~\cite{mitarai2018quantum,cerezo2021variational,kingma2014adam,spall1998overview,powell1994direct}. Given the noisy output state 
$\rho_{\text{out}}$, the receiver applies

\begin{equation}
\rho_{\text{dec}} = U_{\text{VQC}}^\dagger(\phi)\rho_{\text{out}}U_{\text{VQC}}(\phi),
\label{eq:vqc_decoder}
\end{equation}

where $\phi$ denotes the trainable circuit parameters. The adjoint structure reflects the objective of reversing the channel-induced transformation and recovering the transmitted information from the corrupted quantum state.

The VQC architecture is designed to reflect the structure of the quantum channel through layers of Hadamard gates, entangling operations, and parameterized $R_{YY}$ rotations \cite{kandala2017hardware,cerezo2021variational}. By incorporating prior knowledge of the channel dynamics into the receiver design, the ansatz introduces an inductive bias that restricts the optimization to physically meaningful transformations rather than the exponentially large unitary search space. This produces a more structured and trainable optimization landscape, improving parameter efficiency and convergence in noisy variational quantum systems \cite{cerezo2021variational,mcclean2018barren}. The parameterized $R_{YY}$ rotations serve as the primary adaptive components of the receiver, allowing the circuit to capture inter-qubit correlations arising from channel evolution and noise accumulation. Although the $\mathcal{O}(n^2)$ parameter scaling provides a balance between expressivity and trainability, increasing system size simultaneously enlarges the optimization space, increases circuit depth, and amplifies the effects of gate imperfections, finite-shot noise, and measurement uncertainty \cite{preskill2018nisq,mcclean2018barren}.

The decoding process is formulated as a statistical estimation problem in which the objective is to recover the transmitted classical bit string from noisy quantum measurements \cite{cover2006elements,poor1994introduction}. Let $x \in \{0,1\}^n$ denote the transmitted bit string and let $y \in \{0,1\}^n$ denote the measured output after applying both the quantum channel and the variational receiver. Due to the probabilistic nature of quantum measurement, the receiver induces a conditional distribution $P_\phi(y|x)$ over all possible output bit strings, given by the Born rule as

\begin{equation}
P_\phi(y|x)=\mathrm{Tr}\left(|y\rangle\langle y|\,\rho_{\text{rec}}(x;\phi)\right),
\label{eq:born_rule}
\end{equation}

where $\rho_{\text{rec}}$ denotes the recovered quantum state after channel transmission and variational decoding \cite{nielsen2010quantum}.

The role of the VQC is to reshape the induced output distribution such that the probability of recovering the correct transmitted bit string is maximized after measurement. To quantify decoding performance, the receiver optimization is formulated through the expected reconstruction loss over the joint distribution $P_\phi(x,y)=P(x)P_\phi(y|x)$, where a uniform prior distribution over transmitted bit strings is assumed. A natural objective for this reconstruction problem is the mean squared error (MSE),

\begin{equation}
L(\phi)=\mathbb{E}_{x,y}\left[|x-y|^2\right].
\label{eq:mse_loss}
\end{equation}

For binary vectors, the squared reconstruction error expands as

\begin{equation}
\|x-y\|^2=\sum_{i=1}^{n}(x_i-y_i)^2,
\label{eq:mse_expand}
\end{equation}

where each term equals $1$ when the corresponding bits differ and $0$ otherwise. Consequently,

\begin{equation}
\|x-y\|^2=d(x,y),
\label{eq:hamming_equivalence}
\end{equation}

where $d(x,y)$ denotes the Hamming distance between the transmitted and decoded bit strings \cite{cover2006elements}. The expected loss therefore becomes

\begin{equation}
L(\phi)=\mathbb{E}_{x,y}[d(x,y)].
\label{eq:hamming_loss}
\end{equation}

Normalizing by the number of transmitted bits yields

\begin{equation}
L(\phi)=\mathbb{E}_{x,y}\left[\frac{1}{n}d(x,y)\right],
\label{eq:ber_loss}
\end{equation}

which corresponds directly to the Bit Error Rate (BER). Hence, minimizing the MSE is equivalent to minimizing the BER, giving the optimization objective both statistical and communication-theoretic significance \cite{cover2006elements}. The optimal receiver parameters are therefore obtained as

\begin{equation}
\phi^*=\arg\min_\phi \mathbb{E}_{x,y}\left[\|x-y\|^2\right]
=\arg\min_\phi \mathbb{E}_{x,y}\left[\frac{1}{n}d(x,y)\right].
\label{eq:mmse_objective}
\end{equation}

This formulation naturally connects the variational receiver to the minimum mean square error (MMSE) principle from classical estimation theory, where minimizing the expected squared error yields statistically optimal reconstruction under noisy observations \cite{kay1993,poor1994introduction}. In the present setting, however, the receiver does not explicitly compute the conditional expectation $\mathbb{E}[x|y]$. Instead, the VQC implicitly learns a parameterized probabilistic decoding transformation through optimization over the induced quantum measurement distribution. This interpretation is particularly important in realistic noisy quantum systems because decoherence and non-unitary channel dynamics generally prevent exact deterministic inversion, making statistical reconstruction a more physically meaningful objective \cite{preskill2018nisq,nielsen2010quantum}.

In practice, evaluating the expectation over all possible input bit strings becomes computationally intractable due to the exponential growth of the Hilbert space \cite{nielsen2010quantum}. Therefore, the loss is approximated using a randomly sampled subset $S \subset \{0,1\}^n$,

\begin{equation}
\hat{L}(\phi)=\frac{1}{|S|}\sum_{x\in S}\sum_y
P_\phi(y|x)\frac{d(x,y)}{n}.
\label{eq:sampled_loss}
\end{equation}

Furthermore, the probabilities $P_\phi(y|x)$ are estimated through finite quantum measurements, introducing stochastic fluctuations into the objective function. Consequently, the optimization becomes a stochastic empirical risk minimization problem in which randomness arises from both input sampling and quantum measurement uncertainty \cite{bottou2018optimization}. As system size increases, finite-shot noise, accumulated gate errors, and deeper circuit structures further increase the variance of objective evaluations and gradient estimates, thereby complicating optimization and slowing convergence \cite{preskill2018nisq,mcclean2018barren}.

The decoding objective can also be connected to information-theoretic performance measures. Let $I(X;Y)$ denote the mutual information between the transmitted and decoded bit strings. Assuming a uniform input distribution, $H(X)=n$, and the mutual information can be written as

\begin{equation}
I(X;Y)=H(X)-H(X|Y).
\label{eq:mutual_information}
\end{equation}

Using standard entropy bounds, the conditional entropy satisfies

\begin{equation}
H(X|Y)\leq nH_b(\text{BER}),
\label{eq:entropy_bound}
\end{equation}

where $H_b(\cdot)$ denotes the binary entropy function. This yields

\begin{equation}
I(X;Y)\geq n\left(1-H_b(\text{BER})\right),
\label{eq:mi_bound}
\end{equation}

demonstrating that minimizing BER directly maximizes a lower bound on the recoverable mutual information \cite{shannon1948mathematical,holevo1998capacity}. Thus, improved decoding accuracy corresponds directly to improved information recovery capability after transmission through the noisy quantum channel.

The optimization of the receiver parameters is performed using classical optimization algorithms such as SPSA, ADAM, and COBYLA \cite{spall1998overview,kingma2014adam,powell1994direct}. SPSA is particularly suitable for noisy variational quantum optimization because it estimates gradients using stochastic perturbations and requires only two objective evaluations per iteration, making it computationally efficient and robust to finite-shot noise and statistical measurement fluctuations even in high-dimensional parameter spaces \cite{spall1998overview}. ADAM employs adaptive first- and second-moment gradient estimates to stabilize parameter updates and accelerate convergence in relatively smooth optimization landscapes \cite{kingma2014adam}, although its performance can degrade as gradient noise increases in larger systems. COBYLA performs derivative-free optimization using local linear approximations of the objective function \cite{powell1994direct}, but its effectiveness decreases in noisy high-dimensional landscapes where accurate local approximations become more difficult to construct. In larger variational quantum systems, deeper circuits and increasing parameter dimensionality can also lead to highly non-convex optimization landscapes and barren plateau behavior, where gradient magnitudes become exponentially suppressed, thereby reducing trainability and slowing convergence \cite{mcclean2018barren,cerezo2021variational}.

Overall, the proposed VQC-based receiver provides a flexible and scalable framework for decoding in quantum MIMO systems. By combining physically motivated ansatz design, MMSE-inspired optimization, and noise-aware training strategies, the receiver can adaptively learn channel-induced transformations while remaining robust to measurement uncertainty and realistic hardware noise. Rather than relying on explicit channel inversion, which becomes impractical in non-unitary noisy environments, the proposed framework performs statistical reconstruction directly through variational optimization, making it well suited for realistic noisy intermediate-scale quantum (NISQ) communication systems \cite{preskill2018nisq,cerezo2021variational}.

\subsection{Information-Theoretic Limits of the Receiver}

To quantify the fundamental constraints on the VQC-based receiver described in Sec.~\ref{sec:receiver}, we analyze the contraction of quantum-state distinguishability under noisy entangling evolution. The objective of this analysis is not to derive an exact closed-form bit-error rate (BER) expression for arbitrary noise realizations, but rather to establish the dominant asymptotic scaling behavior of decoding performance as the system dimension and circuit complexity increase~\cite{nielsen2010quantum,wilde2013quantum}. The derivation proceeds under four working conditions: $(i)$ the effective per-gate error probability satisfies $p_{\mathrm{eff}} \ll 1$, so that individual noise contributions are perturbative; $(ii)$ successive noise processes accumulate independently at leading order; $(iii)$ all hardware imperfections---single-qubit depolarizing errors, two-qubit gate infidelities, thermal relaxation, and readout asymmetry---are absorbed into the single aggregate parameter $p_{\mathrm{eff}}$; and $(iv)$ the variational ansatz employs pairwise entangling interactions whose total gate count scales as $N_{\mathrm{gates}} \sim c\,n^2$.

Let $\rho_i$ and $\rho_j$ denote two candidate transmitted quantum codewords after propagation through the channel. In the absence of noise, channel evolution is unitary, $\rho \mapsto U\rho U^\dagger$, and unitary operators are isometries of the Schatten $1$-norm~\cite{nielsen2010quantum}, so the geometric separation between states is perfectly preserved:

\begin{equation}
D\!\left(U\rho_i U^\dagger,\, U\rho_j U^\dagger\right)
=
D(\rho_i,\rho_j).
\label{eq:unitary_iso}
\end{equation}

However, under realistic NISQ conditions channel evolution is non-unitary and is described instead by a completely positive trace-preserving (CPTP) map $\mathcal{E}$, so that the received states are

\begin{equation}
\rho_i^{\,\mathrm{out}}
=
\mathcal{E}(\rho_i),
\qquad
\rho_j^{\,\mathrm{out}}
=
\mathcal{E}(\rho_j).
\end{equation}

The distinguishability between the resulting states is quantified through the trace distance

\begin{equation}
D(\rho_i,\rho_j)
=
\frac{1}{2}
\
\rho_i-\rho_j
\|_1,
\label{eq:44}
\end{equation}

which determines the maximum discrimination capability achievable under optimal quantum measurements~\cite{nielsen2010quantum,helstrom1976quantum}.

Since completely positive trace-preserving (CPTP) maps are contractive under the trace norm, the noisy channel satisfies

\begin{equation}
D(\mathcal{E}(\rho_i),\mathcal{E}(\rho_j))
\leq
D(\rho_i,\rho_j).
\label{eq:45}
\end{equation}

Equation~(\ref{eq:45}) implies that successive noisy operations progressively reduce the separation between candidate codewords. In high-dimensional quantum MIMO systems, this contraction becomes increasingly significant because the received states evolve within a Hilbert space of dimension $2^n$. As the number of qubits increases, stochastic perturbations generated by the noisy channel induce diffusion of the transmitted ensemble toward the maximally mixed state

\begin{equation}
\rho_{\mathrm{mix}}
=
\frac{I}{2^n},
\label{eq:46}
\end{equation}

thereby reducing the statistical distinguishability required for reliable decoding~\cite{nielsen2010quantum,wilde2013quantum}.

The contraction of quantum-state distinguishability is fundamentally governed by the density of entangling operations within the variational ansatz. As established in Section II-A, the pairwise interaction structure of the receiver leads to a gate complexity scaling of

\begin{equation}
N_{\mathrm{gates}}
\sim
cn^2,
\label{eq:47}
\end{equation}

where $n$ denotes the number of qubits and $c > 0$ depends on the ansatz depth and entangling topology. In this high-dimensional regime, the non-local propagation of local perturbations through successive entangling layers causes initially localized errors to evolve into correlated multi-qubit errors~\cite{preskill2018nisq}. Consequently, the effective noise parameter $p_{\mathrm{eff}}$ introduced in Section III-B to aggregate gate-level and decoherence effects scales the global system degradation according to the total circuit volume.

Assuming approximately independent accumulation of these noise processes across successive quantum operations, the overlap fidelity between the ideal and noisy states is expected to decay approximately according to the total circuit volume. Under the weak-noise assumption $p_{\mathrm{eff}} \ll 1$, higher-order correlations between individual error channels contribute only perturbatively at leading order, allowing the global fidelity reduction to be modeled through a multiplicative accumulation process~\cite{preskill2018nisq}. Consequently, the effective fidelity scaling may be approximated as

\begin{equation}
F
=
\mathrm{Tr}(\rho_{\mathrm{ideal}}\rho_{\mathrm{out}})
\approx
(1-p_{\mathrm{eff}})^{N_{\mathrm{gates}}},
\label{eq:48}
\end{equation}

Substituting the structural scaling relation of the ansatz yields

\begin{equation}
F
\approx
(1-p_{\mathrm{eff}})^{cn^2},
\label{eq:49}
\end{equation}

For weak-noise regimes satisfying $p_{\mathrm{eff}} \ll 1$, the approximation

\begin{equation}
(1-p)^N
\approx
e^{-Np}
\label{eq:50}
\end{equation}

gives the asymptotic fidelity behavior

\begin{equation}
F
\approx
e^{-cn^2p_{\mathrm{eff}}},
\label{eq:51}
\end{equation}

revealing an exponential decay of fidelity with the combined parameter $n^2p_{\mathrm{eff}}$. The quadratic dependence on $n$ originates from the scaling behavior of correlated entangling interactions within the variational ansatz, while the exponential form captures the cumulative effect of decoherence and gate-level noise across the circuit depth.

The relation in (\ref{eq:51}) characterizes the asymptotic decay of quantum-state overlap under cumulative noisy evolution rather than an exact closed-form description of all channel realizations. More precisely, the fidelity and trace distance are related through the Fuchs--van de Graaf inequalities, which provide upper and lower bounds on quantum-state distinguishability~\cite{fuchs1999cryptographic,nielsen2010quantum}. Although these relations are not exact equalities, they preserve the dominant exponential dependence in the weak-noise asymptotic regime. Consequently, the distinguishability between candidate output states may be modeled to leading order as

\begin{equation}
D(\rho_i^{\mathrm{out}},\rho_j^{\mathrm{out}})
\sim
e^{-cn^2p_{\mathrm{eff}}},
\label{eq:51a}
\end{equation}

which captures the effective contraction of accessible quantum information induced by cumulative entangling noise and decoherence throughout the receiver circuit.

By the Fuchs--van de Graaf inequalities, the trace distance between candidate output states is bounded by the corresponding fidelity decay~\cite{nielsen2010quantum}. Consequently, the exponential reduction in fidelity predicted by (\ref{eq:51}) implies an exponential contraction in the distinguishability of the transmitted codewords as the system dimension increases.

The decoding performance is therefore expected to be fundamentally constrained by the Helstrom bound, which specifies the minimum achievable discrimination error between two quantum states,

\begin{equation}
P_e
=
\frac{1}{2}
\left(
1-D(\rho_0,\rho_1)
\right).
\label{eq:52}
\end{equation}

As the noisy channel progressively contracts the trace distance between transmitted states, the receiver is expected to approach the random-guessing limit. Consequently, the BER of the proposed quantum MIMO receiver is expected to obey the effective asymptotic scaling law

\begin{equation}
\mathrm{BER}(p,n)
\approx
\frac{1}{2}
\left(
1-e^{-cn^2p_{\mathrm{eff}}}
\right).
\label{eq:53}
\end{equation}

Equation~(\ref{eq:53}) predicts that the degradation of decoding performance is jointly governed by entanglement-induced error propagation and the contraction of quantum-state distinguishability under noisy channel evolution. In particular, the quadratic dependence on $n$ emerges from the scaling behavior of correlated entangling interactions, while the exponential term reflects the progressive loss of accessible information caused by cumulative decoherence and depolarization throughout the circuit~\cite{wilde2013quantum,preskill2018nisq}.

The analytical scaling law derived in (\ref{eq:53}) serves as the theoretical benchmark for the numerical simulations and variational training results presented in the following sections.

\section{Results}\label{SecIV}
\subsection{BER Performance Analysis and Theoretical Validation Across Noise Regimes}
\begin{figure*}[!t]
\centering
\begin{subfigure}[t]{0.48\textwidth}
\centering
\includegraphics[width=\linewidth]{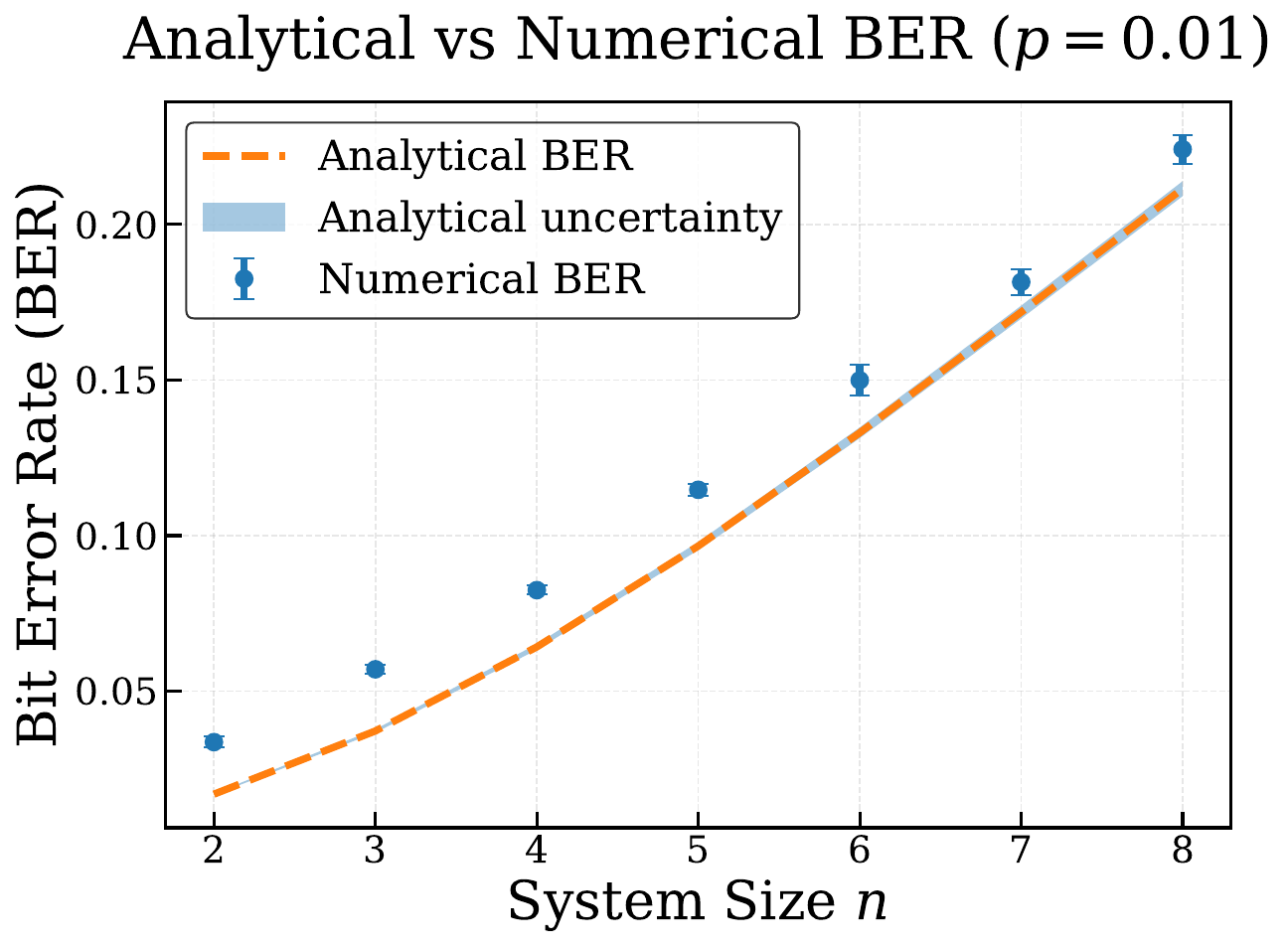}
\caption{}
\end{subfigure}
\hfill
\begin{subfigure}[t]{0.48\textwidth}
\centering
\includegraphics[width=\linewidth]{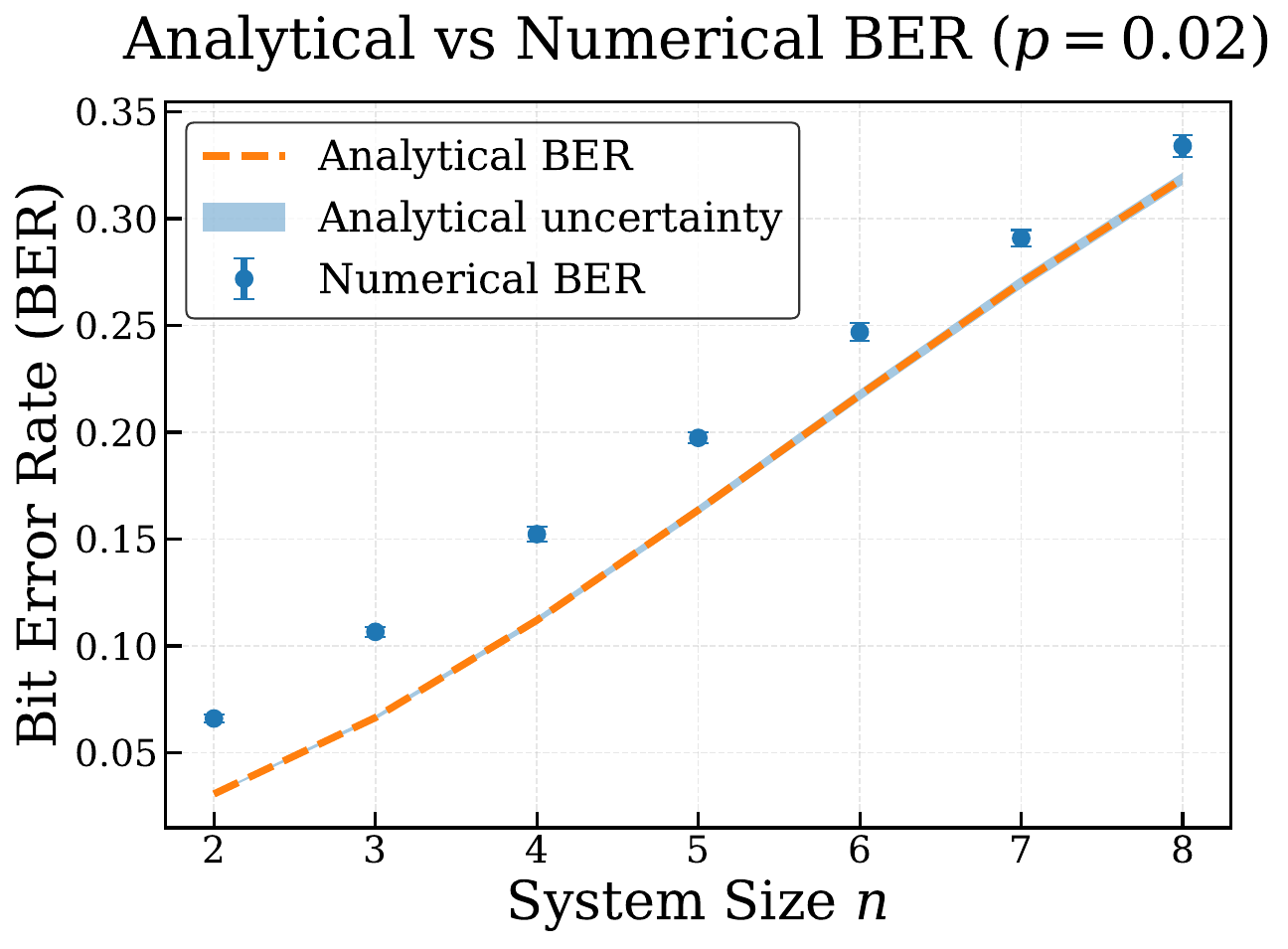}
\caption{}
\end{subfigure}

\vspace{0.3cm}

\begin{subfigure}[t]{0.48\textwidth}
\centering
\includegraphics[width=\linewidth]{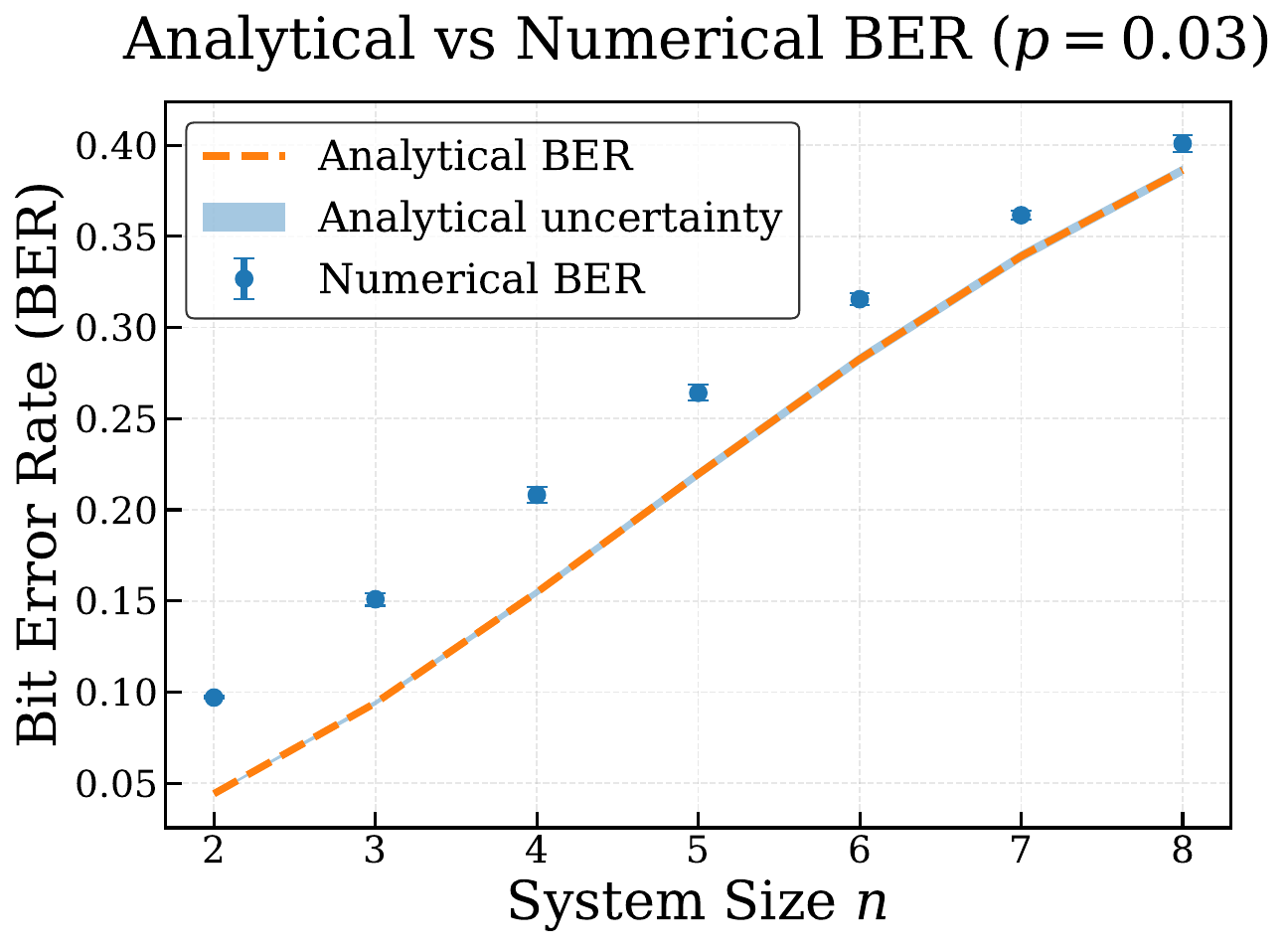}
\caption{}
\end{subfigure}
\hfill
\begin{subfigure}[t]{0.48\textwidth}
\centering
\includegraphics[width=\linewidth]{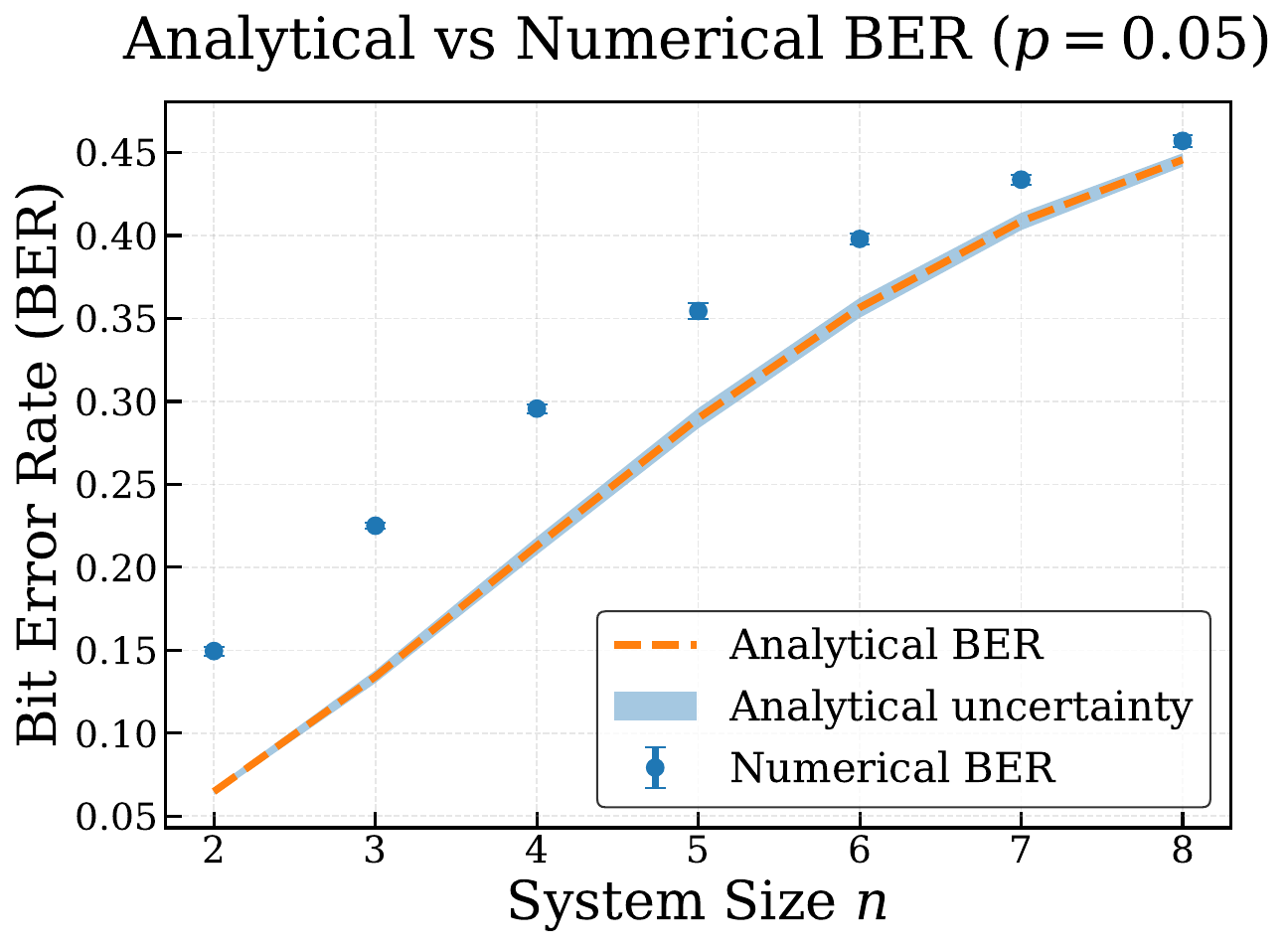}
\caption{}
\end{subfigure}

\caption{
Comparison between numerical BER obtained from noisy quantum simulations and the analytical scaling model for different noise strengths.
(a) $p=0.01$,
(b) $p=0.02$,
(c) $p=0.03$, and
(d) $p=0.05$.
}
\label{fig:ber_theory_validation}
\end{figure*}
Figure~\ref{fig:ber_theory_validation} reveal a systematic increase in BER with both system size and noise strength, indicating that decoding performance degrades as the quantum MIMO architecture becomes larger and more susceptible to noise accumulation. Across all investigated noise regimes, the BER exhibits a monotonic dependence on $n$, with the degradation becoming progressively more pronounced at higher values of $p$. The analytical model captures these trends remarkably well, reproducing both the dimensional scaling and the noise-dependent shift of the BER curves over the entire parameter range considered.

This behavior is consistent with the increasing complexity of the variational receiver as the system dimension grows. Higher-dimensional quantum MIMO systems require a larger number of variational degrees of freedom and quantum operations to process the transmitted states, thereby increasing the cumulative impact of decoherence, gate imperfections, and measurement noise. The resulting reduction in state distinguishability limits the receiver's ability to accurately recover the transmitted information, leading to the observed increase in decoding error probability.

The analytical model captures the observed BER scaling across all investigated noise regimes. At low noise ($p=0.01$), the numerical results closely follow the theoretical prediction over the entire range of system sizes, indicating that the dominant dimensional-scaling behavior is well described by the proposed framework. The narrow uncertainty intervals further suggest that statistical fluctuations contribute only weakly to the observed performance trends.

As the noise strength increases to $p=0.02$ and $p=0.03$, the overall BER shifts upward while preserving the same qualitative dependence on system size. The persistence of this scaling behavior indicates that the underlying mechanisms governing decoding performance remain unchanged across the intermediate-noise regime. The analytical predictions accurately track both the magnitude and growth rate of the BER, demonstrating that the model remains applicable beyond the weak-noise limit.

The strongest deviations from ideal performance occur at $p=0.05$, where BER growth with system size becomes most pronounced. Although modest departures between theory and simulation emerge for larger systems, the overall scaling behavior remains well captured by the analytical description. The observed deviations are confined to the high-noise regime and remain small compared with the overall variation induced by changes in system dimension and noise strength.

These trends are naturally explained by the fidelity-contraction relation
\begin{equation}
F \approx e^{-c n^{2} p_{\mathrm{eff}}},
\end{equation}
which predicts an exponential suppression of state fidelity as both system size and accumulated noise increase. The corresponding reduction in quantum-state distinguishability directly limits the ability of the receiver to infer the transmitted symbol sequence, yielding the BER scaling relation
\begin{equation}
\mathrm{BER}(p,n)
\approx
\frac{1}{2}
\left(
1-e^{-c n^{2} p_{\mathrm{eff}}}
\right),
\label{eq:ber_scaling_validation}
\end{equation}
The agreement between numerical and analytical results therefore indicates that the principal features of the decoding-performance landscape are governed by fidelity loss arising from cumulative noise accumulation. Moreover, the variation associated with repeated optimization runs remains substantially smaller than the systematic changes induced by increasing system size and noise strength, confirming that dimensional scaling constitutes the dominant factor controlling BER performance in the regime studied.

Overall, the results provide strong evidence that the proposed analytical BER model accurately describes the performance scaling of variational quantum MIMO MMSE decoding. The consistent agreement between theory and simulation validates the underlying assumptions of cumulative noise accumulation and fidelity degradation, while confirming that the model remains predictive across a broad range of system dimensions and physical noise strengths.

\subsection{Scaling-Law Validation via Linearized BER Analysis}
\begin{figure*}[!t]
\centering
\begin{subfigure}[t]{0.48\textwidth}
\centering
\includegraphics[width=\linewidth]{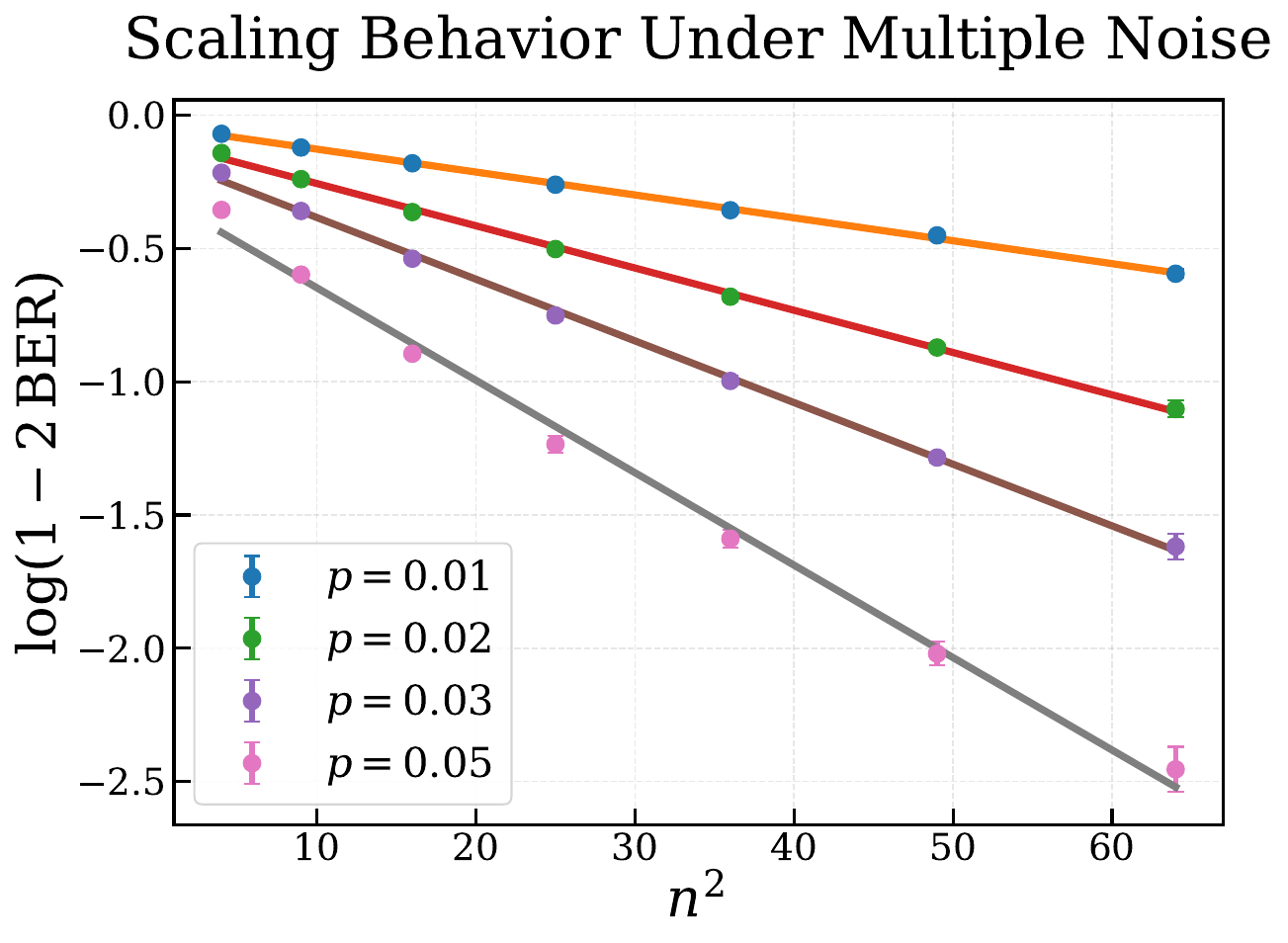}
\caption{Finite-size scaling transformation}
\end{subfigure}
\hfill
\begin{subfigure}[t]{0.48\textwidth}
\centering
\includegraphics[width=\linewidth]{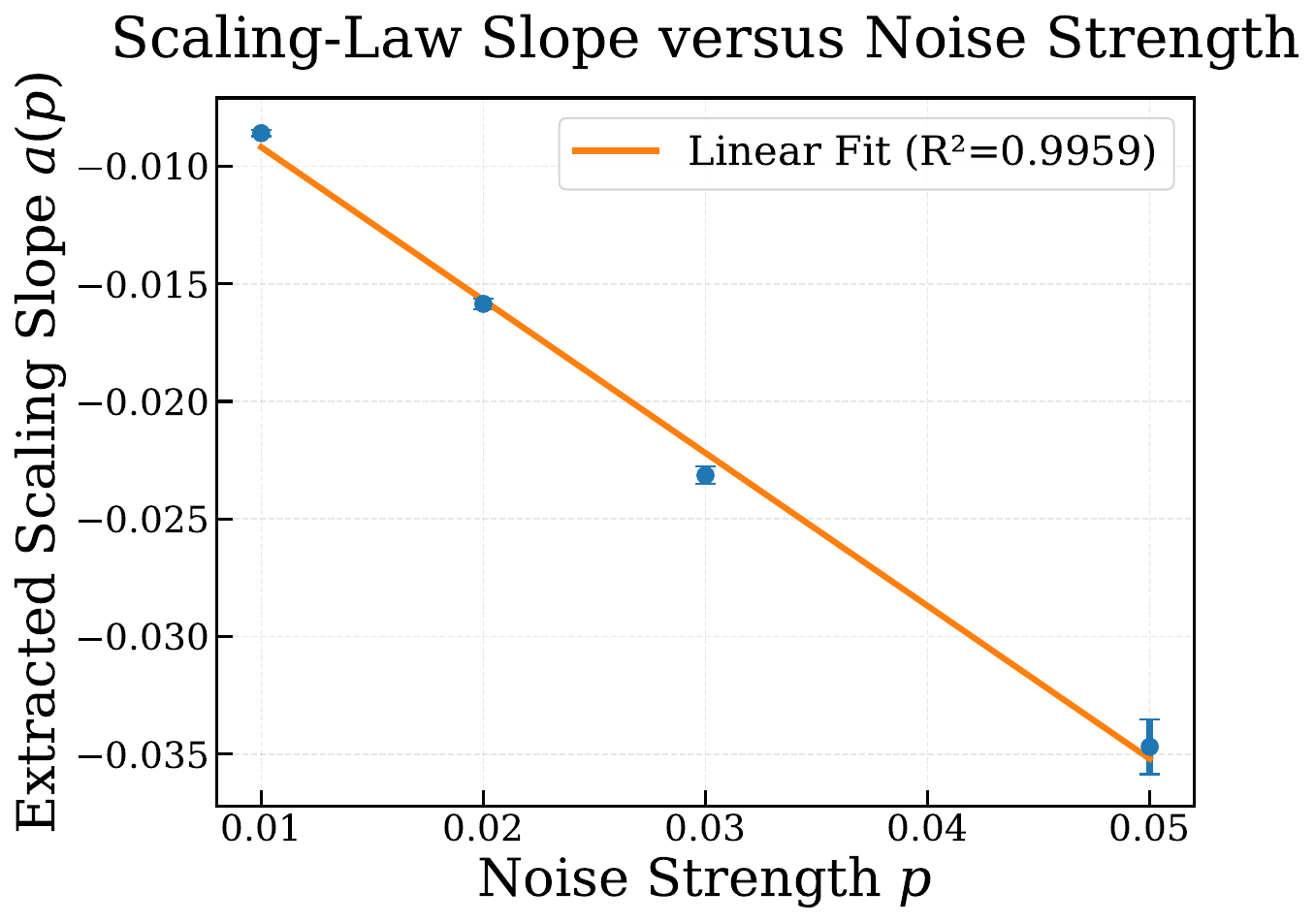}
\caption{Noise dependence of the scaling coefficient}
\end{subfigure}

\caption{
Quantitative validation of the proposed BER scaling law.
(a) Transformed BER quantity $\log(1-2\mathrm{BER})$ as a function of $n^{2}$ for noise strengths b) Extracted scaling coefficient as a function of $p$}
\label{fig:vber_scaling_validation}
\end{figure*}

To validate the proposed BER scaling relation, the transformed quantity $\log(1-2\,\mathrm{BER})$ was analyzed as a function of $n^{2}$ for system sizes ranging from $n=2$ to $n=8$ across multiple noise regimes. The transformed data exhibit an approximately linear dependence on $n^{2}$ for all investigated values of $p$, indicating that the exponential scaling behavior predicted by Eq.~(\ref{eq:ber_scaling_validation}) remains valid throughout the parameter range considered. Increasing noise primarily shifts the curves toward larger negative values while preserving their overall linear structure, suggesting that the underlying scaling mechanism is robust against variations in noise strength.

A systematic dependence on noise is evident from the evolution of the fitted slopes. As $p$ increases, the magnitude of the slope grows monotonically, implying a progressively faster degradation of decoding performance with system size. Importantly, no qualitative change in the scaling behavior is observed across the investigated noise regimes, indicating that the same physical mechanism governs BER growth from the weak- to intermediate-noise regime.

The observed scaling behavior can be understood from the cumulative effect of noise acting on an increasingly large number of interacting quantum degrees of freedom. In the proposed quantum MIMO model, the channel unitary contains pairwise interaction terms whose total number grows as $\mathcal{O}(n^{2})$. Each interaction contributes to the overall entanglement structure of the transmitted state, but also provides an additional pathway through which noise-induced perturbations can propagate. As the system dimension increases, local decoherence processes no longer affect qubits independently; instead, their impact is amplified through the network of coherent interactions, leading to a collective reduction in the fidelity of the received quantum state. Consequently, the probability of successful recovery decreases approximately exponentially with the number of effective interaction channels, naturally producing the observed linear dependence of $\log(1-2\,\mathrm{BER})$ on $n^{2}$.

Figure~\ref{fig:vber_scaling_validation} further shows that the extracted scaling coefficient increases approximately linearly with noise strength. The corresponding fit yields a high coefficient of determination ($R^{2}=0.9959$), indicating a strong and systematic dependence on $p$. This near-linear behavior suggests that, within the investigated parameter regime, the dominant effect of noise is to introduce approximately independent fidelity losses at each interaction layer. Under such conditions, the total degradation accumulates proportionally to the physical noise strength, leading to a scaling coefficient that grows linearly with $p$. The absence of significant nonlinear corrections further indicates that the system remains below a regime where higher-order error correlations or noise-induced phase transitions become dominant.

The persistence of linearity in both $\log(1-2\,\mathrm{BER})$ versus $n^{2}$ and in the extracted scaling coefficient versus $p$ therefore provides evidence that the BER degradation is governed by a stable extensive mechanism associated with noise accumulation across pairwise quantum interactions. Noise modifies the rate at which errors accumulate but does not alter the fundamental scaling structure itself. This observation supports the validity of the proposed exponential scaling model and suggests that the derived relation captures an intrinsic property of the noisy quantum MIMO channel rather than a feature specific to a particular system size or noise realization.

\subsection{Optimization Performance}
\begin{figure*}[!t]
\centering
\begin{subfigure}[t]{0.48\textwidth}
\centering
\includegraphics[width=\linewidth]{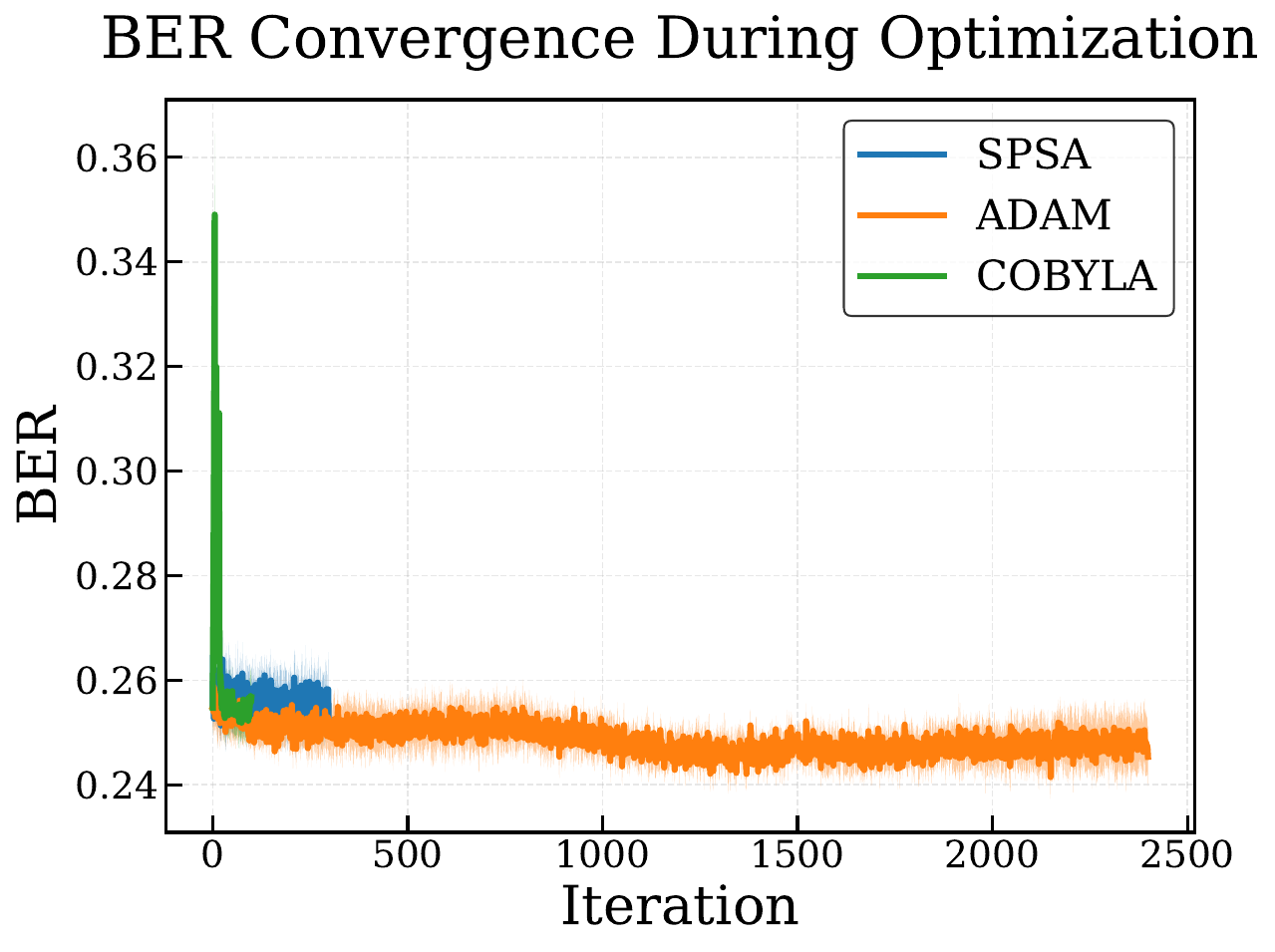}
\caption{BER convergence during optimization}
\end{subfigure}
\hfill
\begin{subfigure}[t]{0.48\textwidth}
\centering
\includegraphics[width=\linewidth]{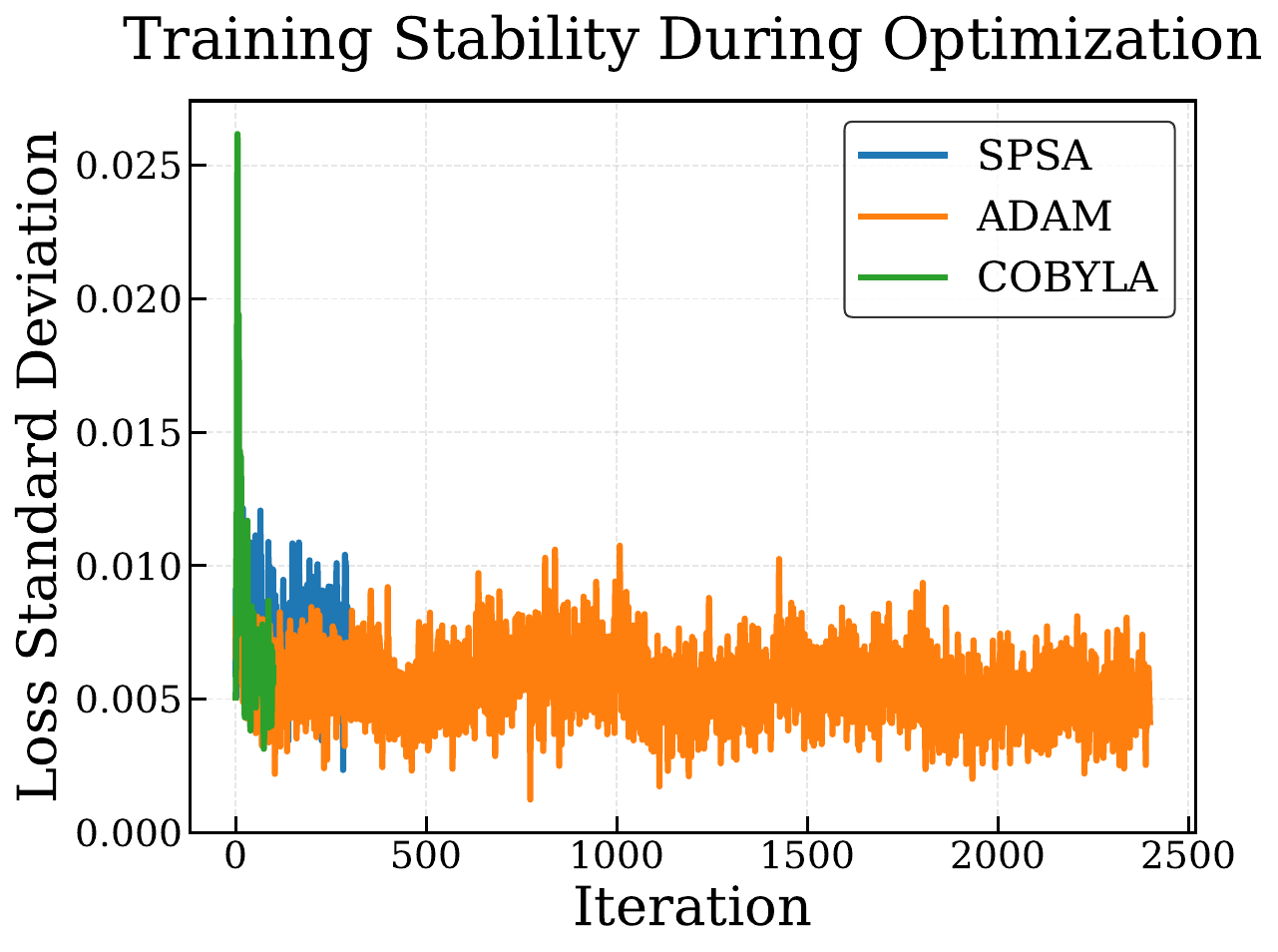}
\caption{Optimization stability during training}
\end{subfigure}

\caption{Optimization dynamics of the variational quantum receiver for SPSA, ADAM, and COBYLA. (a) Evolution of the mean BER during optimization. (b) Cross-initialization standard deviation of the loss as a function of optimization iteration.}
\label{fig:optimization_analysis}
\end{figure*}

\begin{table*}[!t]
\caption{Performance and robustness characteristics of the variational quantum receiver under SPSA, ADAM, and COBYLA optimization. The reported metrics assess decoding accuracy, convergence reproducibility, and sensitivity to initialization across independent optimization runs.
}
\label{tab:optimizer_comparison}
\centering
\begin{tabular}{lccccc}
\hline\hline
Optimizer & Final Mean BER & 95\% CI & Final Stability & Average Stability & Maximum Stability \\
\hline
SPSA   & 0.2527 & $\pm$ 0.0036 & $6.35\times10^{-3}$ & $7.23\times10^{-3}$ & $1.22\times10^{-2}$ \\
ADAM   & 0.2454 & $\pm$ 0.0056 & $4.14\times10^{-3}$ & $5.52\times10^{-3}$ & $1.07\times10^{-2}$ \\
COBYLA & 0.2567 & $\pm$ 0.0048 & $6.18\times10^{-3}$ & $7.37\times10^{-3}$ & $2.62\times10^{-2}$ \\
\hline\hline
\end{tabular}
\end{table*}

The optimization behavior of the variational receiver provides a useful probe of the structure of the decoding landscape and its susceptibility to stochastic noise. To investigate these properties, SPSA, ADAM, and COBYLA were evaluated over ten independent random initializations. The resulting optimization dynamics are shown in Fig.~\ref{fig:optimization_analysis}, while the corresponding performance and robustness metrics are summarized in Table~\ref{tab:optimizer_comparison}.

Although all three optimization strategies converge to similar BER values, ADAM consistently attains the lowest average BER and exhibits the smallest variability across independent initializations. The modest spread in both the final BER and stability metrics indicates that the optimization outcome is relatively insensitive to the choice of initialization within the parameter regime considered. This behavior suggests that the variational receiver can be trained reliably despite the stochasticity introduced by finite-shot measurements and hardware noise.

The improved performance of ADAM is originates from its adaptive gradient-scaling mechanism, which enables more effective navigation of noisy objective functions than either SPSA or COBYLA. In particular, the reduced cross-initialization variability observed for ADAM implies that gradient information remains sufficiently informative throughout the optimization process, allowing the algorithm to converge reproducibly toward high-quality decoding solutions. By contrast, the larger fluctuations observed for SPSA and COBYLA reflect a greater sensitivity to statistical noise in the cost-function evaluations, leading to less consistent optimization trajectories.

More significantly, the observed optimization behavior provides evidence for the trainability of the proposed variational quantum receiver. Despite the presence of channel-induced multipartite correlations, finite-shot sampling effects, and realistic noise processes, all three optimization strategies converge to comparable decoding performance with relatively small variability across independent initializations. This consistency indicates that the optimization process remains robust within the parameter regime considered and is not strongly dependent on either the choice of optimizer or the initial parameter configuration. 

Furthermore, the ability of different optimization methods to achieve similar BER values suggests that the variational ansatz is capable of representing decoding operations that effectively mitigate the channel-induced distortions. Consequently, for the system sizes investigated here, the optimization procedure itself does not appear to be the dominant factor limiting decoding performance. Instead, the observed BER is likely influenced more strongly by the representational capacity of the receiver ansatz and the complexity of the quantum correlations generated during channel evolution.

Overall, these observations suggest that the proposed variational quantum receiver possesses a favorable optimization landscape characterized by reproducible convergence, moderate sensitivity to initialization, and resilience against stochastic noise. Such properties are essential for the scalability of quantum machine-learning-based communication receivers, where reliable training becomes increasingly important as channel complexity and system dimensionality grow.

\subsection{Mutual Information Analysis}
\begin{figure*}[!t]
\centering
\begin{subfigure}[t]{0.48\textwidth}
\centering
\includegraphics[width=\linewidth]{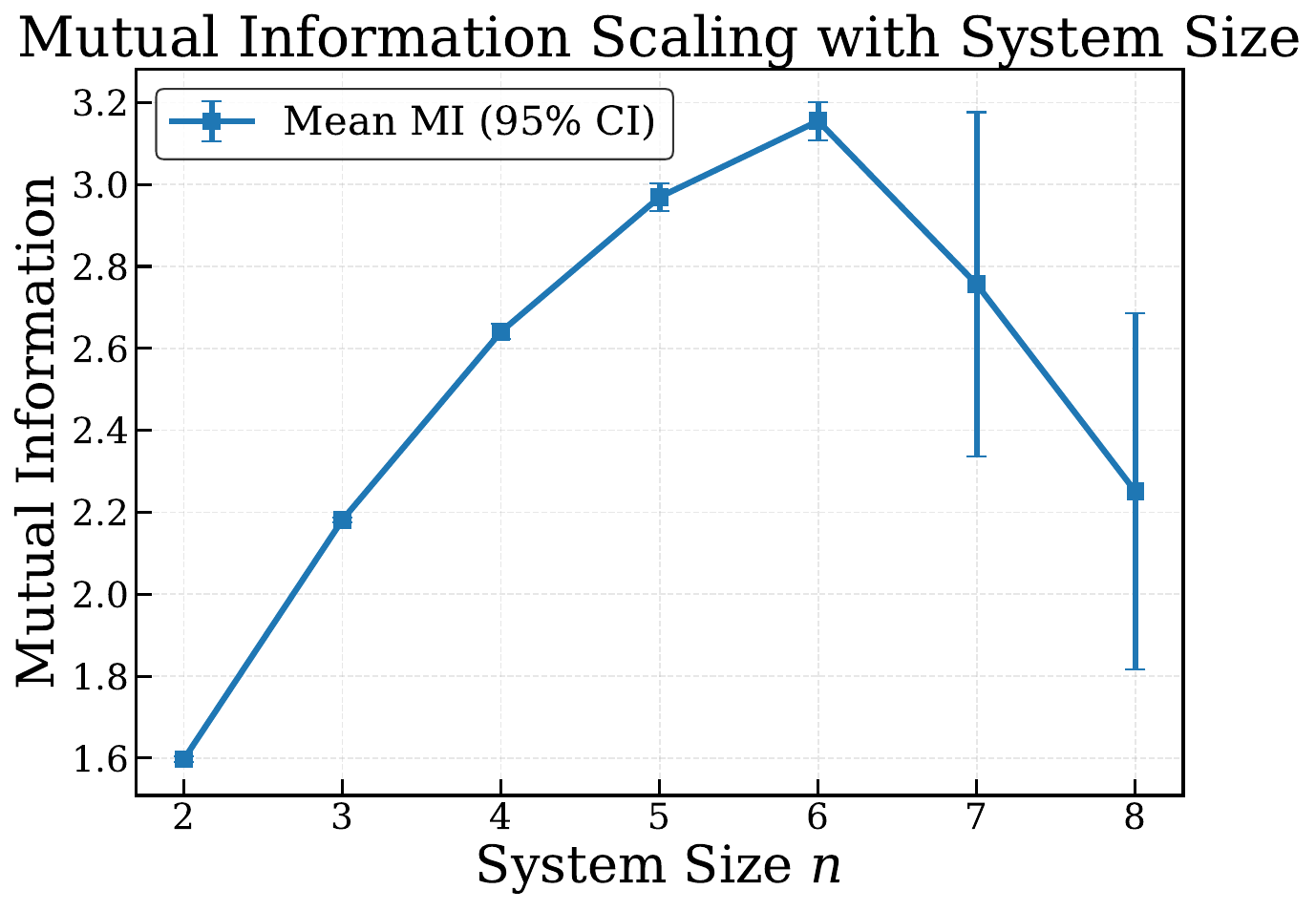}
\caption{Mutual information versus system size}
\label{fig:mi_vs_n}
\end{subfigure}
\hfill
\begin{subfigure}[t]{0.48\textwidth}
\centering
\includegraphics[width=\linewidth]{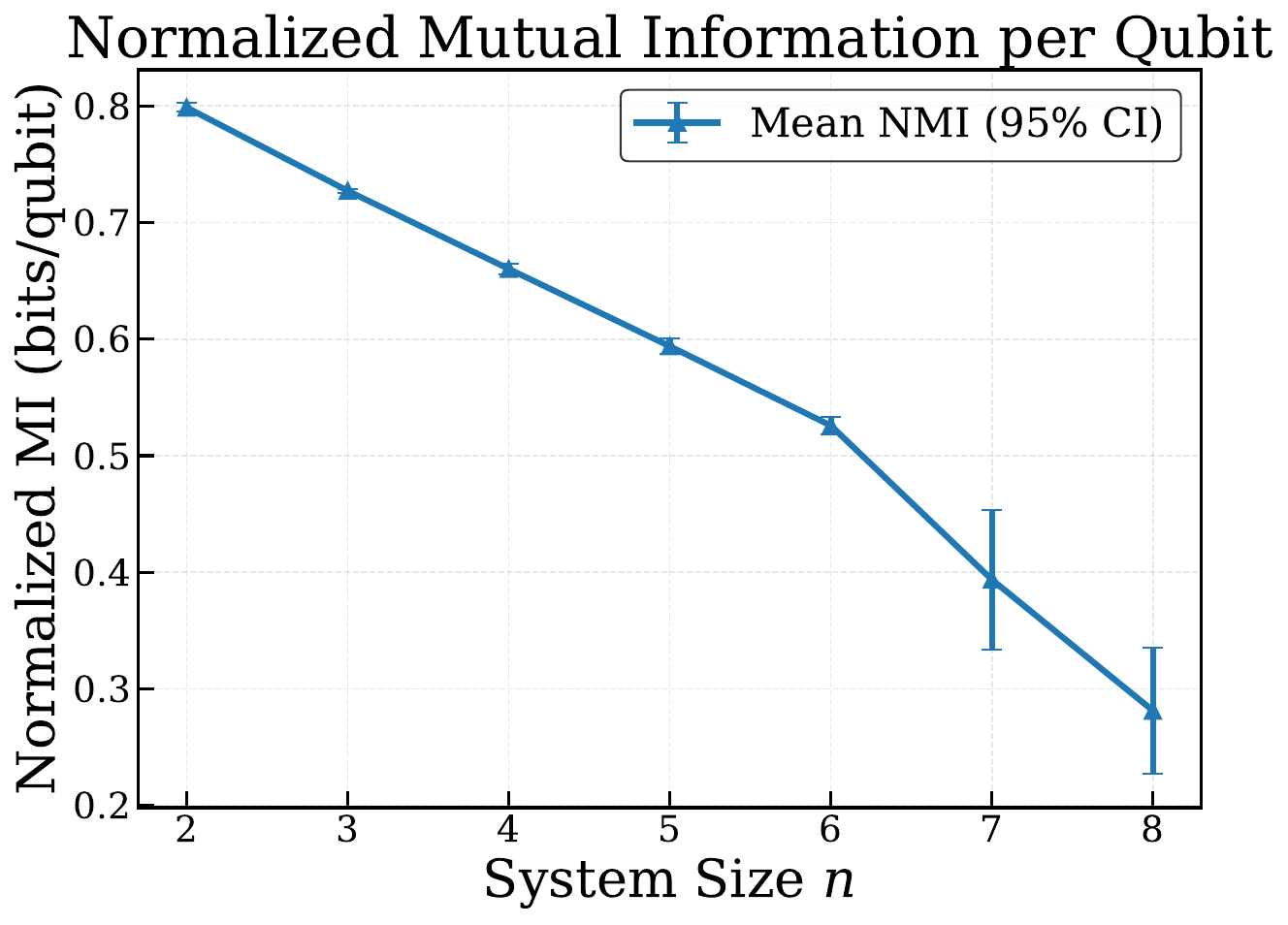}
\caption{Normalized mutual information versus system size}
\label{fig:nmi_vs_n}
\end{subfigure}
\caption{Mutual information characteristics of the variational quantum receiver as a function of system size $n$, evaluated at a fixed noise level of $0.01$.}
\label{fig:mi_statistics}
\end{figure*}
To further assess the communication performance of the proposed receiver, the mutual information was evaluated across different system sizes.Mutual information quantifies the amount of information recovered from the received quantum states and therefore provides an information-theoretic measure of receiver performance.Examining its dependence on system size therefore enables an assessment of how effectively information is preserved as the dimensionality and complexity of the quantum MIMO system increase.

Figure~\ref{fig:mi_vs_n} shows the average mutual information as a function of system size. For system sizes ranging from $n=2$ to $n=6$, mutual information increases, indicating that larger quantum MIMO systems are capable of supporting greater total information transfer. However, the rate of increase gradually diminishes with increasing system size. Beyond $n=6$, the mutual information decreases. This reduction coincides with the increase in BER observed for larger systems, suggesting that the receiver becomes increasingly affected by channel complexity and detection errors as the system dimension grows.

To facilitate meaningful comparisons across different system sizes, the mutual information was further normalized by the number of transmitted qubits,
\begin{equation}
I_{\mathrm{norm}}=\frac{I}{n},
\end{equation}
where $I$ denotes the mutual information and $n$ is the number of transmitted qubits. This normalized mutual information (NMI) provides a measure of communication efficiency on a per-qubit basis.

The resulting NMI values are shown in Fig.~\ref{fig:nmi_vs_n}. In contrast to the total mutual information, the normalized mutual information decreases monotonically with increasing system size. This trend indicates that although larger systems initially support greater aggregate information transfer, the information preserved per transmitted qubit decreases as the system dimension increases. Consequently, the growth of total mutual information is sublinear with respect to system size, implying a gradual reduction in communication efficiency as channel coupling and interference become more pronounced.

The statistical analysis further shows that the variability of both MI and NMI remains relatively small for system sizes up to $n=6$, indicating stable receiver performance across different random channel realizations. For $n=7$ and $n=8$, the confidence intervals increase substantially, reflecting greater sensitivity to channel variations and a broader distribution of achievable information-transfer performance. This increased variability coincides with the observed rise in BER, suggesting that larger system dimensions place the variational quantum receiver in a more challenging operating regime where performance becomes increasingly dependent on the specific channel realization.

The observed behavior reflects the increasing complexity of the quantum MIMO channel with system size. While additional qubits provide more communication degrees of freedom and initially increase the total mutual information, they also introduce stronger channel interactions and interference. As a result, the information contributed by each additional qubit decreases, leading to a monotonic reduction in the normalized mutual information. For larger systems, these effects become sufficiently pronounced to limit the overall information preserved by the receiver, resulting in the decline in mutual information observed for $n \geq 7$.

Overall, the combined MI and NMI analysis demonstrates that the proposed receiver continues to preserve a substantial amount of information as the system size increases, while also revealing a progressive reduction in information efficiency per qubit. These results provide an information-theoretic perspective on the scalability of the receiver that complements the BER and entanglement-based analyzes presented in the preceding sections.

\subsection{Robustness Analysis Under Channel Mismatch}
\begin{figure*}[!t]
\centering
\begin{subfigure}[t]{0.48\textwidth}
\centering
\includegraphics[width=\linewidth]{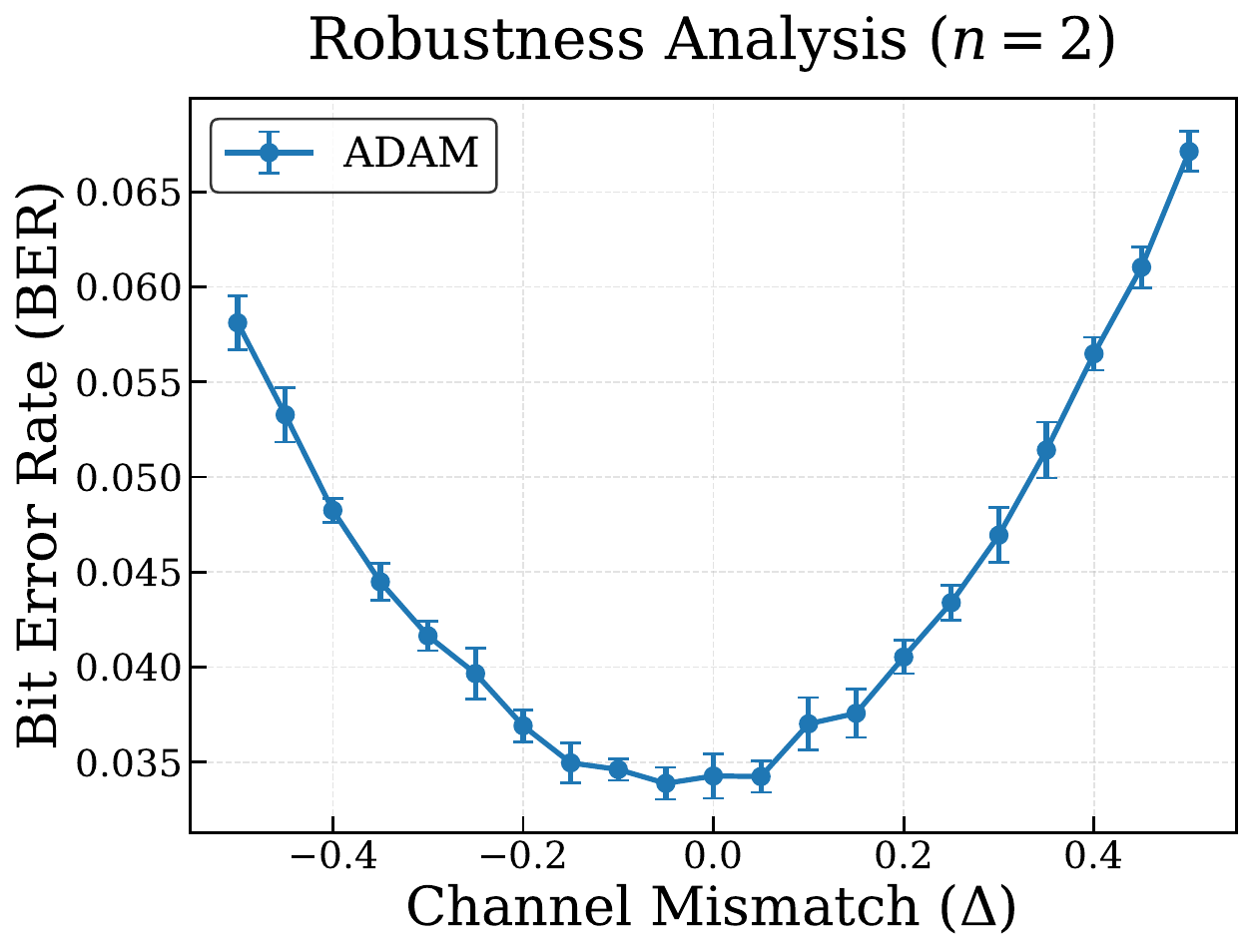}

\end{subfigure}
\hfill
\begin{subfigure}[t]{0.48\textwidth}
\centering
\includegraphics[width=\linewidth]{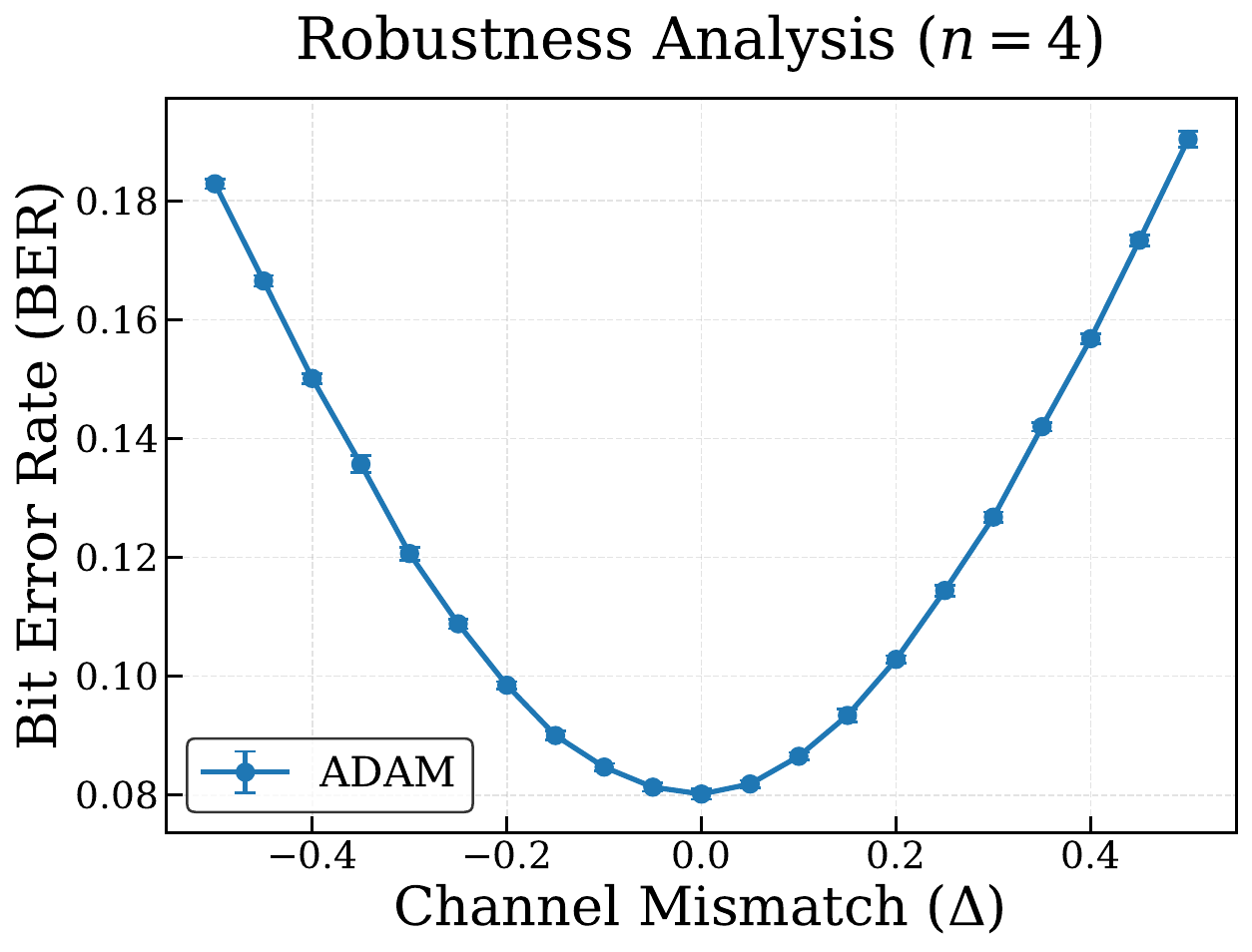}

\end{subfigure}
\vspace{-0.15cm}
\begin{subfigure}[t]{0.48\textwidth}
\centering
\includegraphics[width=\linewidth]{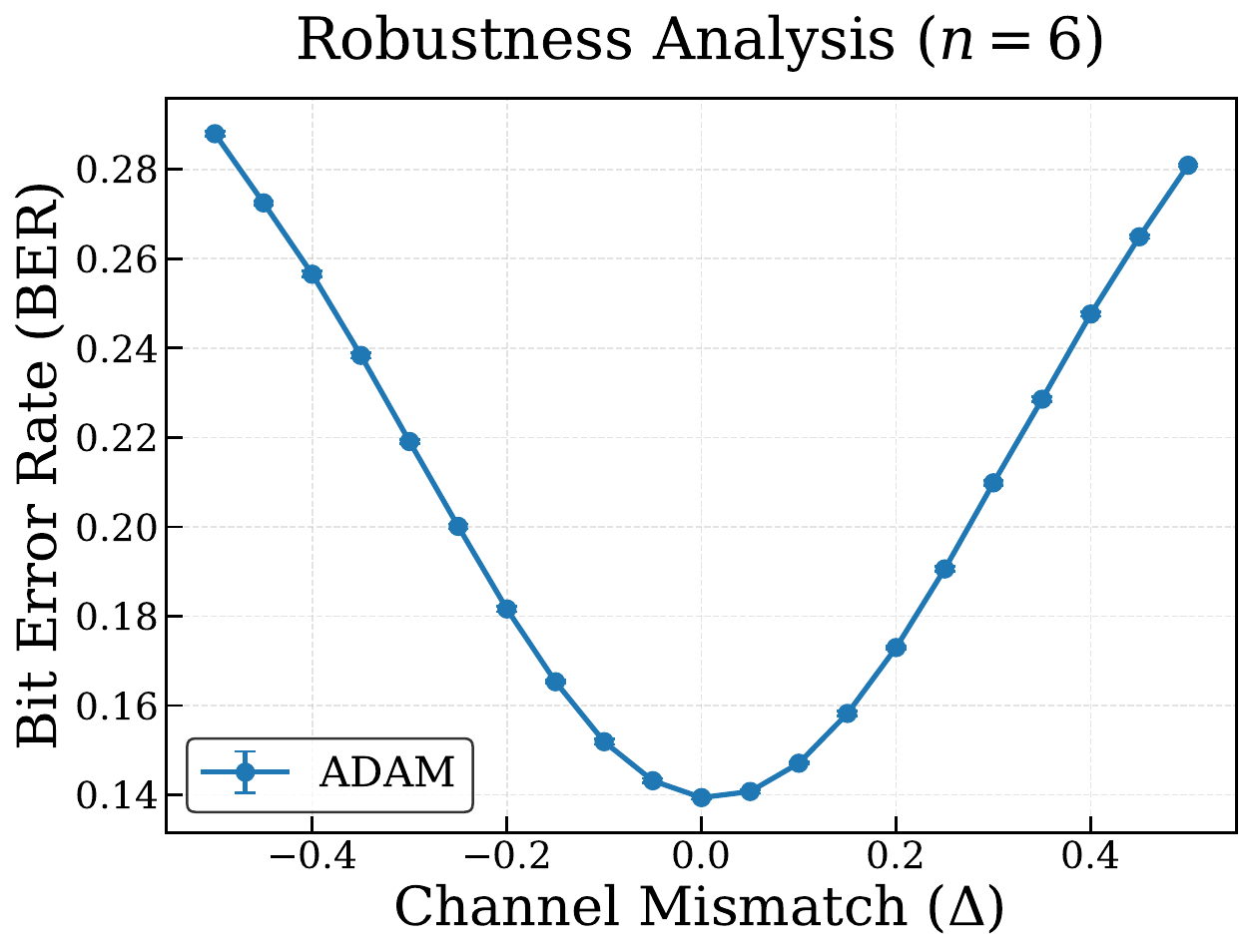}

\end{subfigure}
\hfill
\begin{subfigure}[t]{0.48\textwidth}
\centering
\includegraphics[width=\linewidth]{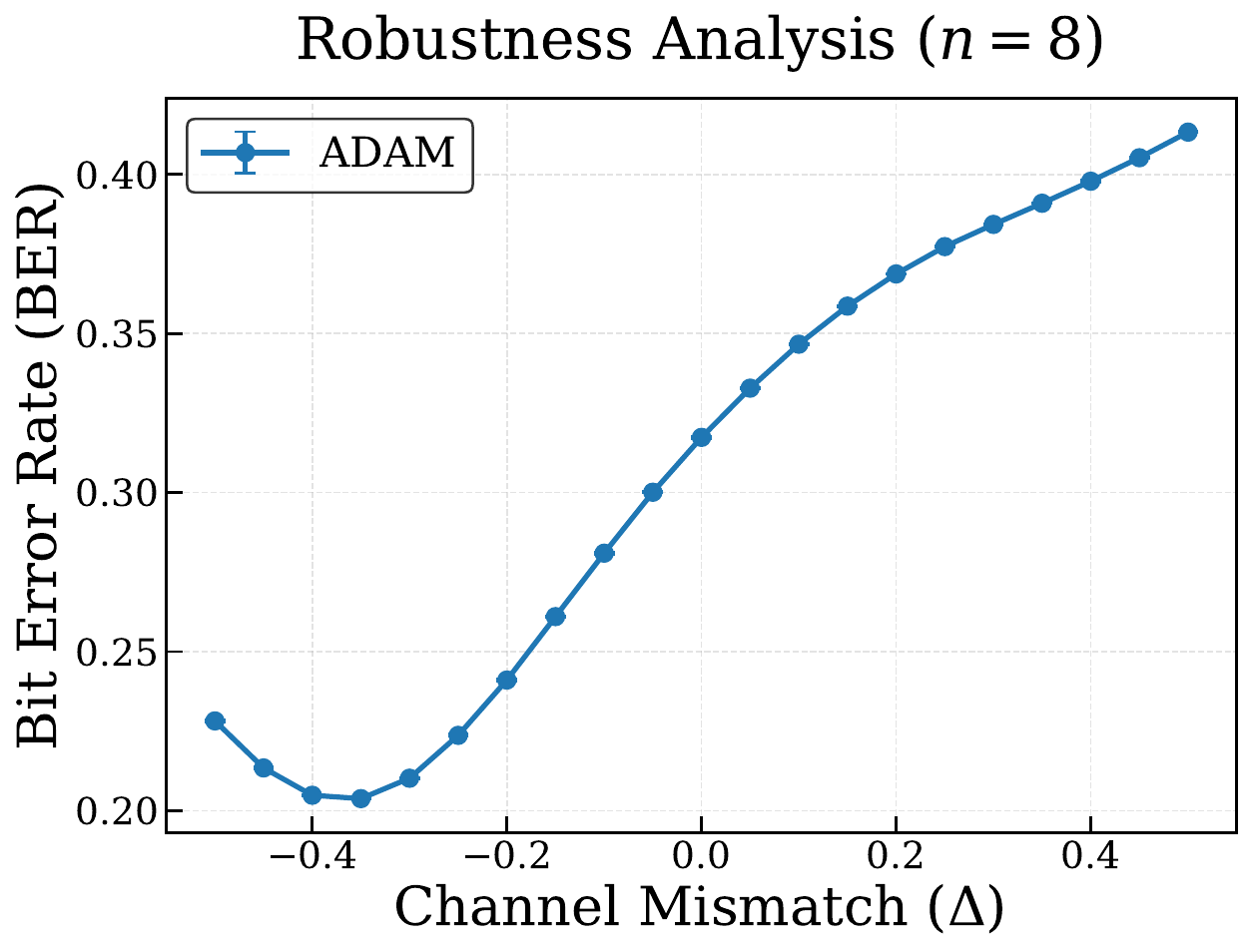}

\end{subfigure}
\caption{:BER performance under channel mismatch for different system dimensions$(n=2,4,6,8)$ using the
ADAM optimizer.}
\label{fig:rber_scaling}
\end{figure*}

The robustness of the proposed variational quantum circuit (VQC)-based MIMO receiver is evaluated under imperfect channel-state information by introducing a channel mismatch parameter $(\Delta)$, which quantifies the deviation between the actual channel interaction and the channel model assumed during decoding. The resulting BER performance for system dimensions $(n=2,4,6,8)$ is shown in Fig.~\ref{fig:rber_scaling}.

In practical quantum communication systems, perfect channel knowledge is rarely available due to estimation errors, calibration imperfections, device fluctuations, and environmental noise. Consequently, the inverse channel operation implemented by the receiver may differ from the true channel transformation. Let the actual channel be represented by the unitary operator $(U(\theta))$, while the receiver assumes the perturbed model $(U(\theta+\Delta))$. The mismatch parameter($\Delta$) therefore characterizes the uncertainty in the channel model used during decoding.

For all investigated system dimensions, the BER exhibits a smooth dependence on the mismatch parameter, with the minimum error occurring near the nominal operating point. This behavior indicates that the learned variational decoder remains well aligned with the underlying channel model and does not experience abrupt performance degradation under moderate perturbations. The absence of sharp transitions or discontinuities demonstrates that the receiver operates within a stable region of the variational parameter landscape and retains reliable decoding capability even when the channel estimate is imperfect.

The observed BER degradation can be understood from the reduction in reconstruction fidelity caused by channel mismatch. When the assumed channel differs from the true channel, the variational recovery circuit no longer implements the optimal inverse transformation. As a result, residual correlations and imperfect interference cancellation remain in the recovered quantum state, reducing state distinguishability at measurement and increasing the probability of decoding errors. Since the mismatch is introduced gradually, the resulting BER increase is also gradual, producing the smooth profiles observed in Fig.~\ref{fig:rber_scaling}.

For the small to moderate system dimensions $(n=2,4,6)$, the BER curves are nearly symmetric around $(\Delta=0)$, and the error probability varies only moderately across the entire mismatch interval. This behavior suggests that the variational decoder captures the dominant channel structure effectively and remains relatively insensitive to moderate overestimation or underestimation of the channel parameters. The low-dimensional Hilbert space and reduced number of entangling interactions contribute to a smoother optimization landscape and improved robustness against model uncertainties.

For the largest system size ((n=8)), asymmetry appears in the BER curve, and the minimum BER is shifted away from the exact zero-mismatch condition toward a small negative value of $(\Delta)$. This shift suggests that the optimized variational decoder is not perfectly centered around the nominal channel model. In larger quantum systems, the increased number of trainable parameters, stronger multi-qubit correlations, and more complex optimization landscape can introduce small biases in the learned recovery transformation. Consequently, the empirically optimal decoding condition may occur at a slightly perturbed channel parameter rather than the ideal matched-channel case.

Despite this increased asymmetry, the BER degradation remains gradual across the entire mismatch range. Even for $(n=8)$, no evidence of catastrophic performance collapse is observed, indicating that the receiver maintains stable operation under channel uncertainty. This controlled degradation is particularly important for realistic quantum communication systems, where channel estimates are inevitably imperfect and may vary over time.

Overall, the BER-versus-mismatch analysis demonstrates that the proposed VQC-based quantum MIMO receiver exhibits strong robustness against moderate channel-estimation errors. Although the BER increases with both mismatch magnitude and system dimension, the smooth error profiles and absence of abrupt performance deterioration indicate that the variational decoding strategy remains reliable under realistic imperfect-channel conditions.

\subsubsection{BER Sensitivity Coefficient Analysis}

\begin{figure*}[!t]
\centering
\begin{subfigure}[t]{0.48\textwidth}
\centering
\includegraphics[width=\linewidth]{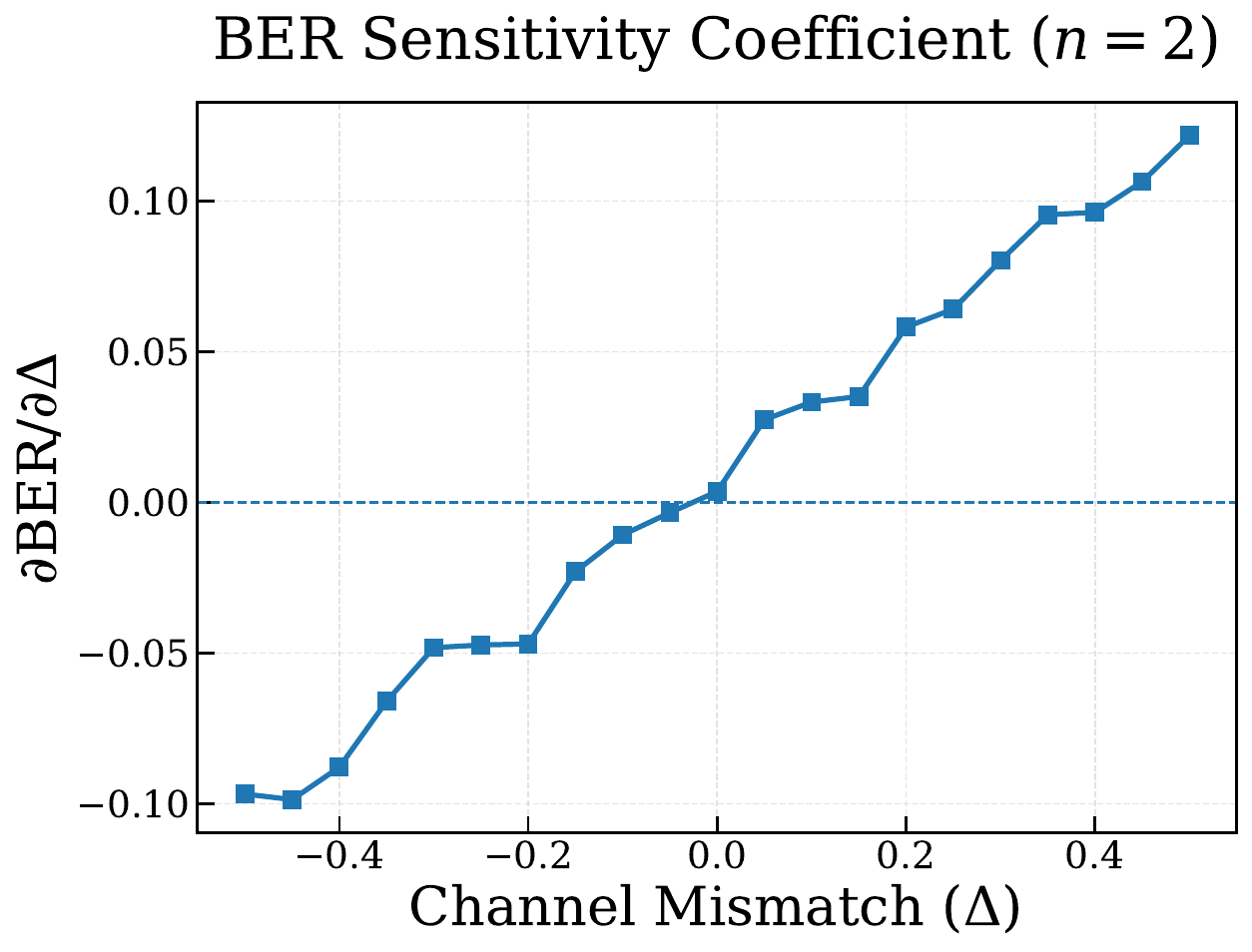}

\end{subfigure}
\hfill
\begin{subfigure}[t]{0.48\textwidth}
\centering
\includegraphics[width=\linewidth]{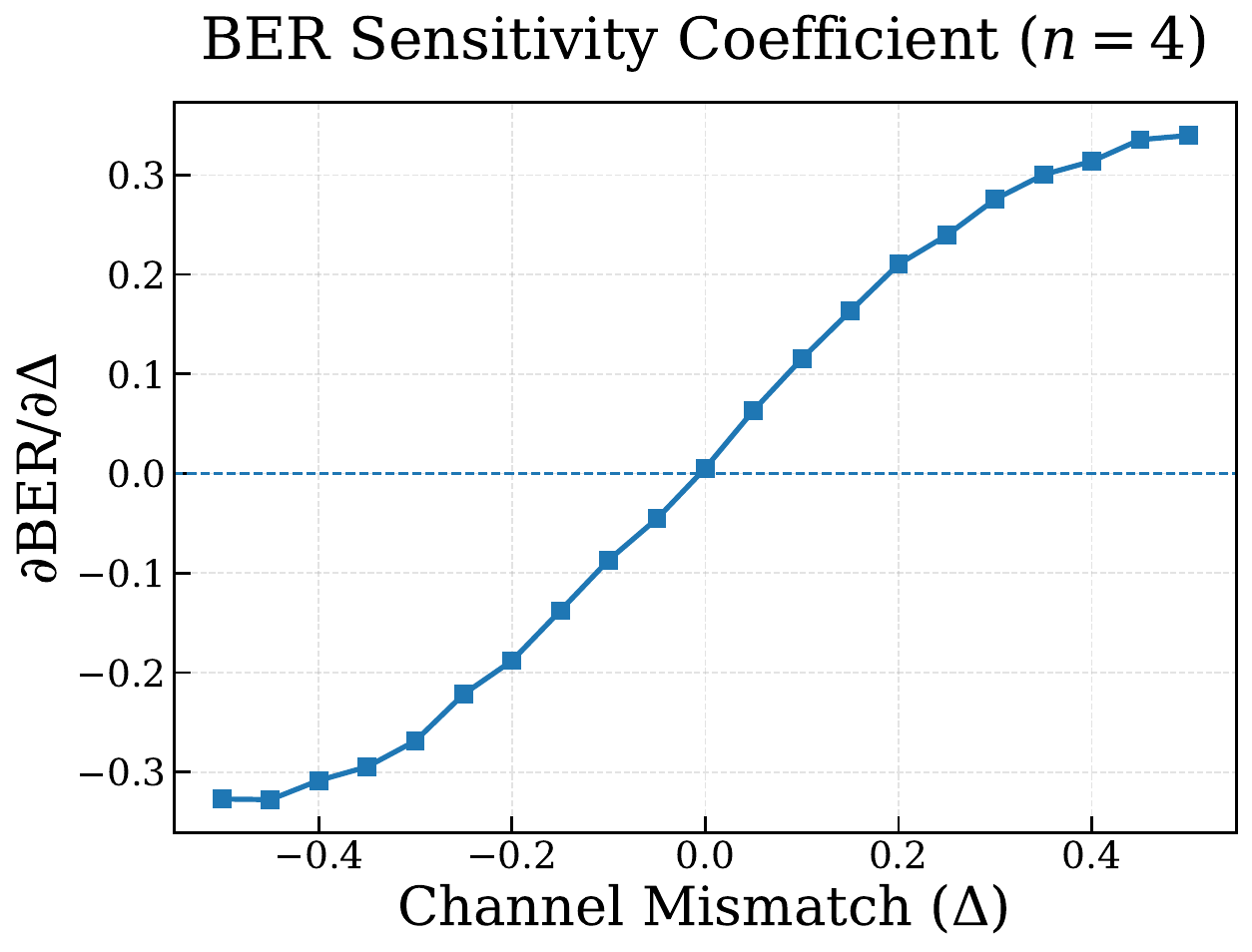}

\end{subfigure}

\vspace{-0.15cm}

\begin{subfigure}[t]{0.48\textwidth}
\centering
\includegraphics[width=\linewidth]{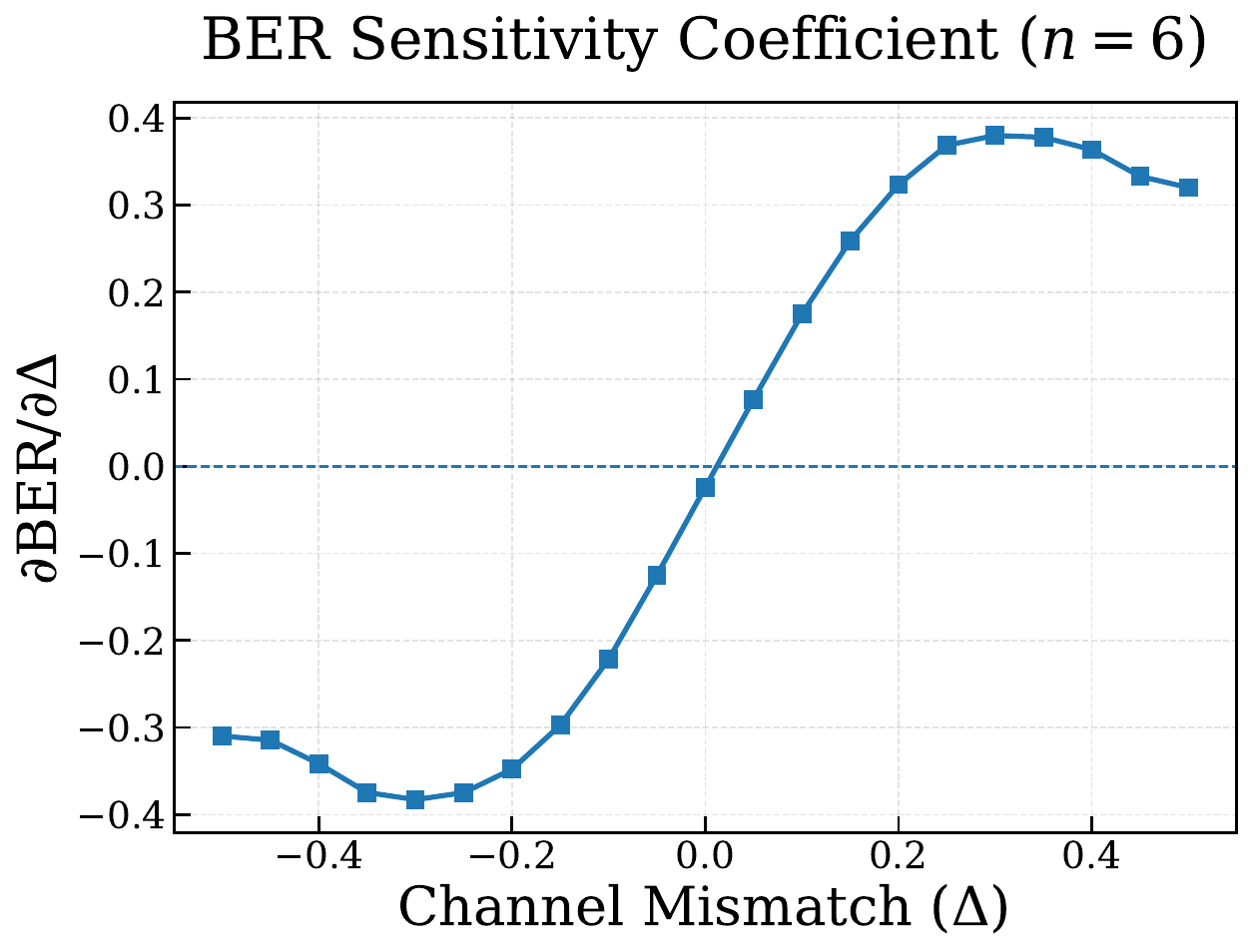}

\end{subfigure}
\hfill
\begin{subfigure}[t]{0.48\textwidth}
\centering
\includegraphics[width=\linewidth]{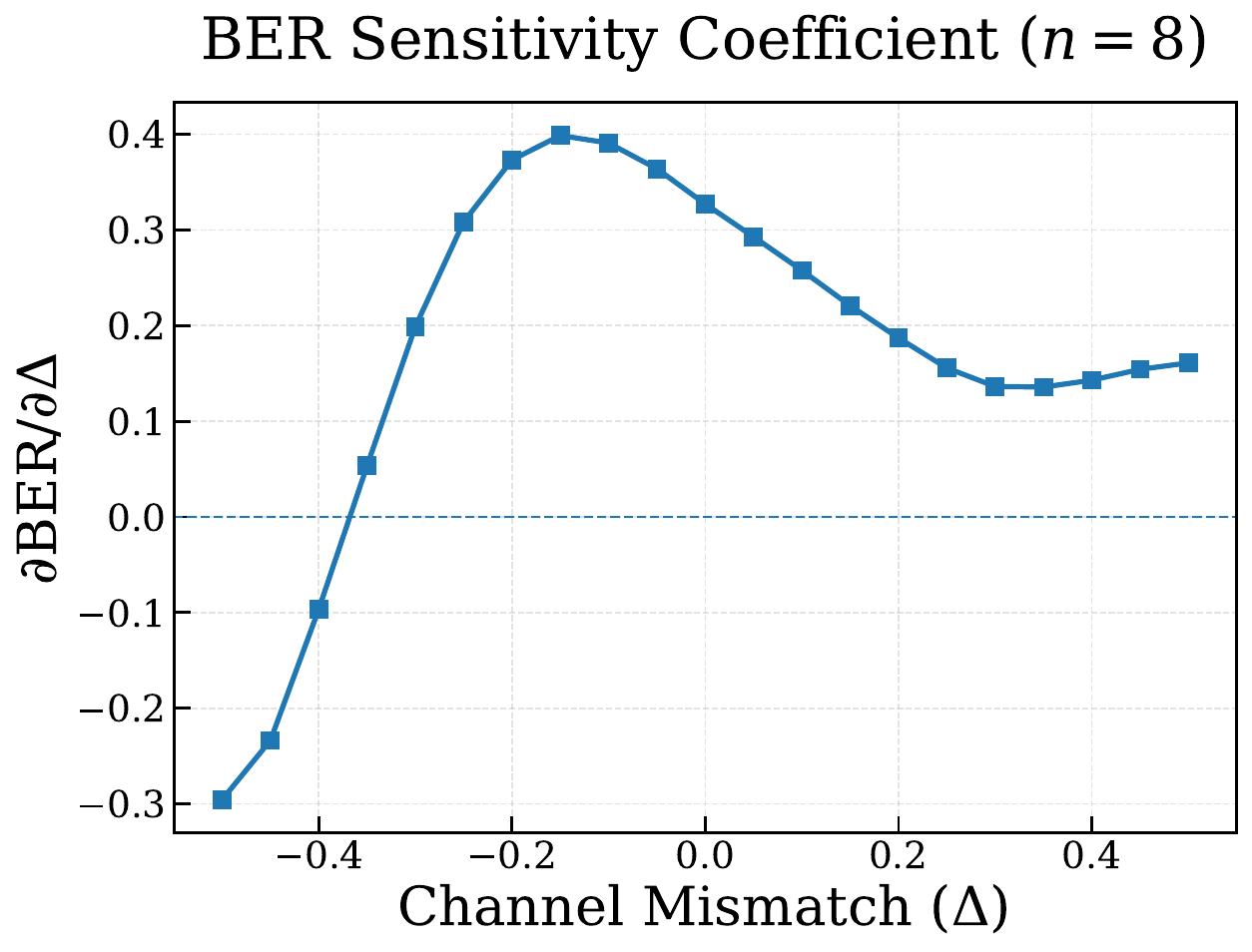}

\end{subfigure}

\caption{BER sensitivity coefficient as a function of channel mismatch for different system dimensions.}
\label{fig:ber_sensitivity}
\end{figure*}
While the BER curves of Fig.~\ref{fig:rber_scaling} provide a direct measure
of performance degradation under channel mismatch, a more rigorous
quantification of robustness can be obtained from the BER sensitivity
coefficient
\begin{equation}
S_{\mathrm{BER}}
=
\frac{\partial \mathrm{BER}}{\partial \Delta},
\label{eq:ber_sensitivity}
\end{equation}
which characterizes the local rate at which the error probability changes
with respect to channel-model perturbations. The corresponding sensitivity
profiles are shown in Fig.~\ref{fig:ber_sensitivity}.

For all system dimensions, the sensitivity coefficient exhibits a smooth
transition from negative values for $\Delta<0$ to positive values for
$\Delta>0$, crossing zero near the BER minimum. This behavior is expected
from the convex structure of the BER curves and indicates that the nominal
operating point corresponds to a stationary point of the mismatch-induced
error landscape. Physically, the vanishing derivative near the optimum
implies that small channel-estimation errors produce only second-order
variations in BER, thereby reducing the impact of imperfect channel
knowledge.

For system dimensions $n=2$, $4$, and $6$, the BER sensitivity coefficient
exhibits an approximately antisymmetric profile around the nominal
operating point, crossing zero near $\Delta=0$. This behavior is consistent with the nearly symmetric BER curves observed in
Fig.~\ref{fig:rber_scaling} and indicates that the optimal decoding
condition occurs close to the matched-channel configuration.

A different behavior emerges for the system size ($n=8$), where thezero-crossing of the sensitivity coefficient is shifted slightly away from $\Delta=0$. This shift directly reflects the displacement of the BER
minimum toward a small negative mismatch value observed in the robustness
analysis. Since the sensitivity coefficient corresponds to the derivative
of the BER curve, the point at which $\partial\mathrm{BER}/\partial\Delta=0$ must coincide with the location of
the minimum BER. Consequently, the displaced zero-crossing provides
quantitative confirmation that the empirically optimal operating point is
no longer exactly aligned with the nominal channel model.

From a physical perspective, this asymmetry arises from the increased
complexity of the variational optimization landscape in larger quantum
MIMO systems. As the Hilbert-space dimension grows, the number of
multi-qubit correlations and entangling interactions increases rapidly,
making the learned recovery transformation more susceptible to small
parameter biases introduced during training. As a result, the optimized
decoder may converge to a nearby local optimum whose maximum-fidelity
condition is slightly offset from the ideal matched-channel case.

Overall, the sensitivity analysis demonstrates that the proposed VQC-based receiver remains robust against moderate channel-estimation errors. The small sensitivity values near the optimal operating point indicate that the BER changes only gradually under channel mismatch, allowing the receiver to maintain stable decoding performance even when the channel model is not perfectly known.

\subsubsection{Stability-Region Analysis}

\begin{figure*}[!t]
\centering

\begin{subfigure}[t]{0.48\textwidth}
\centering
\includegraphics[width=\linewidth]{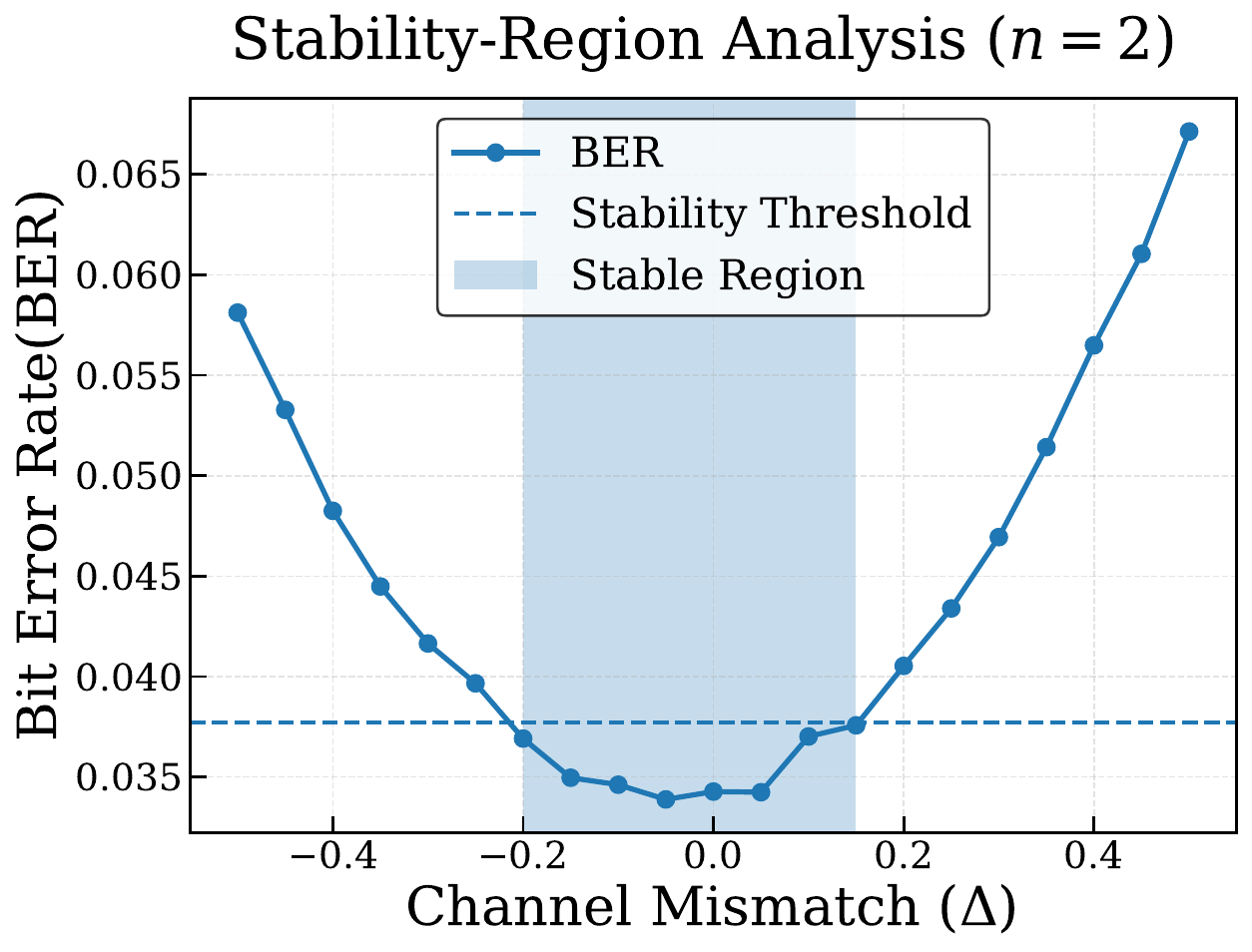}

\end{subfigure}
\hfill
\begin{subfigure}[t]{0.48\textwidth}
\centering
\includegraphics[width=\linewidth]{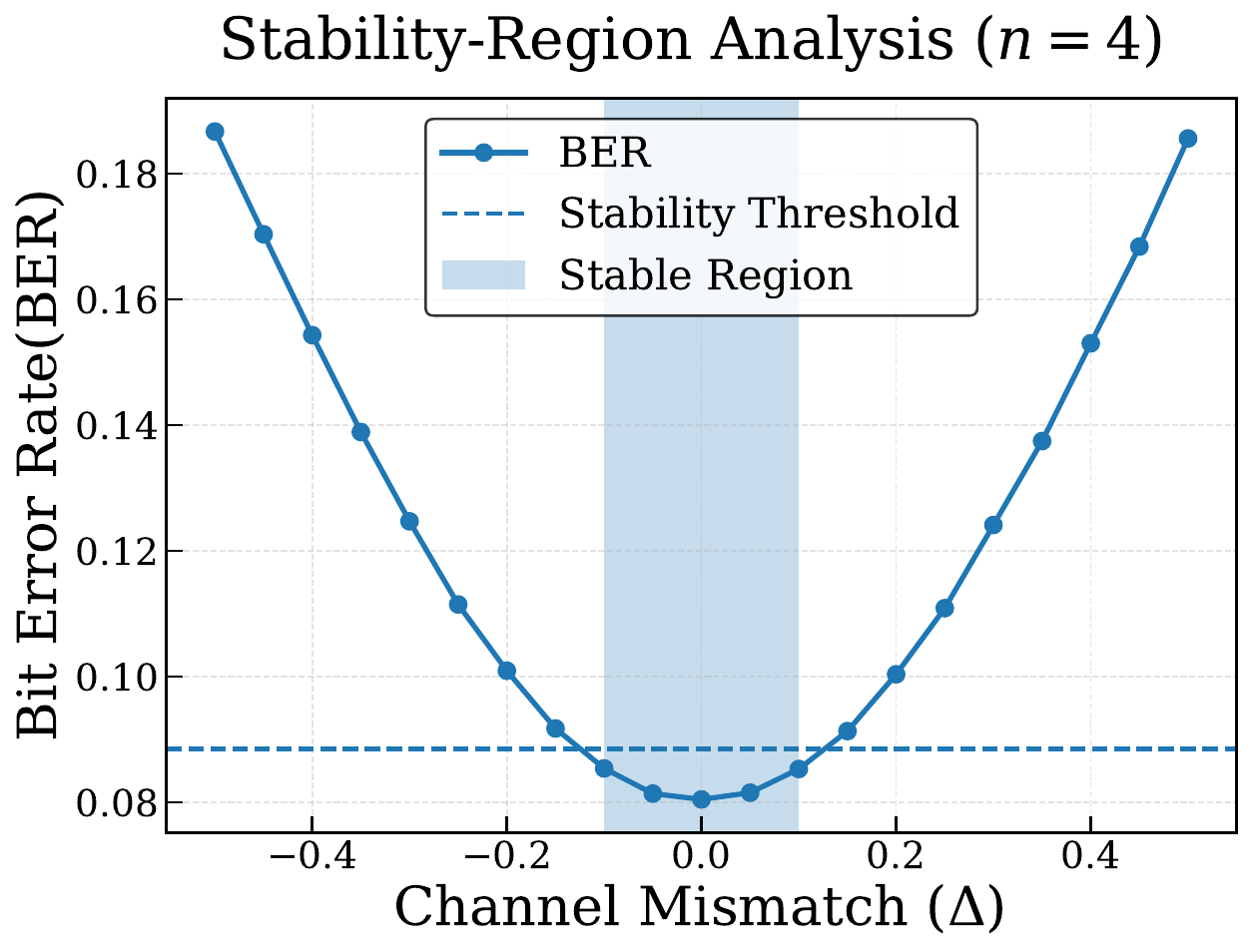}

\end{subfigure}

\vspace{-0.15cm}

\begin{subfigure}[t]{0.48\textwidth}
\centering
\includegraphics[width=\linewidth]{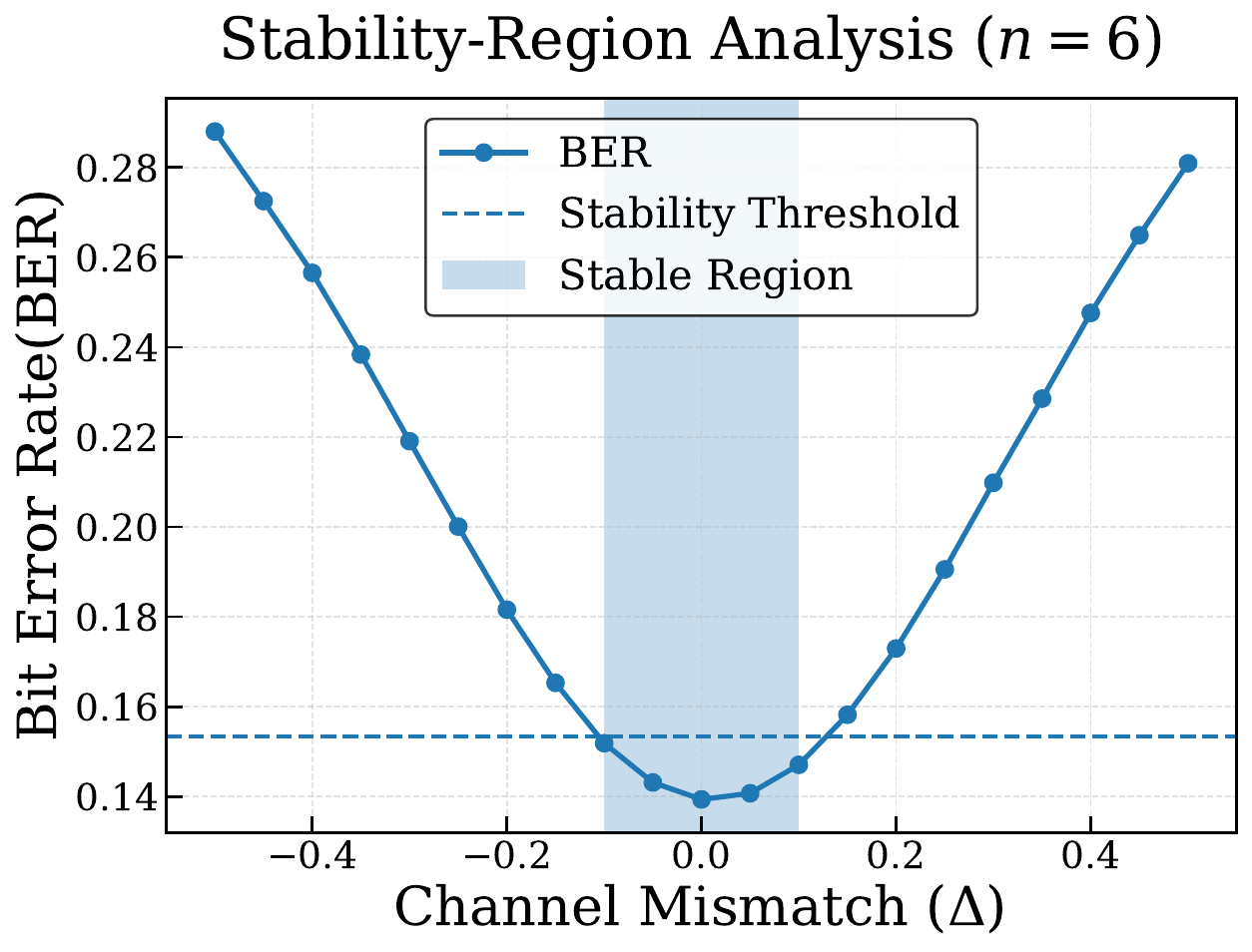}

\end{subfigure}
\hfill
\begin{subfigure}[t]{0.48\textwidth}
\centering
\includegraphics[width=\linewidth]{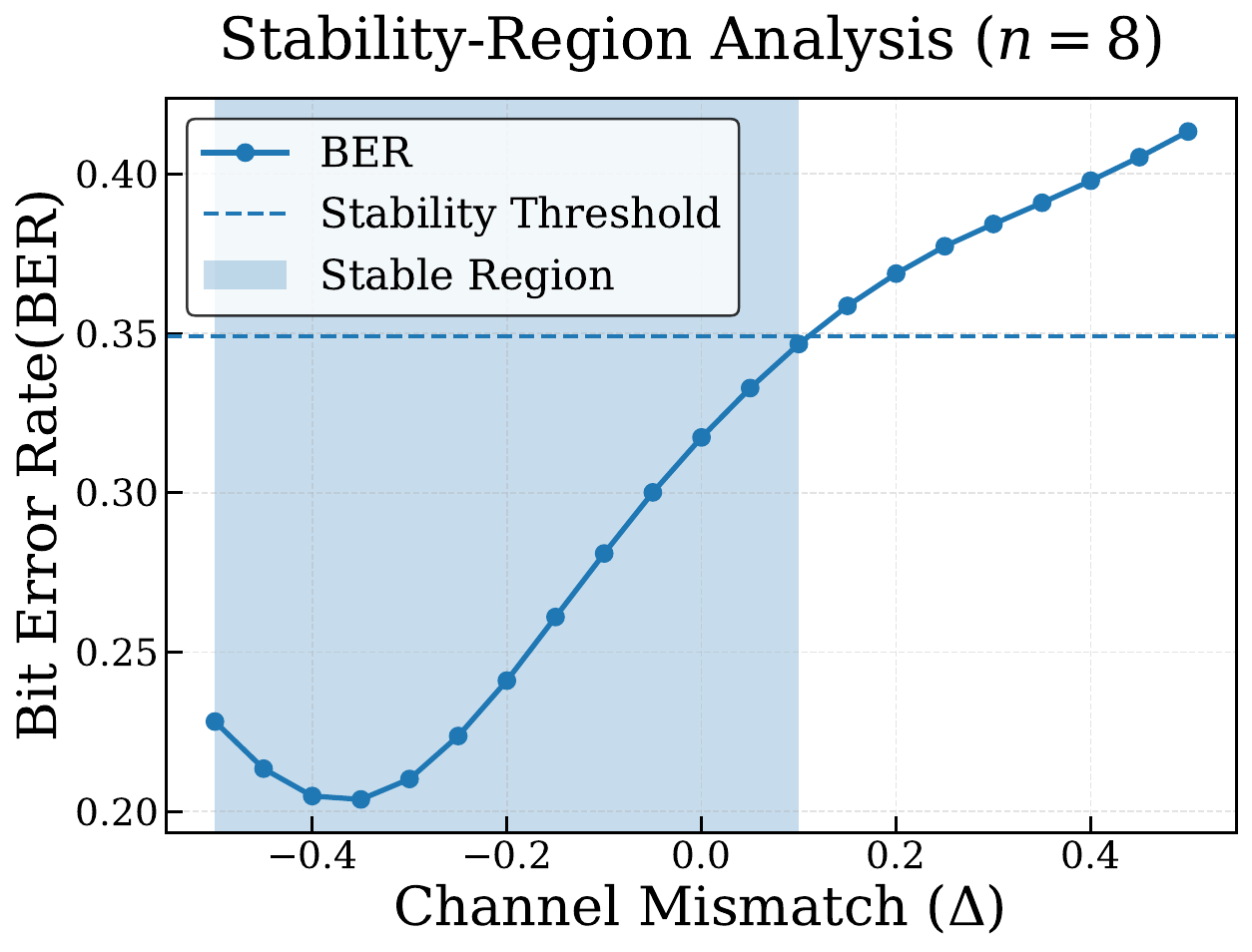}

\end{subfigure}

\caption{Stability-region analysis under channel mismatch}
\label{fig:stability_region}
\end{figure*}

To further quantify the robustness of the proposed VQC-based receiver, a
stability threshold $\mathrm{BER}_{\mathrm{th}}$ is introduced and the
stability region is defined as the range of mismatch values satisfying

\begin{equation}
\mathrm{BER}(\Delta)
\le
\mathrm{BER}_{\mathrm{th}}.
\label{eq:stability_condition}
\end{equation}

The corresponding stability regions for different system dimensions are
shown in Fig.~\ref{fig:stability_region}.

Unlike the BER sensitivity coefficient, which provides a local measure of
robustness, the stability-region width offers a global characterization
of the receiver's tolerance to channel uncertainty. A larger stability
width indicates that acceptable decoding performance can be maintained
over a broader range of channel-estimation errors, thereby reducing the
accuracy requirements imposed on channel estimation and calibration
procedures.

For the smaller and intermediate system dimensions ($n=2$, $4$, and $6$),
the stability-region width gradually decreases as the system size
increases. This behavior is consistent with the larger Hilbert-space
dimension and the increasing complexity of the underlying multi-qubit
interaction structure. As the number of trainable parameters and
entangling correlations grows, the decoding operation becomes more
dependent on accurate channel information, causing the BER to exceed the
stability threshold at smaller mismatch magnitudes. Consequently, the
range of acceptable channel perturbations becomes progressively narrower
from $n=2$ to $n=6$.

Interestingly, this trend does not continue for the largest investigated
system size ($n=8$). Although the $n=8$ receiver exhibits a higher
overall BER level, its stability region becomes broader than that of the
$n=6$ configuration. This observation indicates that stability-region
width is determined not only by the absolute BER value but also by the
shape of the BER landscape around the optimal operating point.

As observed in the robustness analysis, the BER curve for $n=8$ varies
more gradually with channel mismatch and exhibits a relatively flat
profile over a substantial range of $\Delta$. As a result, the BER
remains below the prescribed threshold across a wider mismatch interval
before crossing the stability boundary. In contrast, the $n=6$ system
shows a steeper BER variation around the threshold level, causing the
allowable operating region to contract despite its lower minimum BER.

From a physical perspective, this behavior suggests that the optimized
high-dimensional variational decoder develops a broader tolerance region
around its operating point. While the larger Hilbert-space dimension
increases the overall decoding complexity, it also appears to produce a
more gradual mismatch-induced degradation mechanism. Consequently, the
receiver can tolerate larger channel-estimation errors before its
performance falls below the required threshold.

The non-monotonic behavior of the stability-region width demonstrates
that robustness cannot be assessed solely through minimum BER values.
Rather, both the magnitude and curvature of the BER landscape determine
the practical operating range of the receiver. The comparatively wide
stability region observed for $n=8$ therefore provides additional
evidence that the proposed variational quantum decoder maintains reliable performance even in high-dimensional quantum MIMO configurations under imperfect channel knowledge.

Overall, the stability-region results reveal that the robustness of
the variational decoder is governed by the interplay between system
complexity and the shape of the mismatch-induced BER profile. Although
increasing system dimension generally reduces the tolerance to channel
uncertainty, the broader stable operating region obtained for $n=8$
demonstrates that high-dimensional variational quantum receivers can
still exhibit strong resilience against channel-estimation errors.

\subsection{BER Performance Under Varying Channel Coupling}
\begin{figure*}[!t]
\centering
\begin{subfigure}{0.47\textwidth}
\centering
\includegraphics[width=\linewidth]{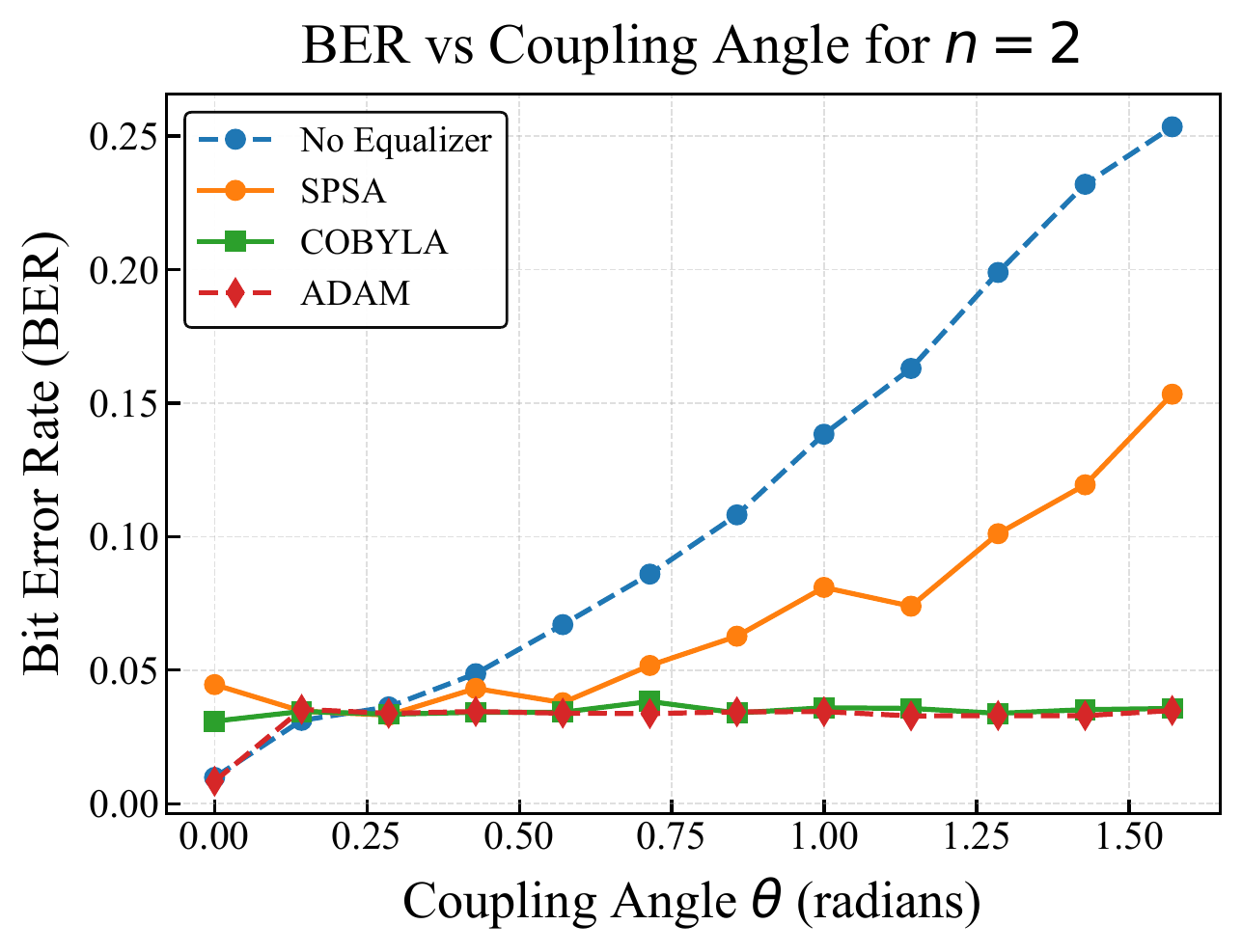}
\caption{}
\end{subfigure}
\hfill
\begin{subfigure}{0.47\textwidth}

\centering
\includegraphics[width=\linewidth]{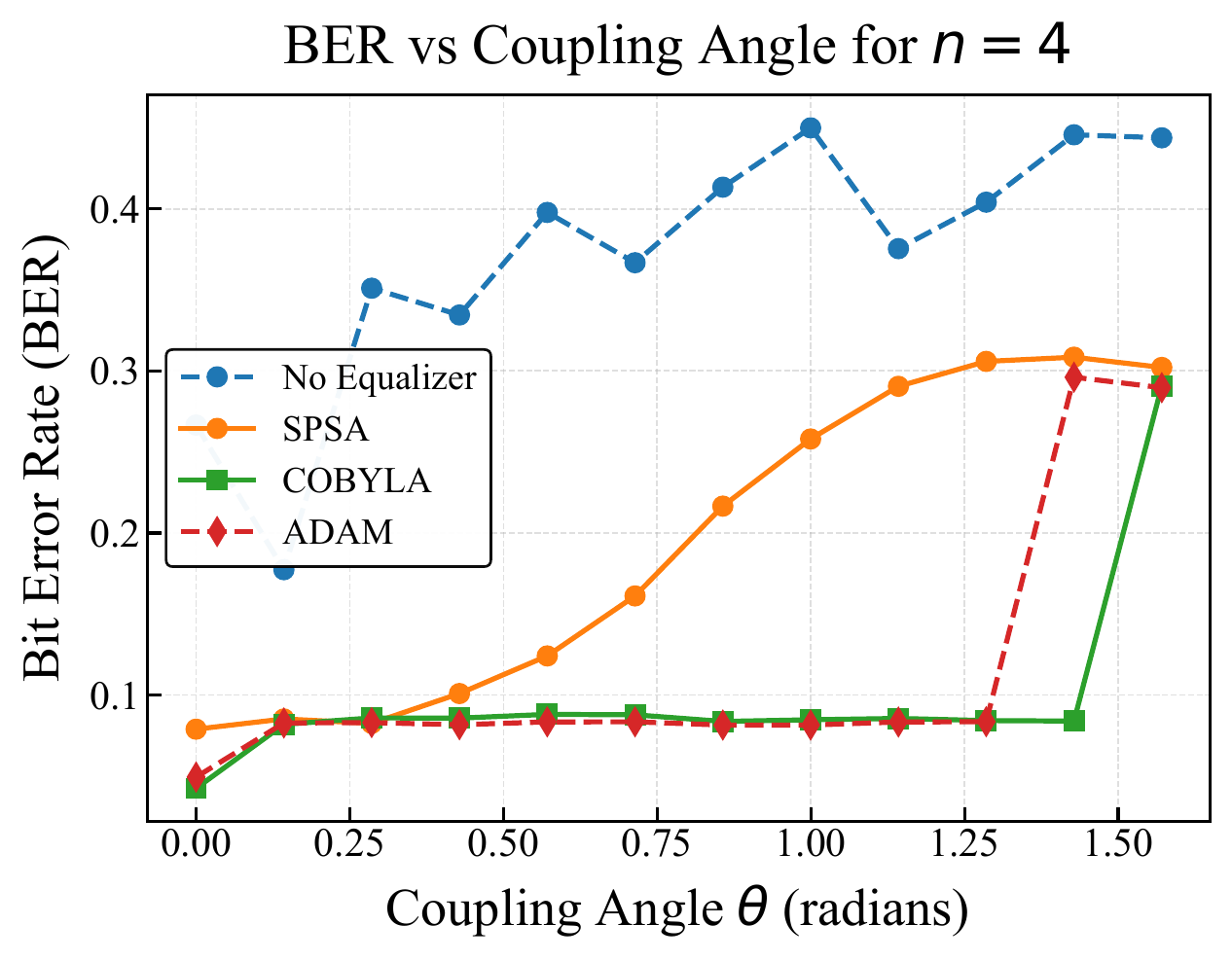}
\caption{}
\end{subfigure}

\vspace{-0.15cm}

\begin{subfigure}{0.47\textwidth}
\centering
\includegraphics[width=\linewidth]{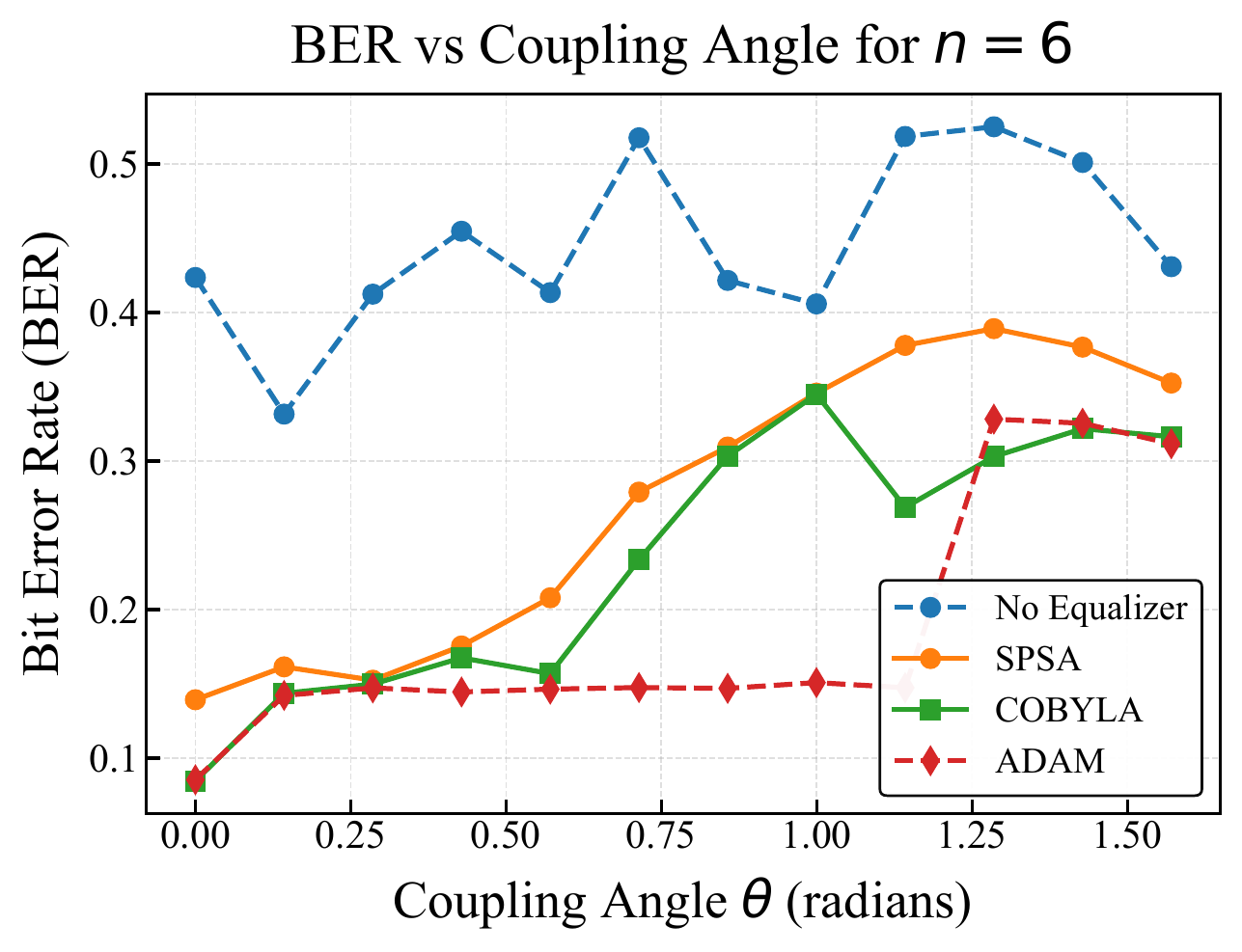}
\caption{}
\end{subfigure}
\hfill
\begin{subfigure}{0.47\textwidth}
\centering
\includegraphics[width=\linewidth]{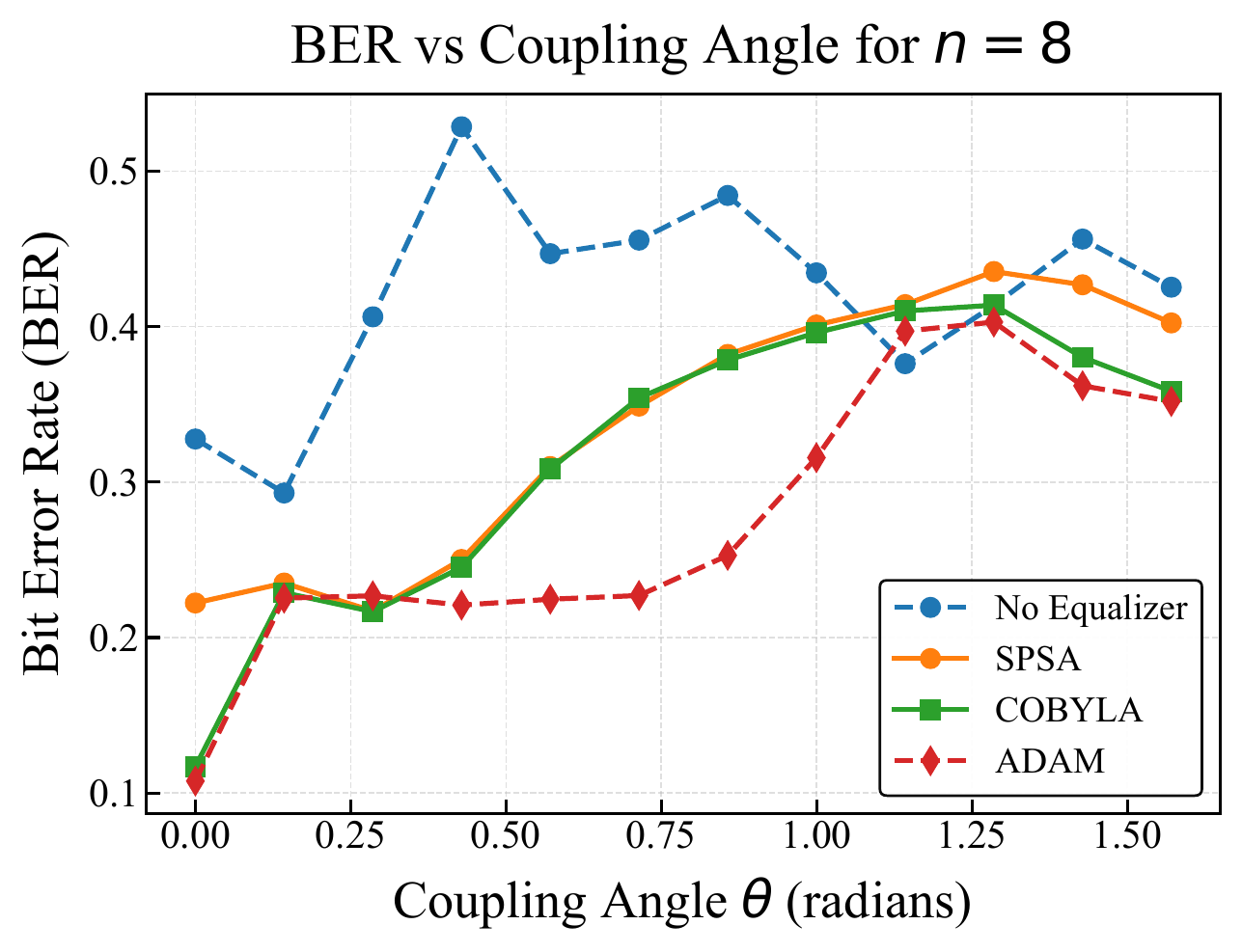}
\caption{}
\end{subfigure}

\vspace{-0.2cm}

\caption{BER performance of the proposed quantum MIMO MMSE receiver using SPSA, COBYLA, and ADAM optimizers compared with the no-equalization (No-Eq) baseline under varying channel coupling strengths $(\theta)$ for system dimensions $n=2,4,6,$ and $8$.}
\label{fig:cber_scaling}

\vspace{-0.35cm}
\end{figure*}

The BER performance of the proposed quantum MIMO framework is evaluated as a function of the channel coupling angle \( \theta \) for different system dimensions. The results are presented in Figs.~\ref{fig:cber_scaling}. A detailed analysis is performed for system sizes \( n = 2, 4, 6, \) and \( 8 \), comparing the un-equalized baseline (No-Eq) with three variational MMSE optimization strategies: SPSA, COBYLA, and ADAM. The coupling parameter is varied over the range \( \theta \in [0,\pi/2] \), corresponding to progressively stronger coherent inter-channel interactions within the quantum channel.

As the coupling strength \( \theta \) increases, the BER performance degrades across all system dimensions due to the stronger coherent interference introduced by the quantum channel. This degradation becomes increasingly severe for larger values of \( n \), where the variational receiver must compensate for more complex multi-qubit interference structures within a rapidly expanding optimization space. Consequently, the effectiveness of the receiver becomes strongly dependent on both the optimizer stability and the trainability of the variational circuit.

For the smallest configuration (\(n=2\)), all three variational receivers substantially outperform the No-Eq system over the entire coupling range. The No-Eq BER increases monotonically from approximately \(0.008\) to nearly \(0.25\) as \( \theta \rightarrow \pi/2 \), confirming that even low-dimensional quantum channels experience severe degradation under strong coherent mixing. In contrast, both COBYLA and ADAM maintain highly stable BER values around \(0.03\!-\!0.04\), indicating that the variational decoder successfully learns an effective compensation manifold capable of suppressing low-dimensional interference. SPSA also provides considerable BER reduction relative to No-Eq; however, its BER increases noticeably at larger coupling strengths, reaching approximately \(0.19\) near \( \theta = 1.428 \). This behavior arises from the stochastic perturbation mechanism employed by SPSA, where gradient estimation variance grows as the optimization landscape becomes increasingly oscillatory and non-convex under strong entanglement.

For \(n=4\), the channel dynamics become considerably more complex due to the rapid growth of pairwise entangling interactions and the corresponding enlargement of the effective Hilbert space. The No-Eq baseline exhibits strong BER fluctuations between approximately \(0.17\) and \(0.45\), indicating that direct symbol recovery becomes highly unstable once coherent interference dominates the channel evolution. In this regime, the variational receiver must compensate for increasingly correlated interference structures distributed across multiple qubits. ADAM demonstrates the strongest robustness, maintaining BER near \(0.08\) over most of the coupling range. Its adaptive moment-based optimization mechanism stabilizes parameter updates within the highly curved variational landscape, allowing efficient convergence even in the presence of strong parameter correlations. COBYLA performs competitively at moderate coupling strengths but experiences abrupt degradation near \( \theta = \pi/2 \), where its BER rapidly approaches \(0.30\). This suggests that local trust-region approximations become insufficient once the optimization surface develops dense local minima and strong nonlinear parameter dependencies. SPSA exhibits a progressive BER increase with coupling strength, reaching values above \(0.31\), which indicates reduced optimization stability in higher-dimensional interference regimes.

For the \(n=6\) system, the degradation trend becomes substantially more pronounced. The No-Eq BER remains consistently high across nearly the entire coupling range, typically between \(0.36\) and \(0.48\), demonstrating that coherent interference accumulation severely limits direct measurement-based detection. From an information-theoretic perspective, the effective separability of transmitted states decreases as entanglement distributes symbol information nonlocally across the quantum register. Consequently, the variational receiver must reconstruct increasingly complex interference correlations using a finite-depth parameterized circuit. SPSA initially achieves moderate BER reduction at low coupling strengths but deteriorates steadily as \( \theta \) increases, eventually approaching the No-Eq regime. This behavior reflects the poor scaling characteristics of stochastic gradient approximation in high-dimensional parameter spaces, where optimization noise increases with circuit complexity. COBYLA exhibits improved robustness relative to SPSA but develops noticeable BER spikes at intermediate and strong coupling values, indicating intermittent convergence toward suboptimal local regions of the variational loss landscape. ADAM again provides the best overall performance, maintaining BER near \(0.15\) over a wide range of coupling strengths before degrading only under extremely strong interference conditions. The stability of ADAM demonstrates the importance of adaptive gradient normalization when optimizing deep entangling quantum circuits with strongly correlated parameters.

For the largest configuration (\(n=8\)), all optimization methods experience significant performance degradation due to the exponential growth of decoding complexity and entanglement-induced interference structure. The No-Eq BER fluctuates between approximately \(0.39\) and \(0.53\), confirming that direct symbol detection becomes increasingly infeasible in strongly coupled high-dimensional quantum channels. Physically, the transmitted information becomes highly delocalized across entangled basis states, causing the received quantum states to overlap substantially during measurement. SPSA and COBYLA both exhibit steadily increasing BER with coupling strength, frequently exceeding \(0.40\) at large \( \theta \). This indicates that the optimization landscape develops severe nonlinearity, strong parameter interdependence, and barren-region tendencies as the variational search space expands. ADAM again achieves the best overall performance, particularly for low and moderate coupling strengths where BER remains significantly below the other optimizers. However, even ADAM begins to degrade at strong coupling, eventually approaching BER values near \(0.35\). This suggests that the expressive capacity and trainability of the finite-depth variational decoder become insufficient to fully invert the highly entangled channel transformation in large-scale systems.

Overall, the results demonstrate that variational quantum MMSE equalization significantly improves BER performance compared to the un-equalized baseline across all system dimensions. However, BER degradation increases with both coupling strength and system size due to stronger coherent interference and the growing complexity of the variational optimization landscape. Among the investigated optimizers, ADAM consistently provides the most stable and scalable performance, while COBYLA becomes less stable for larger systems and SPSA exhibits higher sensitivity to interference-induced optimization noise.

\subsection{Entanglement Dynamics and Communication Performance}

\begin{figure*}[!t]
\centering
\begin{subfigure}[t]{0.48\textwidth}
\centering
\includegraphics[width=\linewidth]{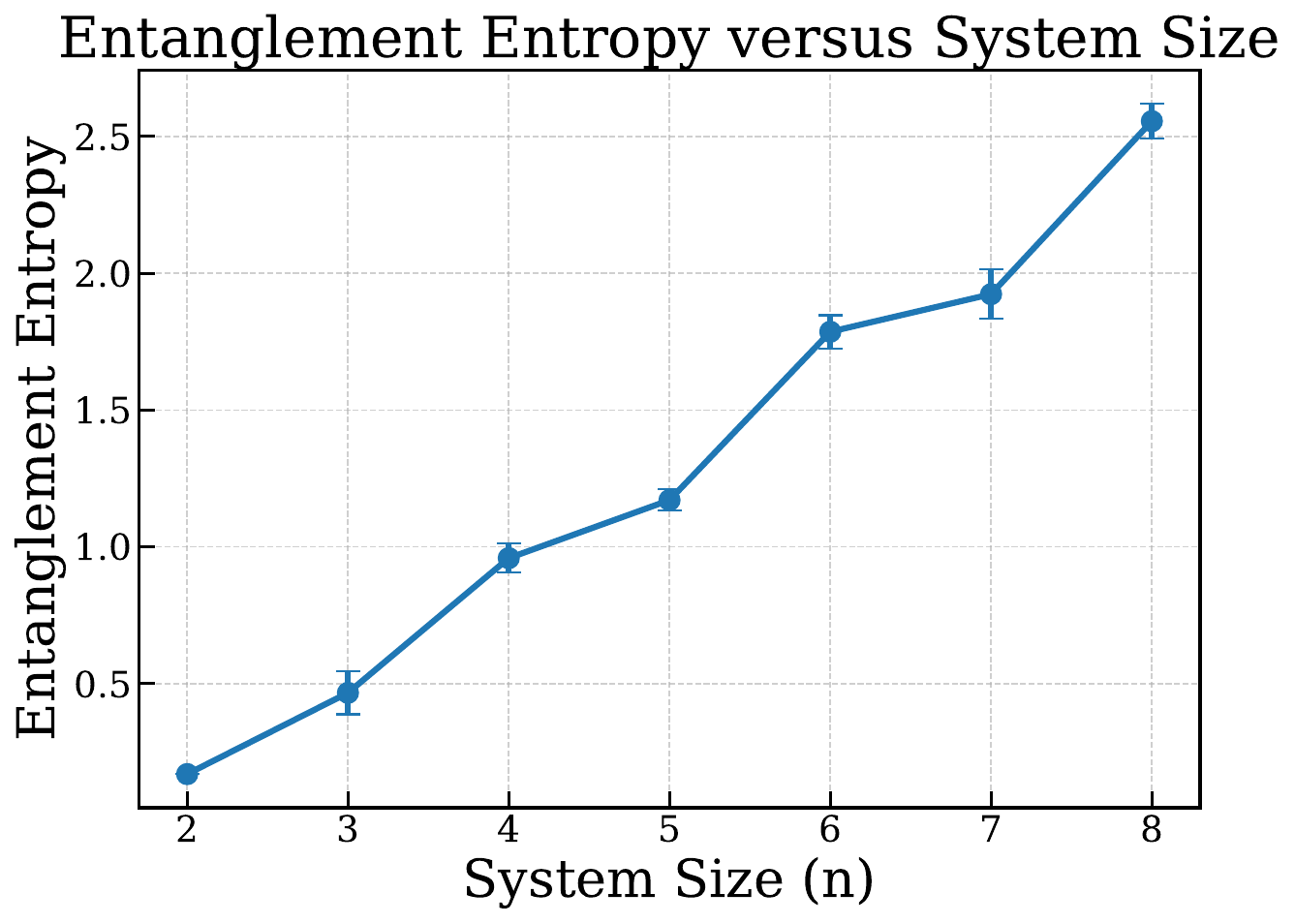}
\caption{}
\end{subfigure}
\hfill
\begin{subfigure}[t]{0.48\textwidth}
\centering
\includegraphics[width=\linewidth]{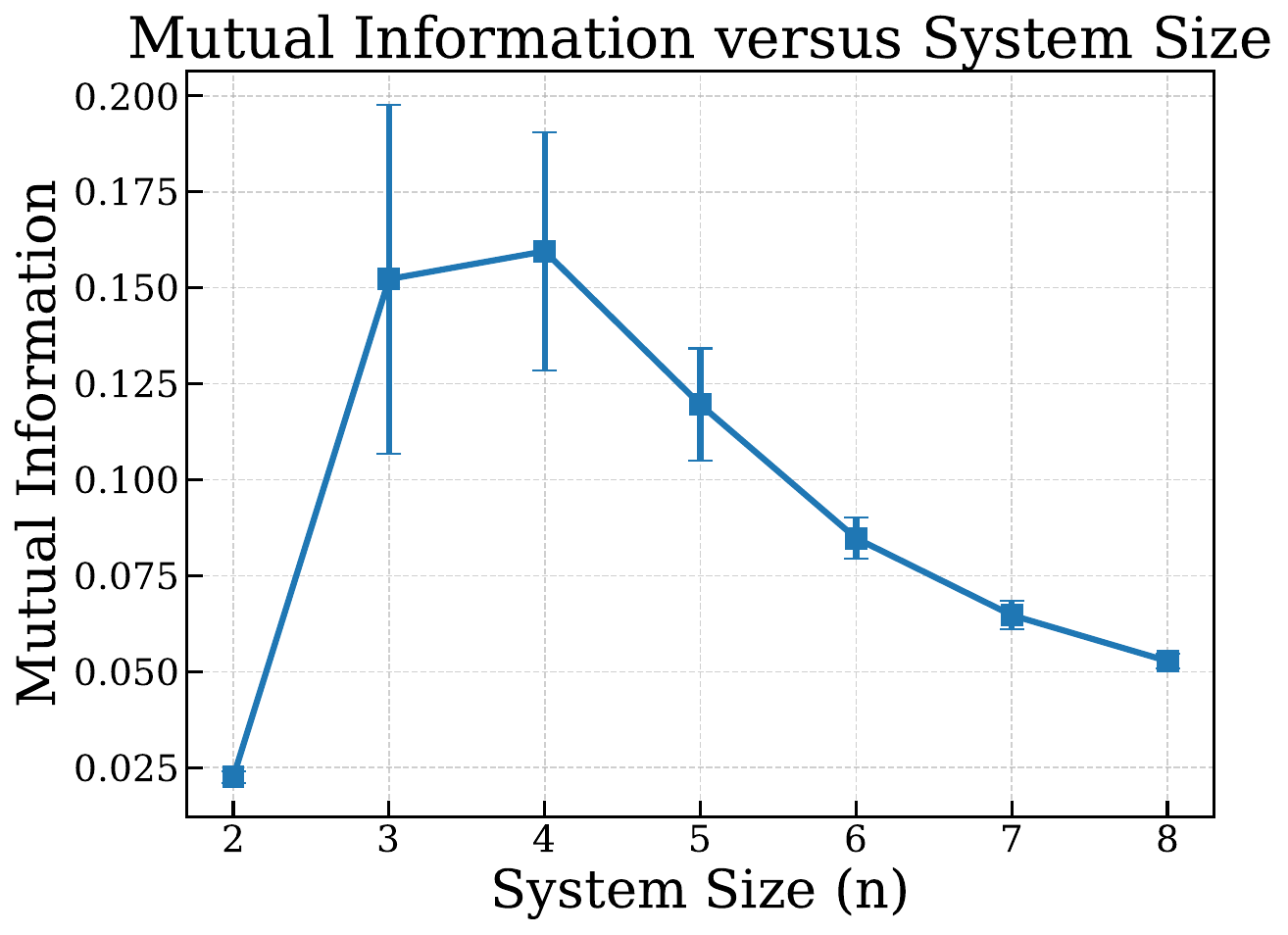}
\caption{}
\end{subfigure}
\vspace{0.3cm}
\begin{subfigure}[t]{0.48\textwidth}
\centering
\includegraphics[width=\linewidth]{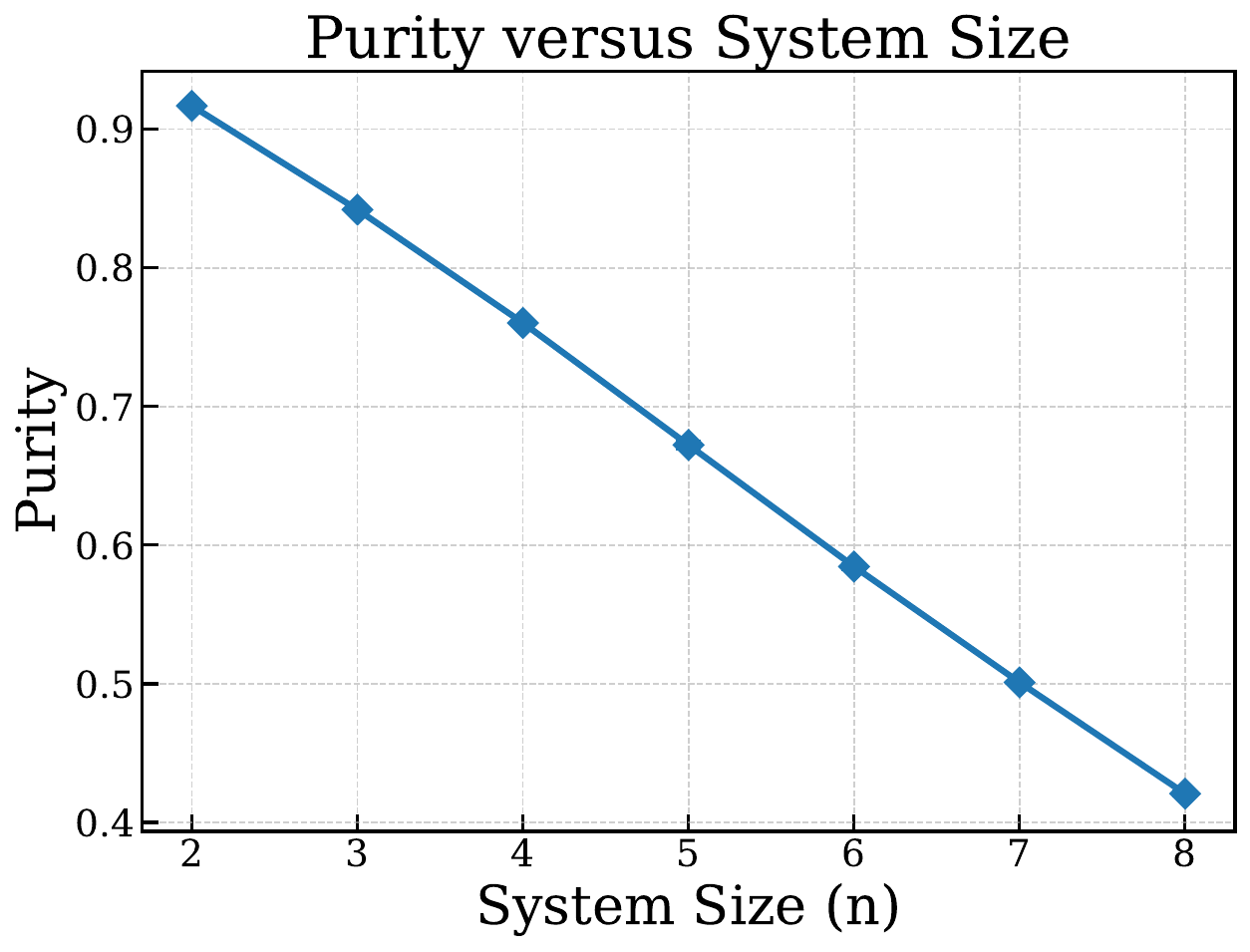}
\caption{}
\end{subfigure}
\hfill
\begin{subfigure}[t]{0.48\textwidth}
\centering
\includegraphics[width=\linewidth]{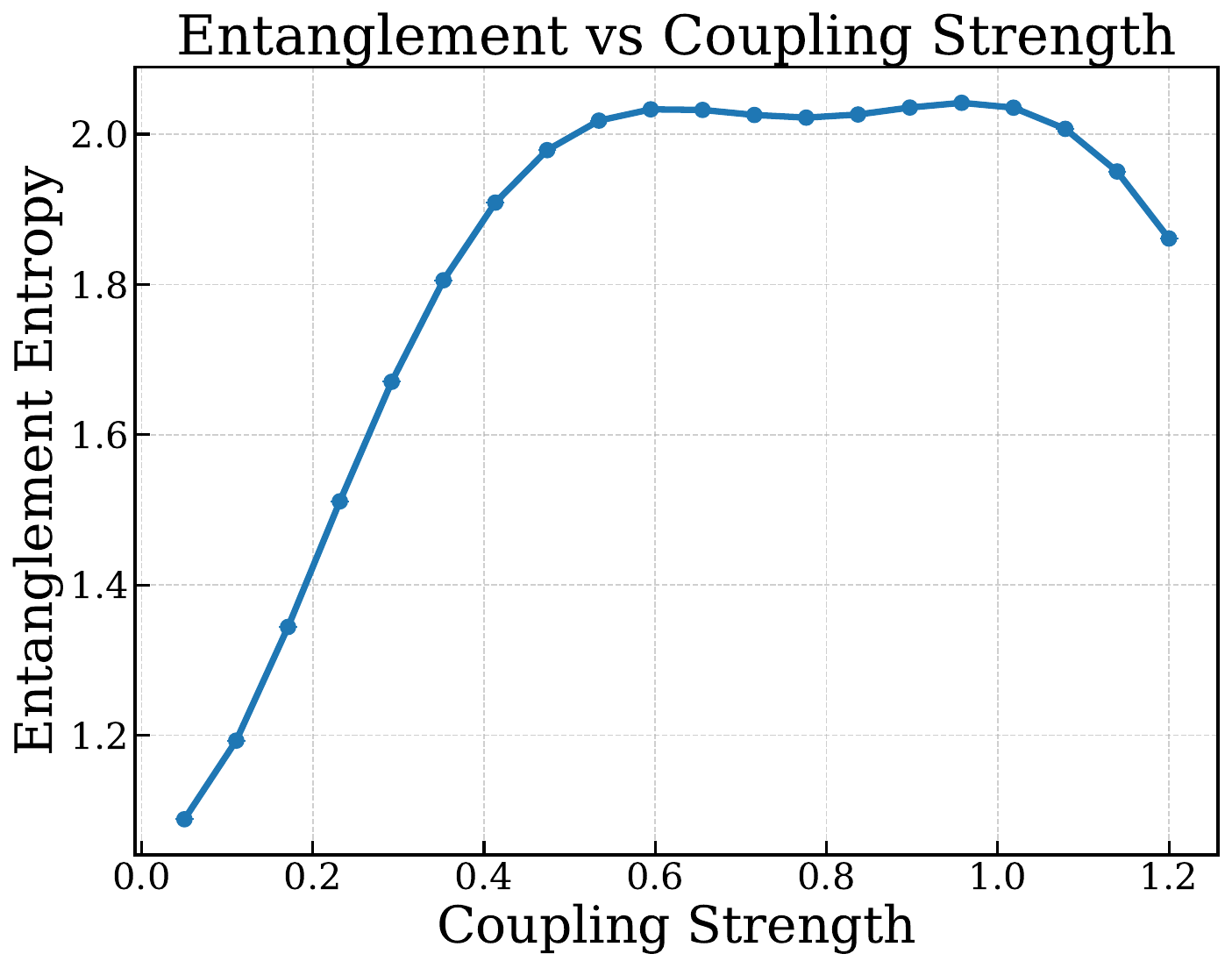}
\caption{}
\end{subfigure}

\caption{
Correlation and information-delocalization properties of the proposed quantum MIMO channel. (a) Bipartite von Neumann entropy as a function of system size. (b) Average pairwise mutual information. (c) Global purity of the noisy channel output state. (d) Bipartite entropy as a function of the interaction strength $\theta$.
}
\label{fig:entanglement_scaling}
\end{figure*}
The degradation of decoding performance observed in larger quantum MIMO systems is hypothesized to originate from the growth of many-body quantum correlations generated during channel evolution. To investigate this mechanism directly, we analyze the correlation structure of the channel output state using three complementary quantities: the bipartite von Neumann entropy, the average pairwise mutual information, and the global purity. Among these measures, the bipartite entropy serves as the primary indicator of entanglement generation and information sharing across the quantum register, while mutual information and purity provide additional insight into the distribution of correlations and the influence of realistic NISQ noise.

The physical origin of the observed scaling behavior can be understood from the progressively increasing connectivity of the quantum channel. As the number of qubits grows, the transmitted information experiences a larger number of coherent interaction pathways, allowing probability amplitudes to spread over an increasingly large fraction of the accessible Hilbert space. Rather than remaining localized within individual qubits or small subsystems, information becomes encoded within collective many-body degrees of freedom. This process promotes the formation of nonlocal correlations and increases the complexity of the resulting decoding problem.

To quantify this behavior, we first compute the bipartite von Neumann entropy of a half-system partition,
\begin{equation}
S_A =
-\mathrm{Tr}
\left(
\rho_A \log_2 \rho_A
\right),
\label{eq:bipartite_entropy}
\end{equation}
where $\rho_A$ denotes the reduced density matrix obtained after tracing out half of the qubits. This quantity measures the amount of information shared across a macroscopic system bipartition and therefore provides a direct probe of large-scale correlation growth.

Figure~\ref{fig:entanglement_scaling}(a) shows that the bipartite entropy increases monotonically with system size. This observation provides direct quantitative evidence that the proposed quantum MIMO channel generates increasingly strong many-body correlations as the interaction network expands. The growth of entropy indicates that information is shared across progressively larger portions of the quantum register, reflecting an increase in nonlocal correlation strength. Consequently, larger quantum MIMO channels exhibit greater information delocalization and a more complex correlation structure.

The influence of interaction strength is examined in Fig.~\ref{fig:entanglement_scaling}(d), where the bipartite entropy is plotted as a function of the coupling parameter $\theta$. For small values of $\theta$, the interactions generate only weak correlations between qubits, resulting in relatively low entropy. As the coupling strength increases, the interaction network becomes more effective at redistributing quantum information throughout the register, leading to a rapid increase in entropy. This behavior indicates that a larger fraction of the transmitted information becomes encoded collectively across multiple qubits rather than remaining localized within individual subsystems.

At larger coupling strengths, the rate of entropy growth gradually decreases and approaches a broad saturation region. This trend suggests that the channel has already generated a highly correlated state, such that further increases in interaction strength contribute only modestly to additional correlation growth. The slight reduction in entropy observed at the largest values of $\theta$ is consistent with coherent interference effects within the underlying unitary dynamics, where increasing interaction strength modifies the distribution of existing correlations rather than continuously increasing their overall magnitude. These results demonstrate that the coupling strength plays a central role in governing the degree of information delocalization within the quantum MIMO channel.

While the entropy quantifies the total amount of information shared across a global partition, it does not reveal how that information is distributed throughout the system. To address this, we evaluate the average pairwise mutual information,
\begin{equation}
I(i:j)
=
S(\rho_i)
+
S(\rho_j)
-
S(\rho_{ij}),
\label{eq:mutual_information}
\end{equation}
averaged over all qubit pairs. This quantity measures the total correlation between individual qubit pairs and therefore probes the locality of the generated correlations.

The results shown in Fig.~\ref{fig:entanglement_scaling}(b) exhibit a behavior that is notably different from that of the bipartite entropy. While the entropy increases steadily with system size, the average pairwise mutual information decreases. This trend indicates that the growth of overall correlation is not driven by stronger correlations between individual qubit pairs. Instead, information becomes distributed across larger subsets of the quantum register, causing the correlation structure to shift from predominantly pairwise correlations to more complex multipartite correlations. Consequently, although the total correlation content of the system increases, a smaller fraction of that correlation can be captured through pairwise measurements alone. The observed reduction in average mutual information therefore provides additional evidence of increasing information delocalization and nonlocal correlation growth within the quantum MIMO channel.

To further investigate the evolution of the channel output state, we evaluate the global purity,
\begin{equation}
\gamma
=
\mathrm{Tr}(\rho^2),
\label{eq:global_purity}
\end{equation}
which measures the degree of mixedness of the density matrix. Unlike the entropy and mutual information, purity is sensitive to both coherent correlation growth and incoherent environmental noise.

As shown in Fig.~\ref{fig:entanglement_scaling}(c), the purity decreases steadily with increasing system size. This behavior originates from two concurrent mechanisms. First, stronger many-body correlations distribute information across increasingly nonlocal degrees of freedom, reducing the amount of information that remains locally accessible. Second, the channel evolution incorporates depolarizing noise, thermal relaxation, and readout imperfections through the CPTP framework. Since the number of entangling operations increases with system size, larger systems experience greater cumulative exposure to decoherence and gate imperfections. The observed reduction in purity therefore reflects the combined effects of coherent information delocalization and incoherent noise accumulation during channel evolution.

Overall, the increase in bipartite entropy, decrease in pairwise mutual information, and reduction in purity establish a consistent physical picture of the channel dynamics. As the system size increases, information becomes progressively distributed across larger portions of the quantum register, leading to stronger multipartite correlations and reduced locality of the encoded information. Simultaneously, the accumulated effect of realistic noise processes increases the mixedness of the channel output state. These effects collectively reduce the effective distinguishability of transmitted quantum states, making the inverse reconstruction problem performed by the variational receiver increasingly difficult.The present results therefore provide a direct physical explanation for the BER degradation observed in larger quantum MIMO systems. Rather than arising solely from an increase in circuit size, the degradation emerges from the interplay between many-body correlation growth, information delocalization, and cumulative decoherence. 

\subsection{Hardware Resource and Scalability Analysis}
\begin{figure*}[!t]
\centering
\begin{subfigure}[t]{0.48\textwidth}
\centering
\includegraphics[width=\linewidth]{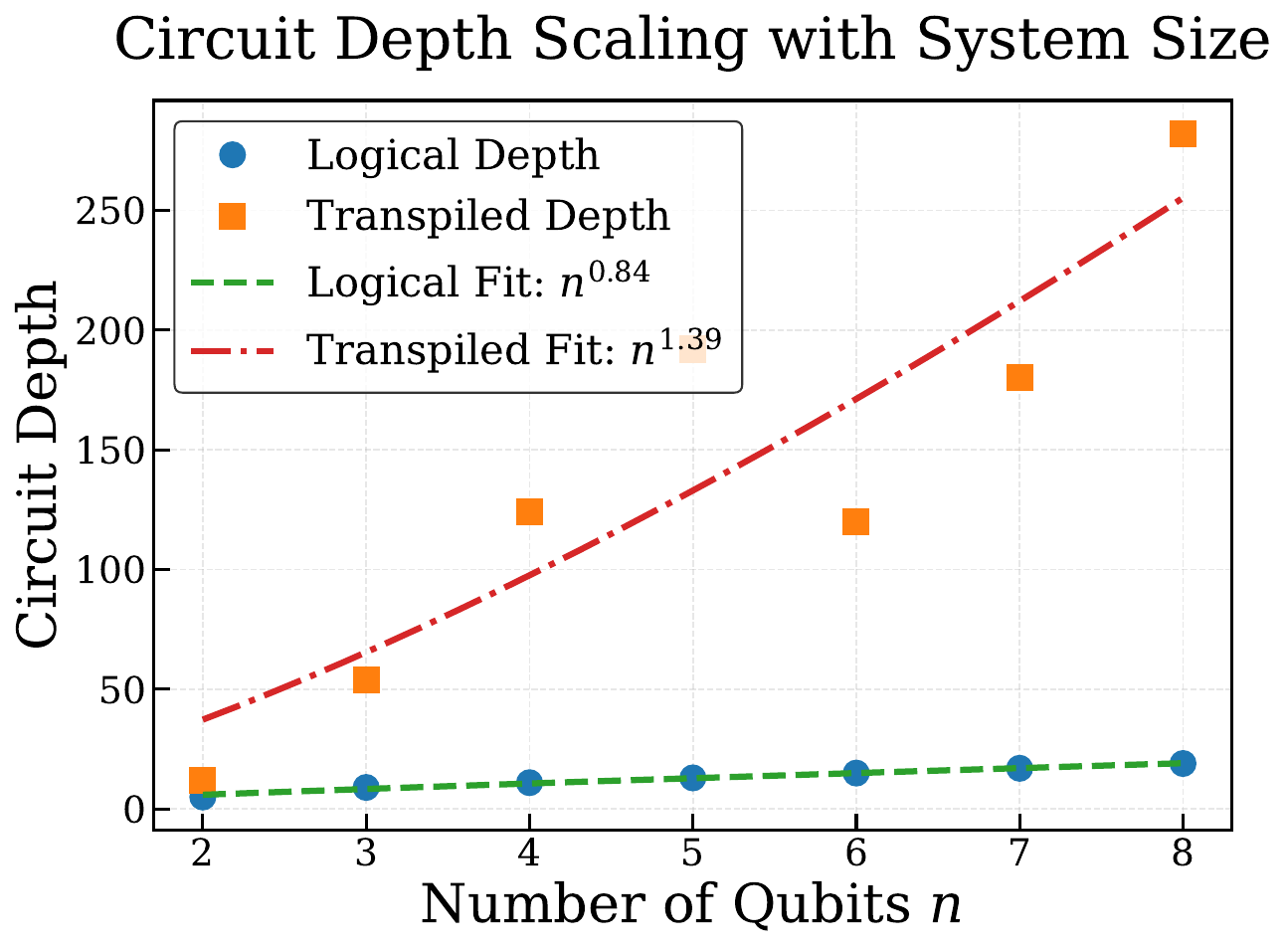}
\caption{}
\end{subfigure}
\hfill
\begin{subfigure}[t]{0.48\textwidth}
\centering
\includegraphics[width=\linewidth]{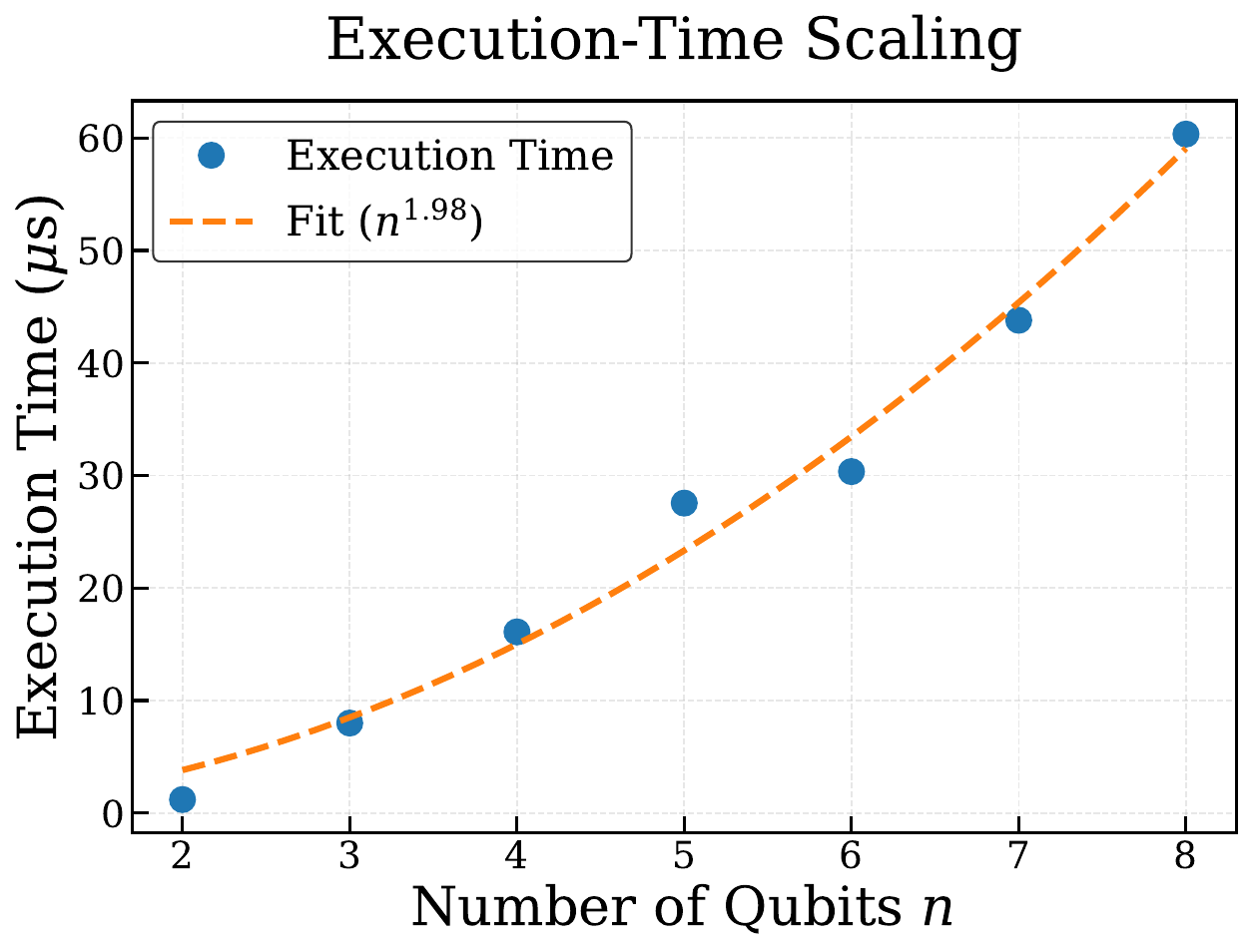}
\caption{}
\end{subfigure}
\vspace{0.3cm}
\begin{subfigure}[t]{0.48\textwidth}
\centering
\includegraphics[width=\linewidth]{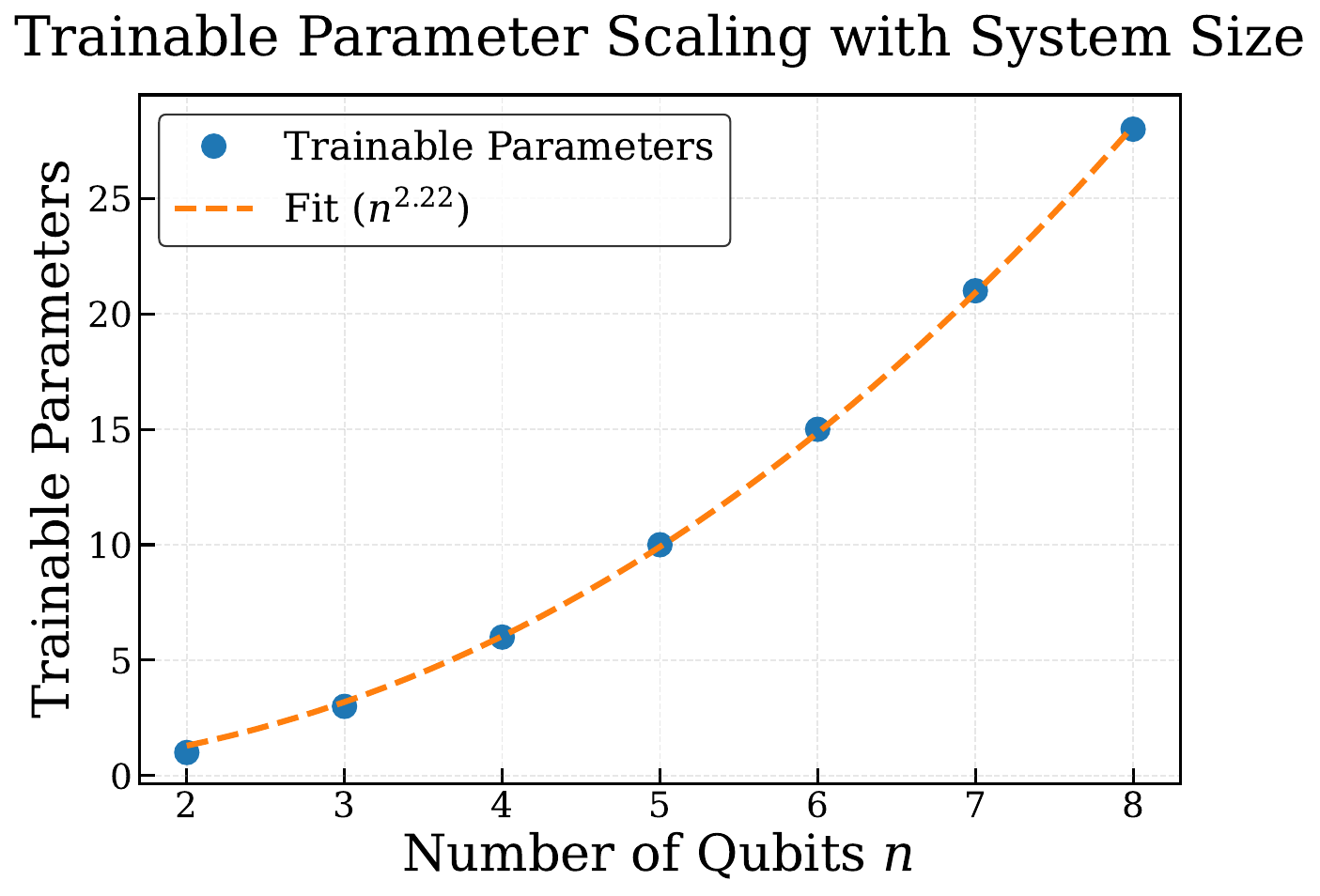}
\caption{}
\end{subfigure}
\hfill
\begin{subfigure}[t]{0.48\textwidth}
\centering
\includegraphics[width=\linewidth]{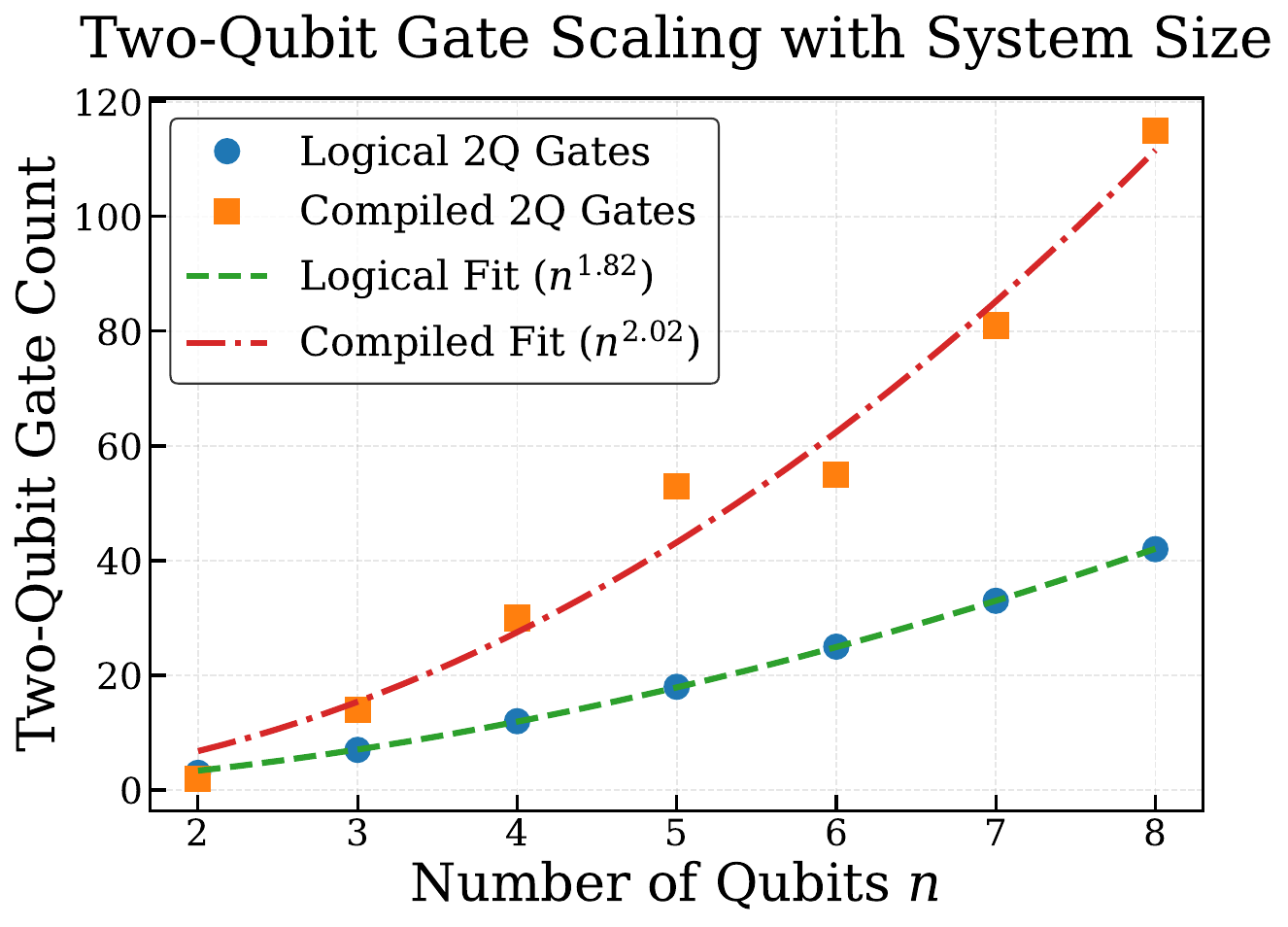}
\caption{}
\end{subfigure}

\caption{Hardware-resource scaling of the proposed variational quantum MIMO receiver. 
(a) Logical vs transpiled circuit depth as system size increases.
(b) Estimated execution time using realistic superconducting gate durations.
(c) Scaling of variational (trainable) parameters with system size.
(d) Logical vs compiled two-qubit gate counts after hardware-aware transpilation.
}
\label{fig:hardware_scaling}
\end{figure*}

\begin{table*}[t]
\caption{
Hardware-resource scaling analysis of the proposed variational quantum MIMO receiver.
}
\label{tab:hardware_resources}
\centering
\renewcommand{\arraystretch}{1.15}
\begin{tabular}{c c c c c c c c c}
\hline\hline
Qubits &
Trainable &
Logical &
Logical &
Logical &
Transpiled &
Compiled &
Compiled &
Estimated Exec. \\
$n$ &
Params &
Depth &
2Q Gates &
Total Gates &
Depth &
2Q Gates &
Total Gates &
Time ($\mu$s) \\
\hline

2 & 1  & 5  & 3  & 7  & 12  & 2   & 16  & 1.30 \\
3 & 3  & 9  & 7  & 13 & 53  & 14  & 89  & 7.95 \\
4 & 6  & 11 & 12 & 20 & 108 & 28  & 163 & 15.15 \\
5 & 10 & 13 & 18 & 28 & 168 & 49  & 279 & 26.20 \\
6 & 15 & 15 & 25 & 37 & 123 & 55  & 336 & 30.55 \\
7 & 21 & 17 & 33 & 47 & 181 & 81  & 476 & 44.05 \\
8 & 28 & 19 & 42 & 58 & 206 & 104 & 607 & 56.35 \\

\hline\hline
\end{tabular}
\end{table*}

To evaluate the practical scalability of the proposed variational quantum receiver, we examine how the required quantum resources grow with system dimension $n$. The analysis considers the scaling of trainable parameters, logical and compiled circuit complexity, execution time, and hardware resource requirements. These quantities collectively determine the feasibility of implementing the receiver on noisy intermediate-scale quantum (NISQ) hardware and provide insight into the factors that ultimately limit performance as the system size increases.

The intrinsic complexity of the variational architecture is governed by its interaction topology. Because the ansatz incorporates pairwise couplings between qubit pairs, the number of independent variational parameters is given by

\begin{equation}
N_{\theta}(n)
=
\binom{n}{2}
=
\frac{n(n-1)}{2},
\label{eq:param_scaling_theory}
\end{equation}

which exhibits quadratic growth with system size. The numerical results summarized in Table~\ref{tab:hardware_resources} and Fig.~\ref{fig:hardware_scaling}(c) are well described by

\begin{equation}
N_{\theta}(n)
=
0.278\,n^{2.220},
\qquad
R^2 = 0.9997.
\end{equation}

The fitted exponent is consistent with the asymptotic quadratic behavior predicted by Eq.~(\ref{eq:param_scaling_theory}). The slight deviation from the ideal quadratic dependence is attributed to finite-size effects over the range of system dimensions investigated and does not indicate a departure from the underlying theoretical behavior. These results confirm that the dominant contribution to the variational complexity originates directly from the all-to-all interaction structure of the ansatz.

The same interaction topology determines the scaling of the entangling resources required by the receiver. For the considered architecture, the logical two-qubit gate count may be approximated as
\begin{equation}
N_{2q}^{\mathrm{logical}}
=
\frac{n(n-1)}{2}
+
2(n-1),
\label{eq:logical_2q_theory}
\end{equation}
where the first term arises from the pairwise interaction network and the second term originates from the nearest-neighbor entangling layers employed in the channel and receiver circuits. In the large-system limit the quadratic contribution dominates, yielding
\begin{equation}
N_{2q}^{\mathrm{logical}}
=
\mathcal{O}(n^2).
\end{equation}
The measured resource data follow
\begin{equation}
N_{2q}^{\mathrm{logical}}
=
0.956\,n^{1.820},
\qquad
R^2 = 0.9999,
\end{equation}
demonstrating close agreement with the theoretical expectation. The fitted exponent remains sufficiently close to two to indicate that the asymptotic resource growth is governed by the interaction graph itself rather than by implementation-specific details. Since entangling operations dominate both circuit complexity and hardware error accumulation, this scaling law establishes the primary resource trend that governs the receiver as the system dimension increases.

Although the total number of entangling operations grows rapidly with system size, the circuit depth exhibits a qualitatively different behavior. Figure~\ref{fig:hardware_scaling}(a) shows that the logical depth is described by
\begin{equation}
D_{\mathrm{logical}}
=
3.324\,n^{0.842},
\qquad
R^2 = 0.9896.
\end{equation}
The sublinear exponent indicates that the growth in circuit complexity is accommodated primarily through an increase in circuit width rather than through a proportional increase in sequential operations. In other words, a significant fraction of the pairwise interactions can be executed in parallel. Consequently, the temporal complexity of the receiver increases substantially more slowly than the total gate count. This separation between depth scaling and gate-count scaling is an important characteristic of the architecture because it limits the growth of coherence-time requirements even as the interaction network becomes increasingly dense.

The favorable logical-depth scaling demonstrates that the intrinsic algorithmic complexity of the receiver remains moderate. However, practical implementation is ultimately determined by the compiled circuit rather than the logical representation. Mapping the dense interaction graph onto hardware with restricted qubit connectivity introduces additional routing and decomposition operations, generating an implementation overhead that grows with system size.

This behavior is evident from the transpiled-depth scaling shown in Fig.~\ref{fig:hardware_scaling}(a), which follows
\begin{equation}
D_{\mathrm{transpiled}}
=
14.263\,n^{1.387},
\qquad
R^2 = 0.8092.
\end{equation}
Compared with the logical-depth scaling, both the larger exponent and the substantially increased prefactor indicate that hardware constraints become progressively more important as the system size increases. The widening gap between logical and transpiled depth quantifies the cost of embedding a densely connected variational architecture onto a processor with sparse physical connectivity. Consequently, the dominant scalability challenge gradually shifts from the intrinsic complexity of the ansatz to the limitations imposed by the hardware architecture.

A similar trend is observed in the compiled two-qubit gate count shown in Fig.~\ref{fig:hardware_scaling}(d). The numerical results are well described by
\begin{equation}
N_{2q}^{\mathrm{compiled}}
=
1.673\,n^{2.020},
\qquad
R^2 = 0.9770,
\end{equation}
which preserves the asymptotic quadratic scaling predicted by the logical interaction topology while exhibiting a noticeably larger prefactor. The increase reflects the additional entangling operations introduced during compilation through gate decomposition and routing procedures. Because two-qubit gates constitute the dominant source of physical error on contemporary quantum processors, the compiled gate count provides a direct measure of the cumulative hardware exposure experienced during circuit execution.

The impact of these additional resources is reflected in the execution-time analysis shown in Fig.~\ref{fig:hardware_scaling}(b). Using realistic gate durations, the estimated runtime follows
\begin{equation}
T(n)
=
0.970\,n^{1.976},
\qquad
R^2 = 0.9844,
\end{equation}
which is effectively quadratic in system size. The runtime scaling closely mirrors the growth of the compiled entangling-gate resources, indicating that the execution latency is determined primarily by the increasing number of two-qubit operations. The observed behavior therefore confirms that the computational cost of the receiver remains polynomial throughout the investigated system range.

The resource trends revealed by the scaling analysis have direct implications for communication performance under realistic hardware conditions. As the compiled gate count and execution time increase, the circuit experiences greater cumulative exposure to depolarization, relaxation, and measurement errors. The resulting reduction in quantum-state fidelity degrades the distinguishability of the decoded codewords and consequently increases the probability of decoding error. From this perspective, the hardware-resource scaling provides a direct link between circuit complexity and the noise-induced performance degradation observed in larger systems.

Overall, the results presented in Table~\ref{tab:hardware_resources} and Fig.~\ref{fig:hardware_scaling} reveal a consistent hierarchy of scaling behaviors. The variational architecture exhibits an intrinsic near-quadratic growth in parameter count and entangling-gate complexity, reflecting the dense interaction topology employed by the receiver. At the same time, extensive gate parallelism suppresses the growth of logical depth, producing a substantially weaker dependence on system size than would be expected from the total gate count alone. Hardware compilation preserves the underlying polynomial scaling but introduces an increasingly significant overhead in both depth and entangling resources. Consequently, the principal scalability limitations originate from present-day hardware constraints rather than from the variational architecture itself. These observations suggest that improvements in processor connectivity, compilation efficiency, and fault-tolerant control will directly extend the accessible operating regime of the proposed quantum receiver while preserving its favorable algorithmic scaling characteristics.

\subsection{Trainability Analysis of the Variational Quantum Decoder}

\begin{figure*}[!t]
\centering

\begin{subfigure}[t]{0.48\textwidth}
\centering
\includegraphics[width=\linewidth]{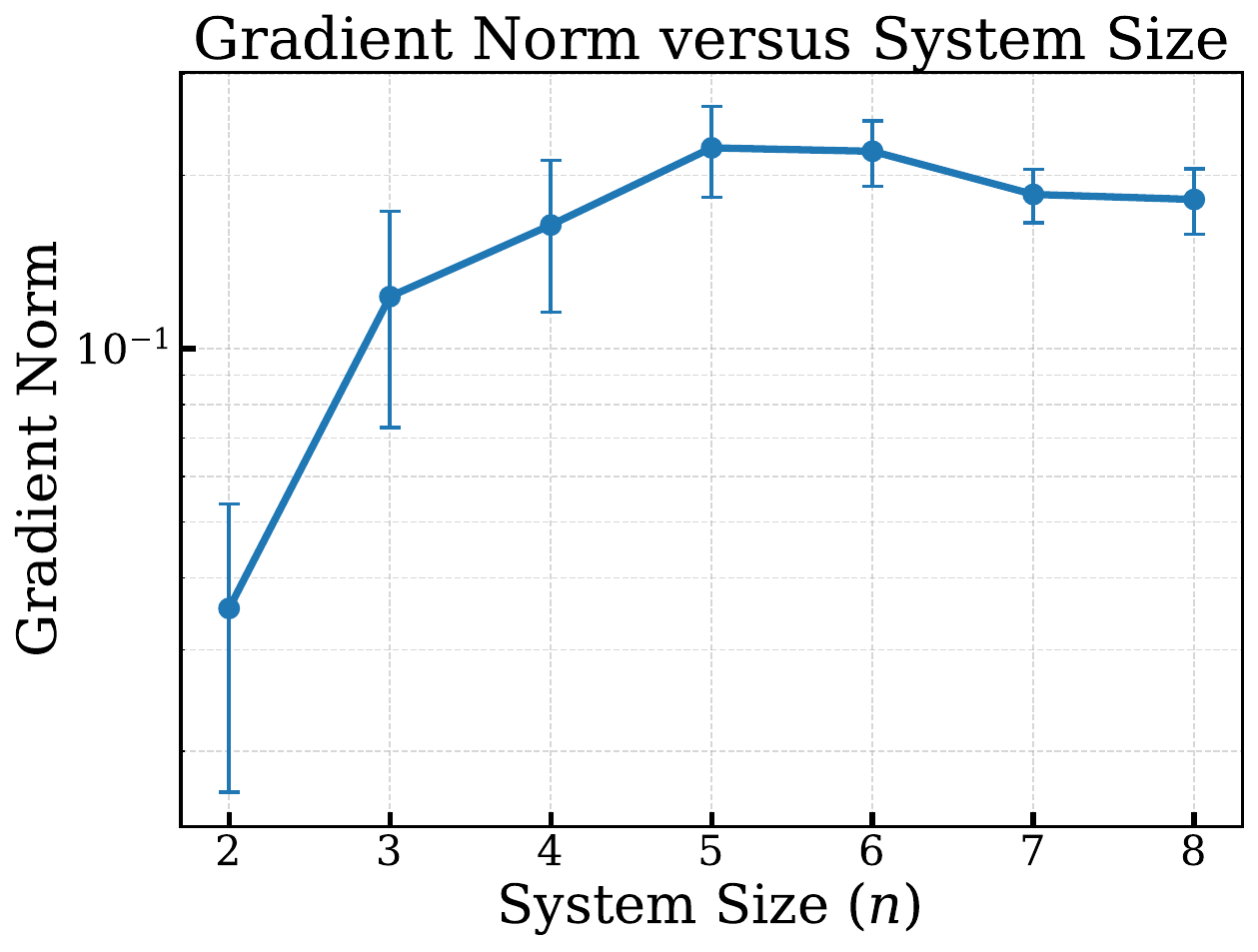}
\caption{}
\label{fig:grad_norm}
\end{subfigure}
\hfill
\begin{subfigure}[t]{0.48\textwidth}
\centering
\includegraphics[width=\linewidth]{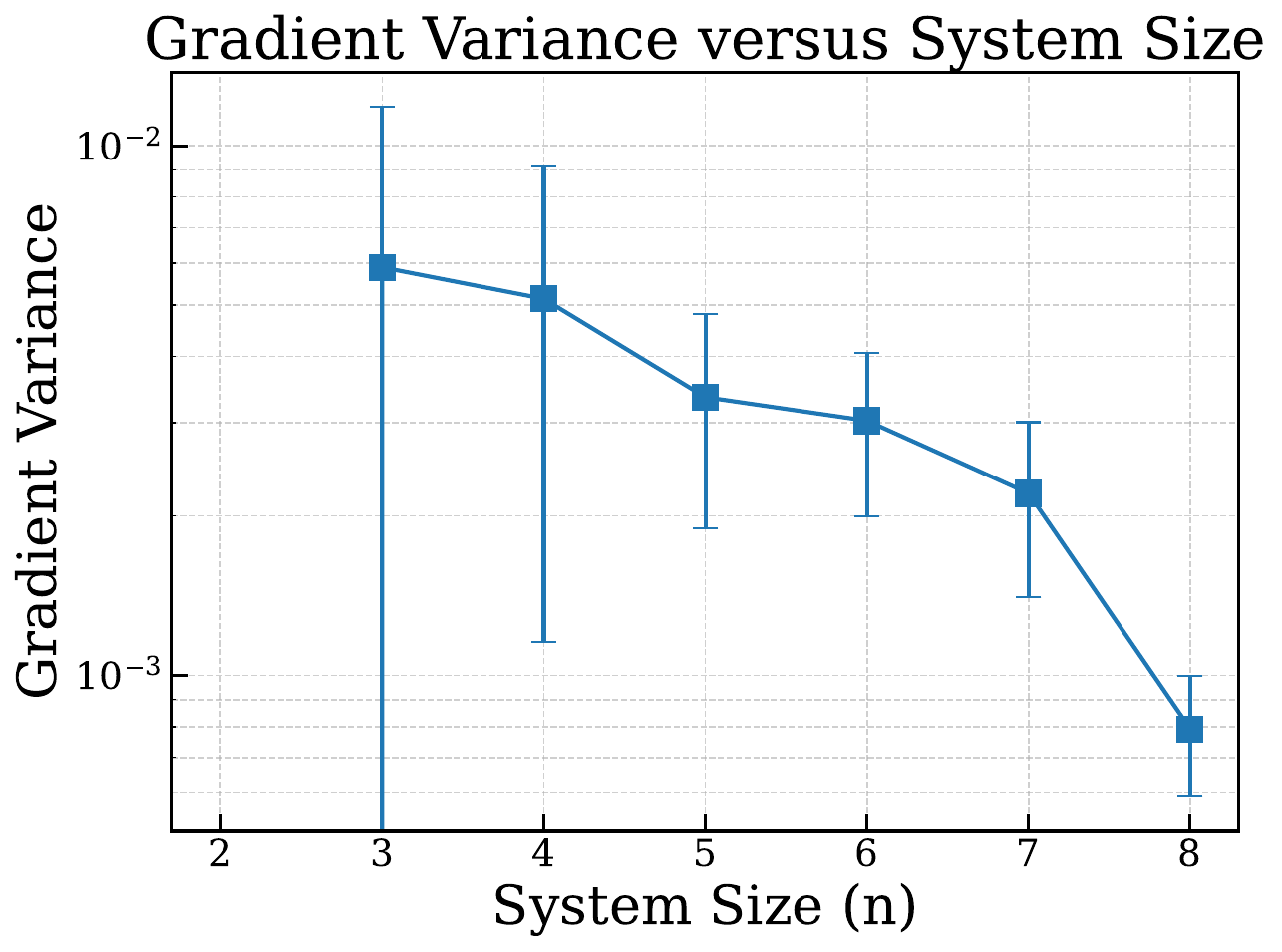}
\caption{}
\label{fig:grad_variance}
\end{subfigure}

\vspace{0.3cm}

\begin{subfigure}[t]{0.48\textwidth}
\centering
\includegraphics[width=\linewidth]{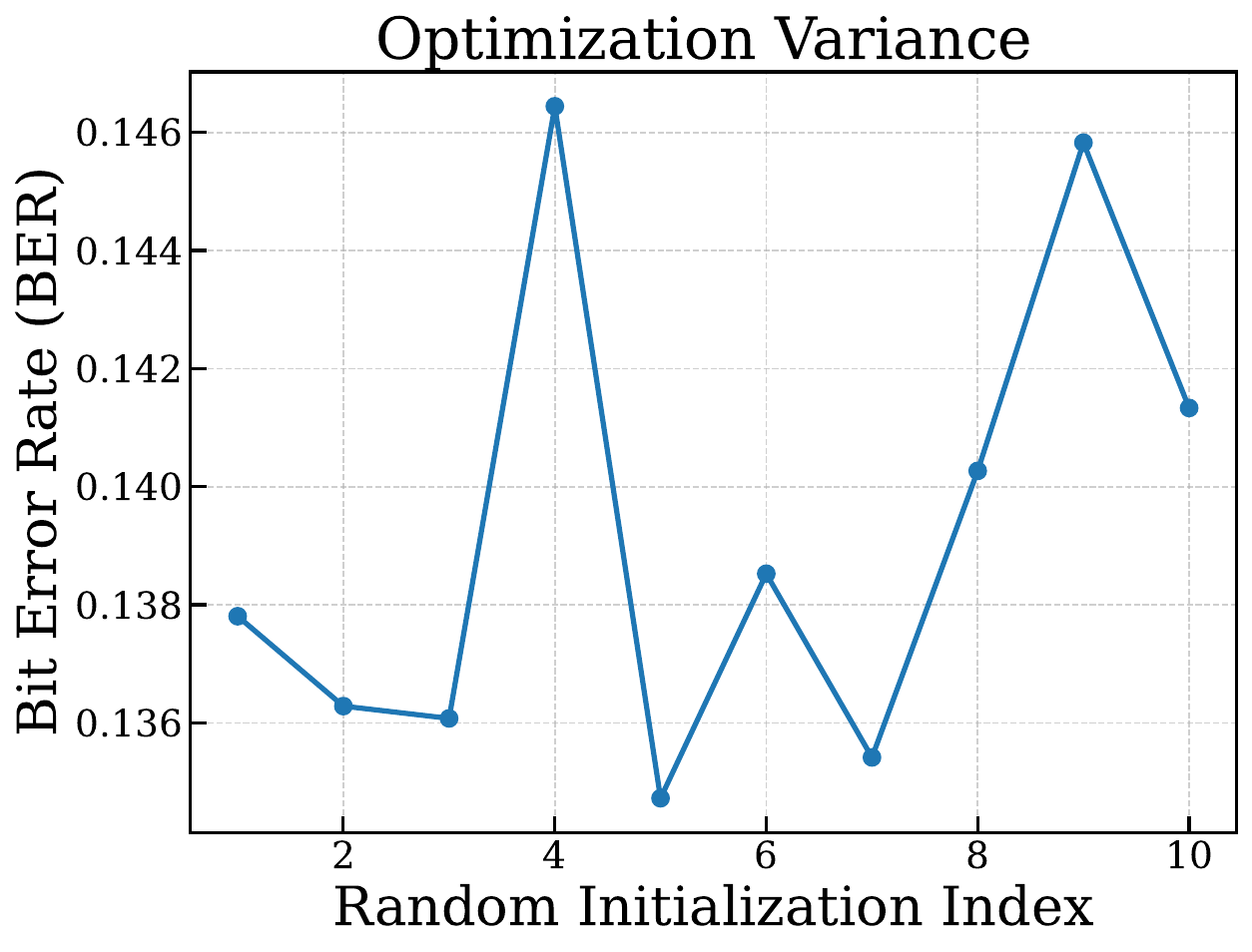}
\caption{}
\label{fig:optimization_variance}
\end{subfigure}
\hfill
\begin{subfigure}[t]{0.48\textwidth}
\centering
\includegraphics[width=\linewidth]{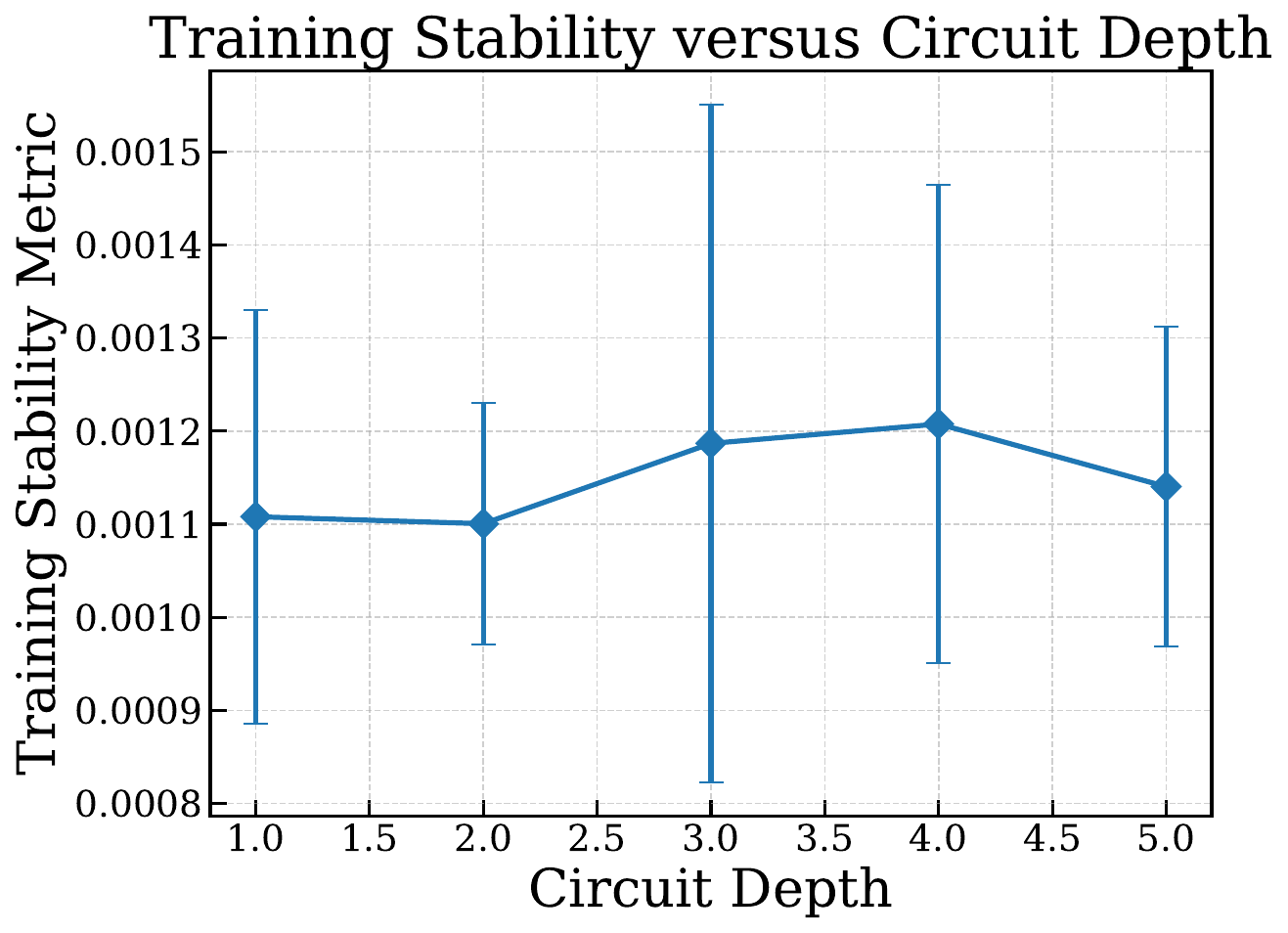}
\caption{}
\label{fig:training_stability}
\end{subfigure}

\caption{
Trainability analysis of the proposed variational quantum decoder. (a) Average gradient norm of the BER objective as a function of system size. (b) Gradient variance across variational parameters. (c) Final BER obtained from multiple optimization runs with different random initializations. (d) Training-stability metric as a function of circuit depth.
}
\label{fig:trainability_analysis}
\end{figure*}
To assess the trainability of the proposed variational decoder, we analyze the scaling of gradient norms, optimization variability under random initialization, and the sensitivity of the BER objective to parameter perturbations. All results are obtained from the complete noisy receiver model and therefore reflect the optimization landscape of the physically relevant decoding objective.

Figure~\ref{fig:grad_norm} presents the average gradient norm of the BER objective as a function of system size. Despite the increase in both Hilbert-space dimension and variational parameter count, the gradient norm remains finite throughout the investigated range. No systematic exponential suppression is observed, indicating that optimization signals remain accessible as the system size increases.

From a trainability perspective, the results indicate that the investigated decoder does not enter a pronounced barren-plateau regime within the considered system sizes. Although a gradual reduction in gradient magnitude is observed with increasing $n$, the gradients remain well above the level expected for optimization landscapes dominated by exponential gradient concentration. The preservation of finite gradients suggests that the dominant effect of increasing system size is not a loss of trainability but rather the accumulation of physical noise within the communication channel.

Additional information regarding the structure of the optimization landscape is provided by the gradient-variance analysis shown in Fig.~\ref{fig:grad_variance}. Whereas the gradient norm characterizes the overall magnitude of the optimization signal, the gradient variance quantifies how that signal is distributed across the variational parameter space. A rapid collapse of the variance would indicate that the gradients become increasingly concentrated around their mean value, reducing the ability of individual parameters to contribute distinct optimization directions.

As shown in Fig.~\ref{fig:grad_variance}, the gradient variance decreases gradually with increasing system size but remains finite throughout the investigated regime. The reduction is considerably weaker than would be expected from an optimization landscape undergoing strong gradient concentration, indicating that the influence of different variational parameters remains statistically distinguishable even for larger systems. Consequently, the optimization process continues to receive meaningful directional information from multiple regions of parameter space rather than being dominated by uniformly small gradient contributions.

Taken together, the gradient-norm and gradient-variance results indicate that the optimization landscape retains both finite optimization signals and meaningful directional information as the system size increases. The persistence of nonvanishing gradient magnitudes and finite gradient variance suggests that the optimization process continues to receive informative updates across the variational parameter space, rather than becoming dominated by uniformly small gradients. This behavior can be attributed to the structured nature of the decoder architecture, in which each variational parameter is associated with a physically motivated pairwise interaction inherited from the communication channel. Because these parameters retain distinct contributions to the decoding process, the optimization landscape preserves a nontrivial distribution of gradient components even as the dimensionality of the problem increases. Consequently, the available optimization information remains sufficiently rich to support effective training throughout the investigated regime, providing no evidence of a pronounced barren-plateau phenomenon within the considered system sizes.

To evaluate the robustness of the optimization process, Fig.~\ref{fig:optimization_variance} shows the final BER obtained from multiple independent optimization runs initialized with different random parameter vectors. Despite the nonconvex nature of the optimization problem, all runs converge to similar BER values with only minor variations across initializations. This behavior indicates that the optimization outcome is relatively insensitive to the choice of initial parameters and that the decoder consistently converges to solutions of comparable quality. The small spread in the final BER values therefore demonstrates the reproducibility and robustness of the proposed variational receiver.

Additional insight into the optimization behavior is provided by the training-stability analysis shown in Fig.~\ref{fig:training_stability}, which examines the dependence of the training dynamics on the depth of the variational circuit. The average stability metric remains approximately constant across the investigated depth range, exhibiting only minor fluctuations around its mean value. A modest increase is observed for intermediate depths, followed by a slight reduction at the largest depth considered. However, the variation remains small relative to the overall scale of the metric, indicating that increasing circuit depth does not substantially alter the optimization dynamics.

These observations are consistent with the gradient-based analyses and suggest that the additional circuit layers primarily enhance the expressivity of the decoder without significantly increasing the complexity of the optimization landscape. Although deeper circuits contain more variational parameters, the structured receiver architecture preserves accessible optimization directions, preventing the emergence of unstable training behavior. As a result, the optimization process remains well behaved across the investigated depth range, indicating that the considered circuit depths do not introduce significant trainability limitations for the proposed variational decoder.

Overall, the trainability metrics reveal a consistent picture of the optimization landscape. The BER objective maintains finite gradient signals as both the Hilbert-space dimension and parameter-space dimension increase, optimization outcomes remain largely independent of initialization, and local parameter perturbations produce only modest variations in the objective value. These observations indicate that the optimization problem remains well conditioned throughout the investigated regime and that the variational decoder can be trained reliably under realistic noisy conditions.

\subsection{Comparison with Classical MIMO Systems}

\begin{figure*}[!t]
\centering

\begin{subfigure}[t]{0.48\textwidth}
\centering
\includegraphics[width=\linewidth]{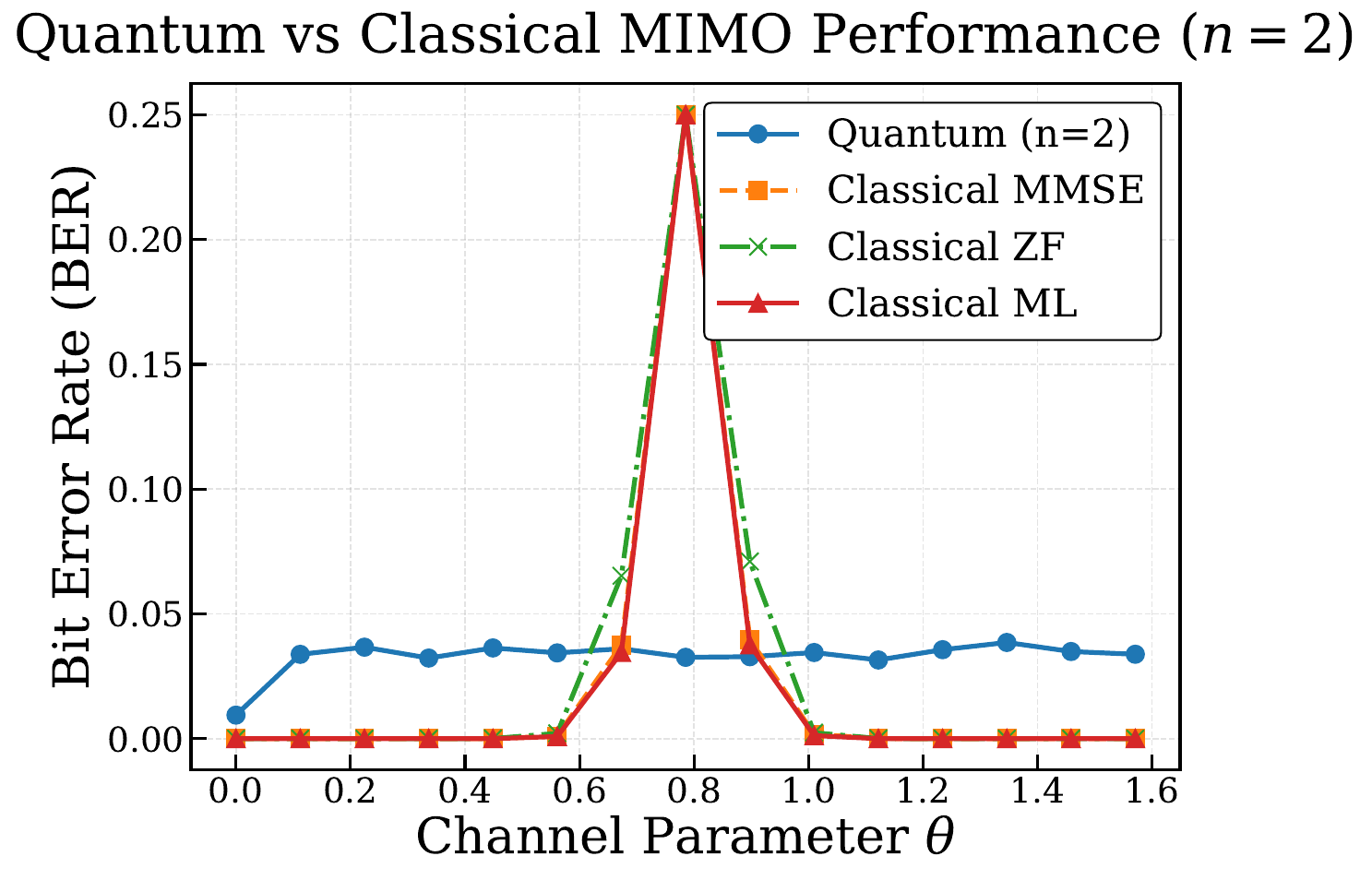}
\caption{$n=2$}
\end{subfigure}
\hfill
\begin{subfigure}[t]{0.48\textwidth}
\centering
\includegraphics[width=\linewidth]{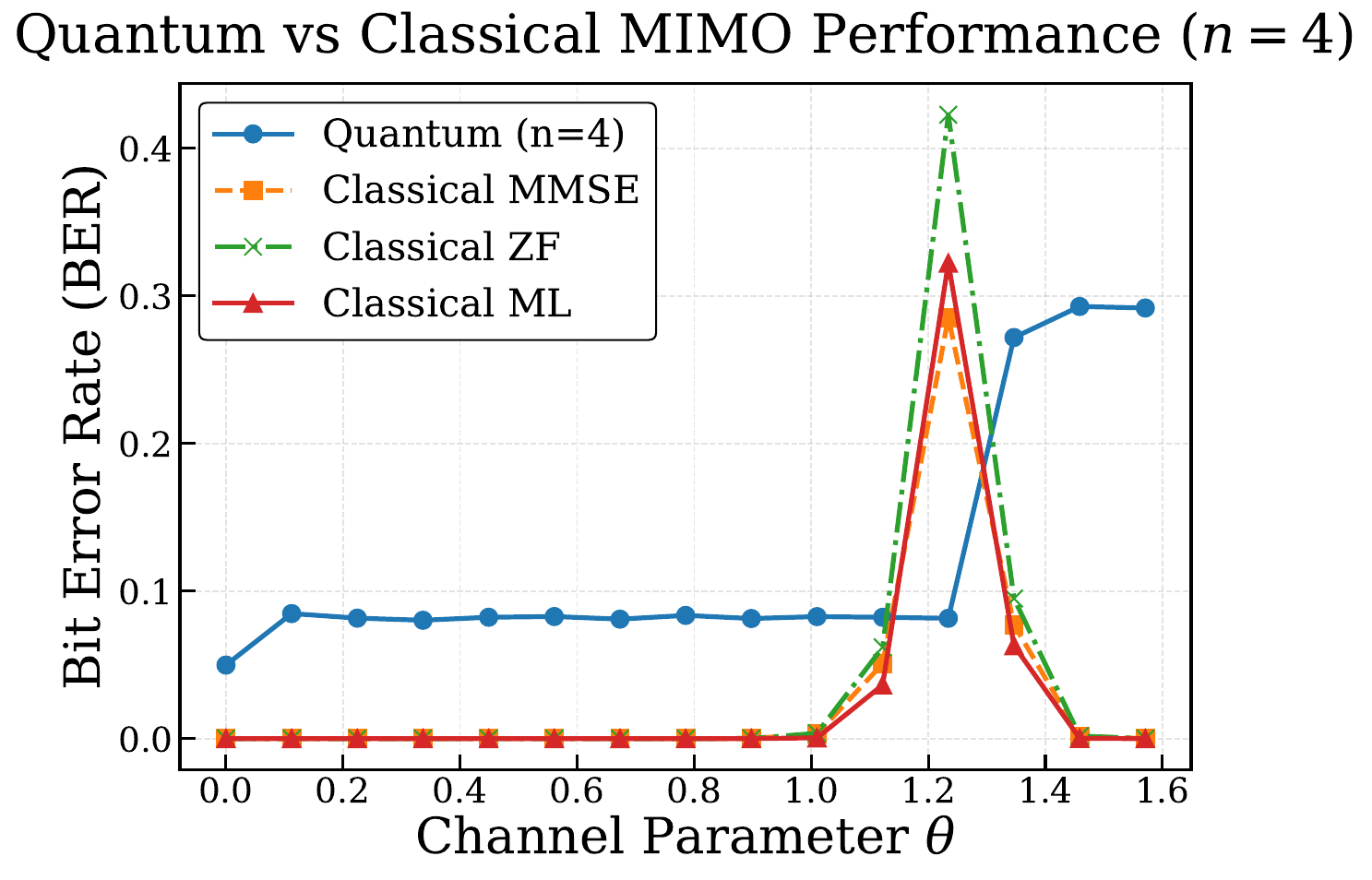}
\caption{$n=4$}
\end{subfigure}

\vspace{0.3cm}

\begin{subfigure}[t]{0.48\textwidth}
\centering
\includegraphics[width=\linewidth]{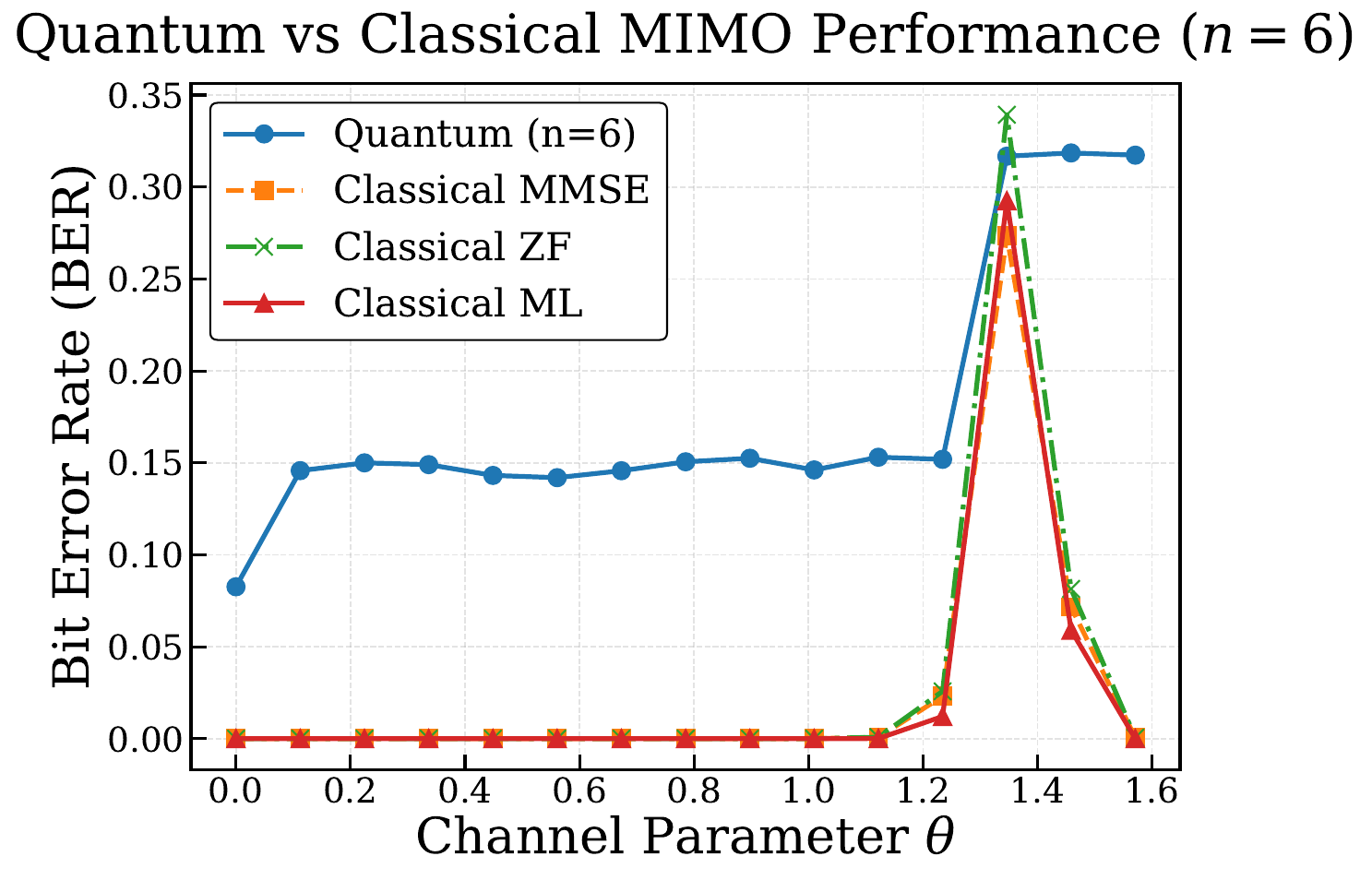}
\caption{$n=6$}
\end{subfigure}
\hfill
\begin{subfigure}[t]{0.48\textwidth}
\centering
\includegraphics[width=\linewidth]{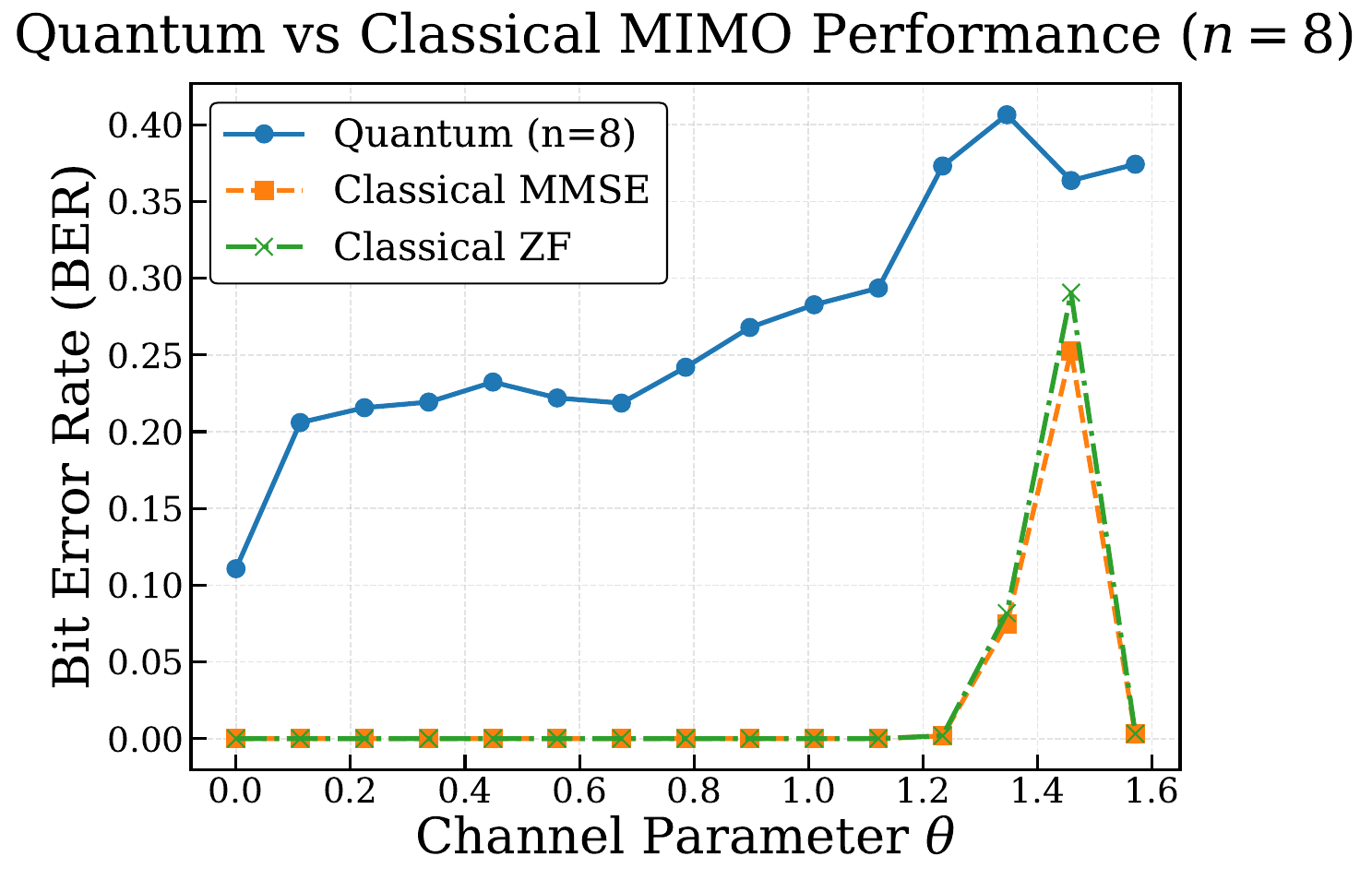}
\caption{$n=8$}
\end{subfigure}

\caption{
Bit error rate comparison between the proposed quantum MIMO framework and classical MMSE, ZF, and ML detectors for different system sizes.
}
\label{fig:qber_scaling}
\end{figure*}

The performance of the proposed quantum MIMO receiver is compared with several classical detection techniques, including Minimum Mean Square Error (MMSE), Zero-Forcing (ZF), and Maximum Likelihood (ML), across different system dimensions. The corresponding BER performance as a function of the channel coupling parameter $\theta$ is shown in Fig.~\ref{fig:qber_scaling}.

A comparative analysis for system sizes $(n=2,4,6,8)$ reveals a clear distinction between the behavior of the variational quantum receiver and conventional classical detectors. For small values of the coupling parameter $\theta$, the effective channel matrix remains well-conditioned and inter-stream interference is limited. In this regime, MMSE, ZF, and ML achieve nearly error-free detection across all considered dimensions, reflecting the ability of classical algorithms to accurately invert or decode the channel when the transmission modes remain weakly correlated. The variational quantum receiver, however, exhibits a finite residual BER even in these favorable conditions, indicating that the learned variational transformation does not fully reproduce the optimal inverse channel operation.

As the coupling strength increases, coherent mixing between transmission modes generates increasingly correlated interference patterns. The resulting channel matrix develops regions of poor conditioning in which multiple communication paths become strongly coupled. These interference-induced spectral distortions produce localized degradation in classical detector performance, particularly for inversion-based methods such as ZF. The severity and location of these degradation regions depend on the dimensionality of the system and the structure of the effective channel spectrum.

For the $n=2$ system, all classical detectors maintain near-zero BER across most of the coupling range. A localized degradation region appears near $\theta \approx 0.8$, where coherent channel mixing reduces symbol separability and produces a temporary BER increase for MMSE, ZF, and ML detection. In contrast, the VQC receiver exhibits an approximately constant BER throughout the entire range. Although this avoids sharp performance fluctuations, the quantum receiver remains consistently less accurate than the classical detectors.

For the $n=4$ system, the instability region shifts toward larger coupling strengths, appearing near $\theta \approx 1.2$--$1.3$. Classical detectors again achieve near-zero BER across most of the parameter range before undergoing a localized performance deterioration. The VQC receiver exhibits a qualitatively different trend, maintaining an approximately constant BER for moderate coupling strengths before degrading more noticeably for $\theta \gtrsim 1.4$. This behavior indicates that the learned variational transformation becomes increasingly challenged as the number of coupled transmission modes grows and the interference structure becomes more complex.

The behavior becomes more pronounced for the $n=6$ system. Classical detectors remain nearly error-free until the channel approaches a strongly correlated configuration near $\theta \approx 1.35$, where a sharp BER increase is observed. Although the VQC receiver also exhibits degradation in the vicinity of this coupling region, its BER remains significantly larger throughout the entire parameter space. Consequently, while the variational decoder appears sensitive to the same challenging channel configurations as the classical detectors, its overall decoding accuracy remains substantially lower.

The localized BER spikes observed for the classical detectors across $n=2$, $4$, and $6$ originate from the same underlying physical mechanism. As $\theta$ increases, coherent mixing progressively couples the transmission modes, generating highly correlated channel configurations in which symbol distinguishability is reduced. These regions correspond to particularly challenging effective channel realizations where multiple transmission paths become strongly intertwined. Such configurations increase the difficulty of reliably recovering the transmitted bit strings and lead to transient BER degradation even for optimal ML detection. The effect is particularly severe for ZF because its direct inversion procedure amplifies small perturbations associated with poorly conditioned channel realizations, while MMSE partially mitigates this behavior through regularization. In contrast, the VQC receiver does not exhibit narrow instability regions but instead experiences a broader loss of decoding accuracy, suggesting that its performance is governed by a more global sensitivity to increasing channel complexity rather than by localized inversion-related failures.

For the $n=8$ system, the disparity between the quantum and classical approaches becomes most significant. MMSE and ZF continue to achieve nearly perfect detection over most values of $\theta$, with degradation confined to a narrow region near $\theta \approx 1.45$ where strong mode coupling produces substantial interference. As in lower dimensions, ZF experiences the largest BER increase due to its sensitivity to difficult channel realizations. The ML detector is omitted for $n=8$ because its computational complexity scales exponentially with system size, requiring evaluation of $2^n$ possible symbol configurations. The variational quantum receiver, however, exhibits a steadily increasing BER with system dimension and coupling strength, reaching its worst performance in the strongly mixed regime. Unlike the classical detectors, whose degradation remains localized around specific interference configurations, the quantum receiver experiences a broad reduction in decoding accuracy across the entire parameter range.

Overall, the results indicate that the proposed variational quantum receiver exhibits qualitatively different behavior from conventional classical detectors. Across all investigated system dimensions, MMSE, ZF, and ML generally achieve lower BER and remain highly effective in well-conditioned channel environments. In particular, ML provides the best detection performance whenever its computational cost remains tractable, while MMSE and ZF continue to offer competitive performance across a broad range of coupling strengths.

The objective of this study is therefore not to demonstrate a universal replacement for established classical detection methods, but rather to investigate the feasibility and behavior of variational quantum equalization within quantum-channel environments characterized by coherent interference and strong coupling. Although the proposed VQC receiver does not achieve the detection accuracy of the classical benchmarks in the considered scenarios, it provides an alternative receiver architecture based on variational quantum inference and exhibits a distinct response to increasing channel complexity. These observations suggest that variational quantum equalization may represent a useful complementary framework for studying signal recovery in strongly coupled quantum communication systems and motivate further investigation of more expressive quantum receiver designs and training strategies.

\subsection{Classical-Quantum Complexity Comparison}
\begin{figure*}[!t]
\centering

\begin{subfigure}[t]{0.48\textwidth}
\centering
\includegraphics[width=\linewidth]{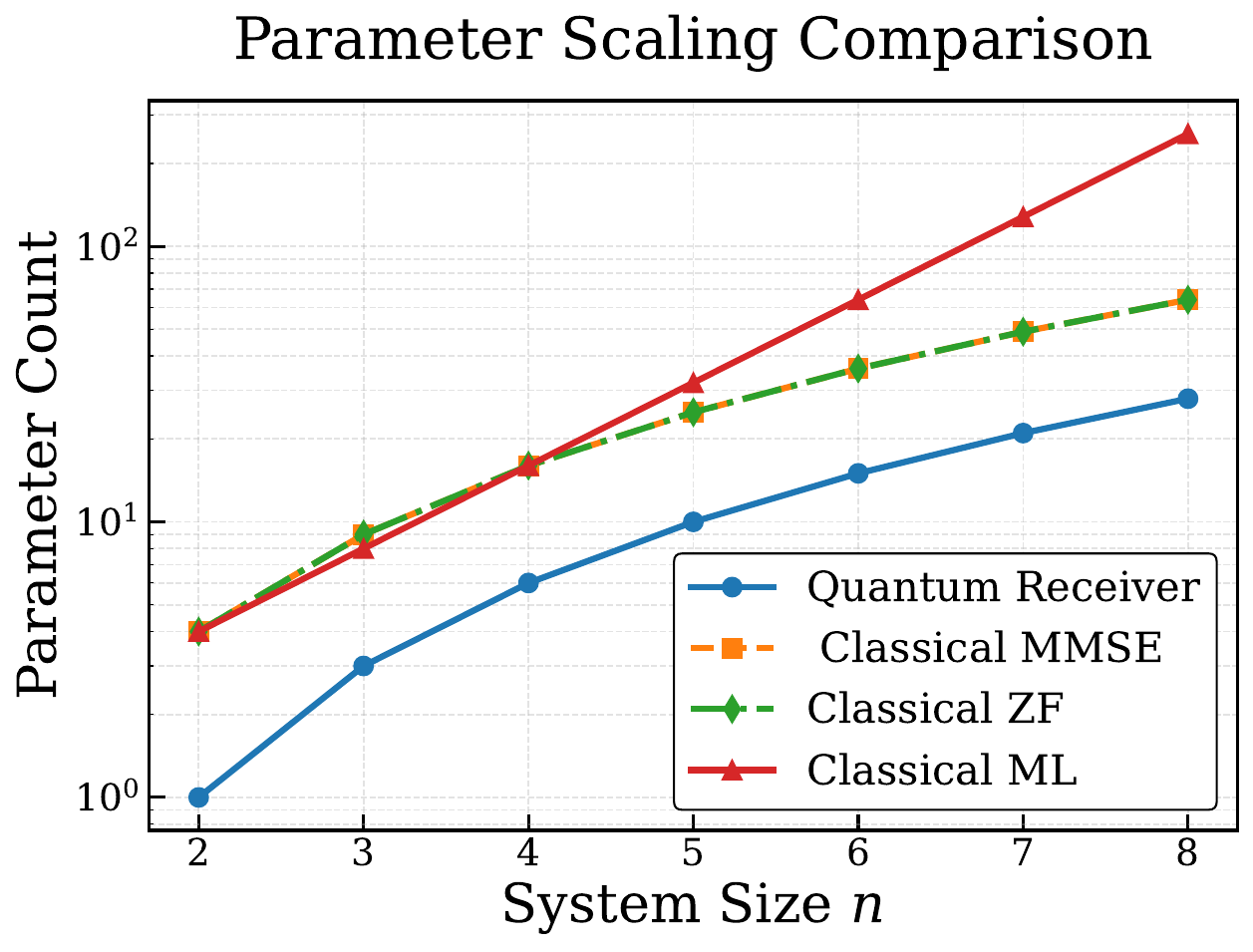}
\end{subfigure}
\hfill
\begin{subfigure}[t]{0.48\textwidth}
\centering
\includegraphics[width=\linewidth]{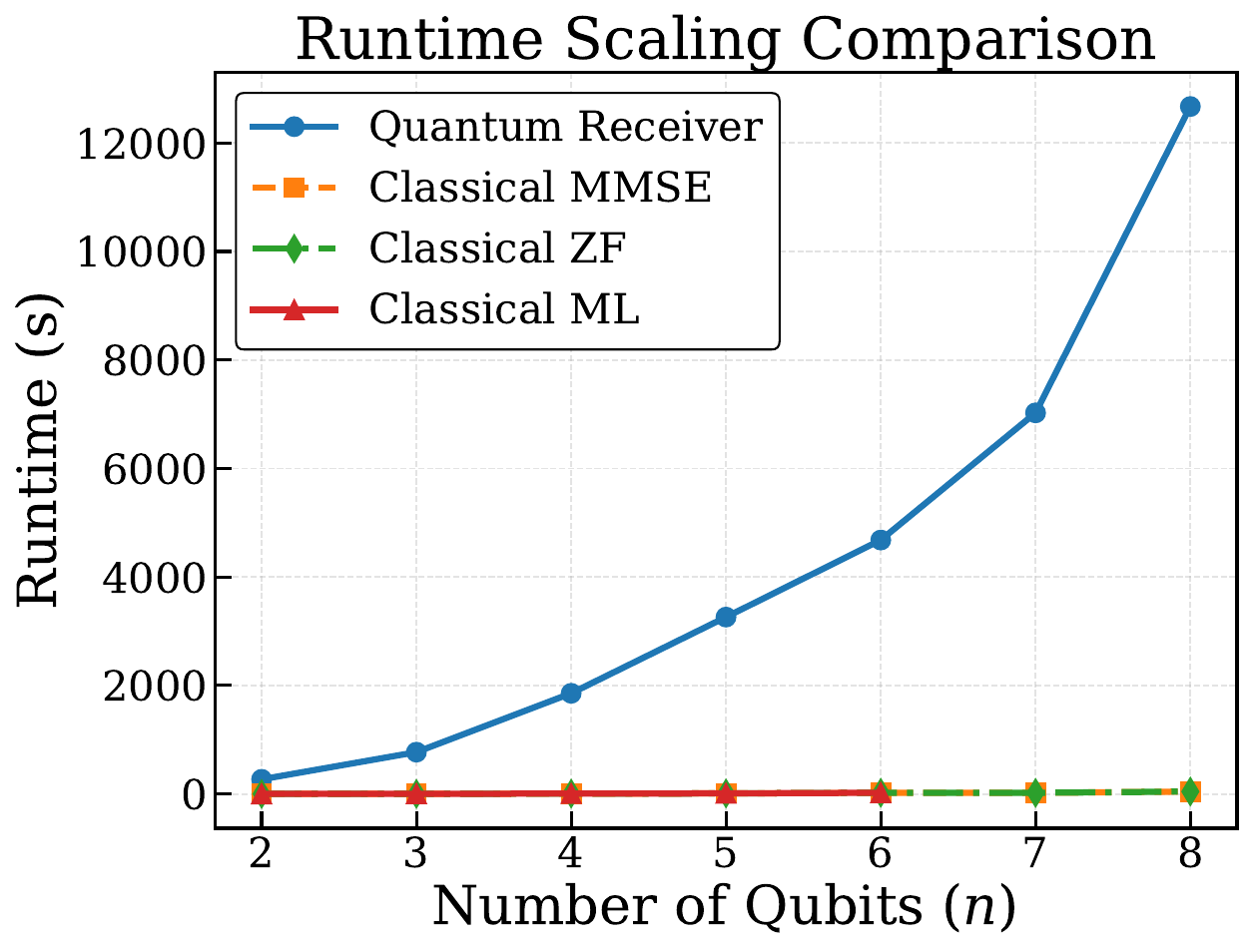}
\end{subfigure}
\vspace{0.5 cm}

\begin{subfigure}[t]{0.48\textwidth}
\centering
\includegraphics[width=\linewidth]{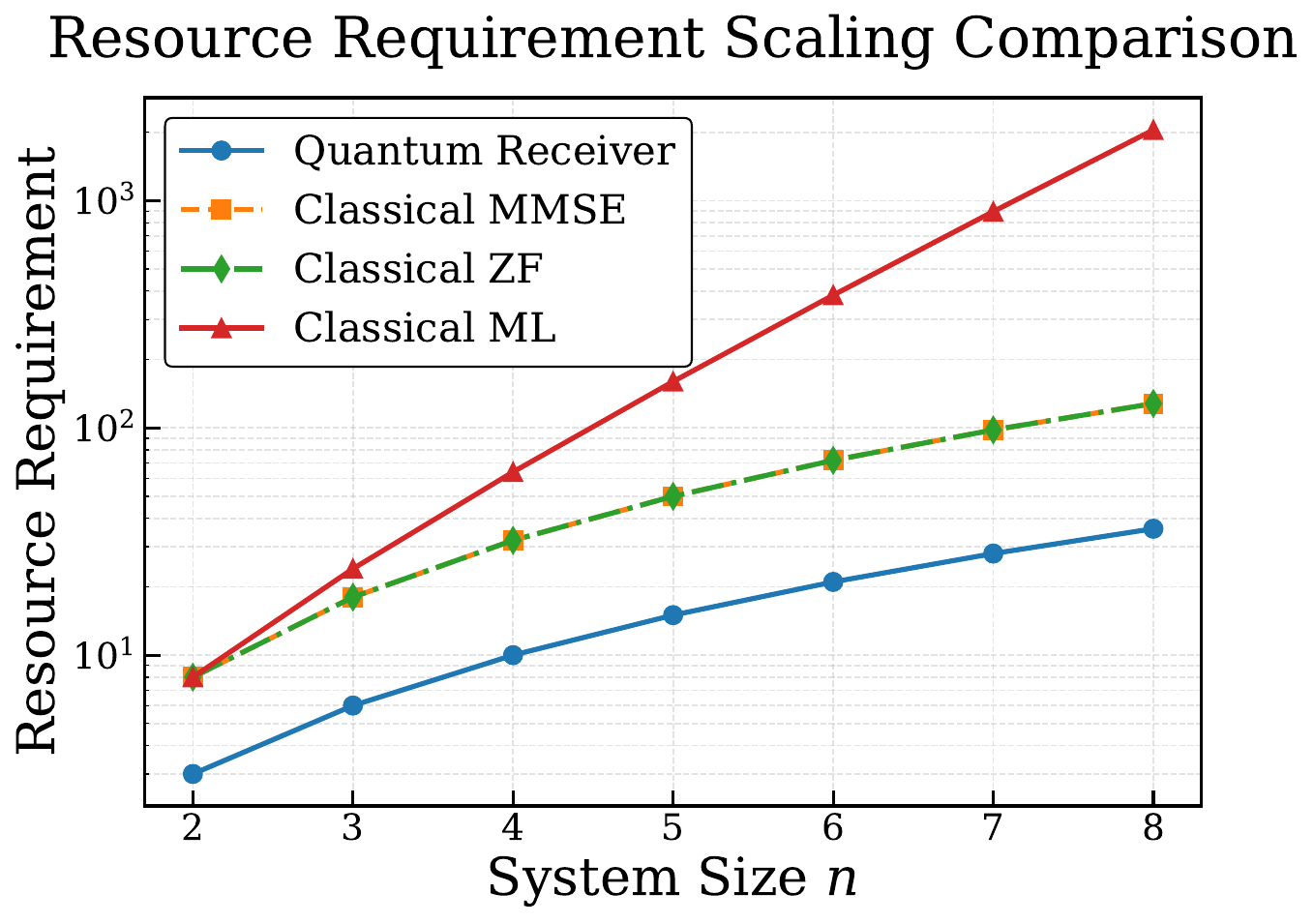}

\end{subfigure}
\hfill
\begin{subfigure}[t]{0.48\textwidth}
\centering
\includegraphics[width=\linewidth]{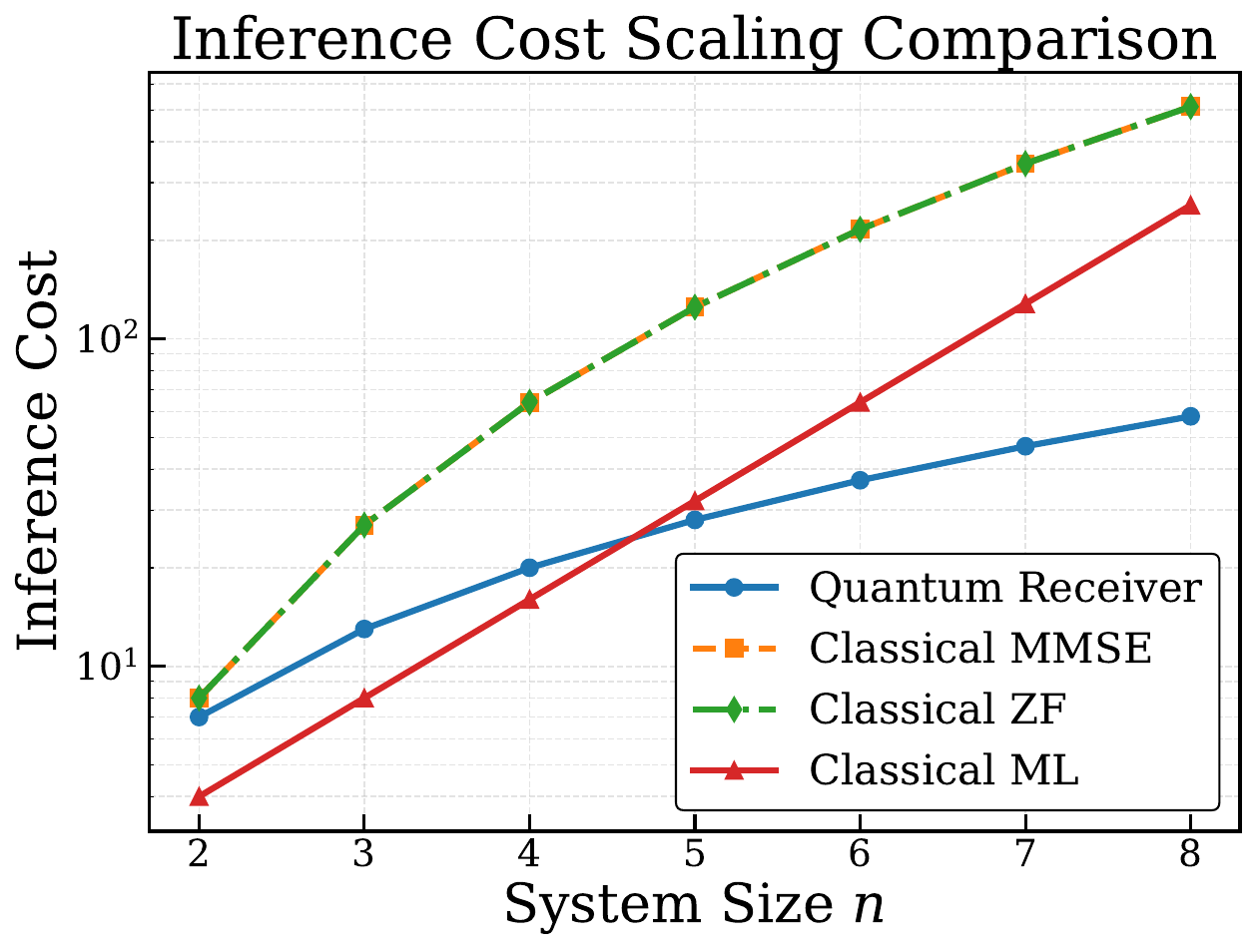}

\end{subfigure}

\caption{
Computational scalability comparison between the proposed VQC-based
quantum receiver and classical MMSE, ZF, and ML detectors.
(a) Resource requirement scaling.
(b) Runtime scaling.
(c) Inference-cost scaling.
(d) Parameter scaling.
}
\label{fig:computational_scalability}
\end{figure*}
\begin{table*}[t]
\caption{
Resource scaling comparison between the proposed variational quantum receiver
and conventional classical MIMO detection algorithms for an $n \times n$
system.
}
\label{tab:resource_scaling}
\centering
\begin{ruledtabular}
\begin{tabular}{ccccc}
System Size $n$
& Quantum Receiver
& MMSE
& ZF
& ML
\\
\hline
2 & 3  & 8   & 8   & 8    \\
3 & 6  & 18  & 18  & 24   \\
4 & 10 & 32  & 32  & 64   \\
5 & 15 & 50  & 50  & 160  \\
6 & 21 & 72  & 72  & 384  \\
7 & 28 & 98  & 98  & 896  \\
8 & 36 & 128 & 128 & 2048 \\
\hline
Asymptotic Scaling
& $\mathcal{O}(n^2)$
& $\mathcal{O}(n^2)$
& $\mathcal{O}(n^2)$
& $\mathcal{O}(n\,2^n)$
\\
\label{tab:complexity_comparison}
\end{tabular}
\end{ruledtabular}
\end{table*}

To further assess the practicality of the proposed VQC-based quantum MIMO receiver, we examine its computational scalability and compare its resource requirements with those of conventional classical detection schemes. Beyond communication performance, large-scale deployment requires efficient utilization of computational resources, favorable runtime characteristics, manageable inference costs, and controlled model complexity. The corresponding results are presented in Figs.~\ref{fig:computational_scalability} and Table~\ref{tab:complexity_comparison}.

\subsubsection{Parameter Scaling and Model Complexity}

The proposed quantum receiver requires significantly fewer trainable parameters than the corresponding classical detector models across the entire range of system dimensions. Moreover, the parameter count grows gradually as the number of qubits increases, indicating that the variational circuit maintains a compact representation even for larger systems.

By comparison, the parameter requirements of the classical ML detector increase much more rapidly with system size. This reflects the growing complexity of representing and processing high-dimensional signal constellations within conventional detection frameworks. MMSE and ZF exhibit intermediate scaling behavior but still require a larger parameter count than the quantum receiver for the investigated system sizes.

The compact parameterization of the VQC is advantageous from both optimization and implementation perspectives. Fewer trainable parameters generally reduce memory requirements, simplify optimization, and improve model interpretability while simultaneously lowering the hardware resources required for deployment.

\subsubsection{Runtime Scaling Analysis}

The runtime of the proposed quantum receiver increases significantly with system dimension, reflecting the growing complexity of variational optimization, quantum-circuit simulation, and parameter-training procedures. Unlike analytical detectors such as MMSE and ZF, whose execution primarily consists of deterministic matrix operations, the quantum receiver requires repeated circuit evaluations throughout the optimization process. Consequently, the measured runtime in the simulation environment exceeds that of the classical baselines.

From a practical implementation perspective, the observed runtime is
primarily influenced by the variational training process and the
classical simulation of quantum circuits. Consequently, the reported
computational cost reflects the overhead associated with the current
simulation framework rather than the inference stage alone. These
results suggest that improvements in quantum hardware and execution
platforms could further enhance the efficiency of the proposed
variational detection scheme.

It should also be noted that the reported runtime includes the complete variational-learning workflow, including repeated quantum-circuit evaluations, classical parameter optimization, and circuit simulation overhead. Therefore, the measured execution times should not be interpreted as evidence of quantum computational speedup, but rather as an assessment of the practical computational requirements associated with implementing variational quantum equalization within the present simulation framework.

\subsubsection{Resource Requirement Scaling}

For the investigated system sizes, the resource requirements of the proposed quantum receiver exhibit a slower empirical growth trend than those of the considered classical detection schemes. While MMSE and ZF exhibit approximately polynomial growth with increasing system dimension, the ML detector experiences the most rapid increase due to the combinatorial nature of exhaustive symbol-search operations. In contrast, the quantum receiver demonstrates a comparatively moderate growth rate, indicating improved scalability for larger MIMO configurations.

This behavior originates from the variational representation of the receiver. Instead of explicitly performing large matrix inversions or exhaustive likelihood evaluations, the VQC learns an approximate decoding transformation through a compact set of trainable quantum operations. Consequently, the number of computational resources required to represent and execute the detection process increases more gradually with system size than in conventional classical approaches.

The separation between the quantum and classical curves becomes increasingly pronounced as the system dimension grows, suggesting that the relative advantage of the variational quantum receiver improves in larger communication systems. This observation is particularly significant because scalability remains one of the primary limitations of optimal classical detection algorithms in massive MIMO deployments.

\subsubsection{Inference Cost Analysis}
The MMSE and ZF detectors exhibit cubic inference-cost scaling with system dimension due to the matrix inversion required during detection. In contrast, the proposed VQC receiver exhibits quadratic scaling, as the circuit depth is determined by the fixed pairwise interaction structure of the variational ansatz.

This behavior arises from the separation between parameter optimization and detection. The variational parameters are obtained during the training stage, whereas symbol detection is subsequently performed using a fixed quantum circuit. Consequently, the inference complexity is governed primarily by circuit execution and measurement operations rather than by parameter optimization.

The ML detector exhibits a fundamentally different complexity scaling. Because likelihood evaluation requires consideration of all possible transmitted symbol configurations, its inference cost grows exponentially with system dimension as $2^n$. For small systems ($n \leq 4$), the ML detector incurs a lower inference cost than the proposed VQC receiver. However, the exponential growth of the ML complexity leads to a crossover near $n=5$, beyond which the VQC receiver becomes increasingly favorable. At $n=8$, the ML inference cost exceeds that of the VQC receiver by approximately a factor of $4.3$.

From an implementation perspective, the operational cost is determined by the inference stage, since the learned variational parameters remain fixed after training. The runtime complexity of the receiver is therefore dominated by circuit execution rather than parameter learning.

Overall, the results demonstrate that the proposed VQC receiver exhibits more favorable complexity scaling than the MMSE, ZF, and ML benchmark detectors. Combined with its BER performance, this suggests that variational quantum detection may offer a computationally attractive solution for larger quantum communication systems, where the cost of conventional detection methods increases rapidly with system dimension.

\subsection{Discussion}

The results demonstrate that variational quantum receivers provide a feasible framework for adaptive decoding in quantum MIMO communication systems operating under realistic NISQ noise conditions. Across the investigated system sizes, the receiver successfully recovered transmitted information while retaining the ability to recover transmitted information despite the presence of depolarization, relaxation, and measurement noise.These findings suggest that variational quantum algorithms may serve as useful communication-theoretic tools in noisy quantum environments where explicit channel inversion is difficult to realize.

A key observation is the trade-off between communication capacity and decoding reliability as the system size increases. Larger quantum systems offer access to higher-dimensional Hilbert spaces and increased information-carrying capability, but they also generate more complex correlation structures that become increasingly difficult to decode in the presence of noise. Consequently, the benefits of increased information capacity are partially offset by the growth of decoding errors and optimization complexity.

The entanglement analysis provides insight into the physical origin of this behavior. The observed increase in bipartite entropy and mutual information indicates that transmitted information becomes progressively distributed across correlated many-body quantum modes as the interaction network expands. From a communication-theoretic perspective, these findings identify entanglement as a fundamental resource within the quantum MIMO channel, enabling information to be encoded across multipartite quantum correlations rather than localized pairwise states. While such distributed encoding enhances the channel's information-carrying capability and expressive capacity, it also increases the complexity of reliable decoding under realistic noise conditions, contributing to the performance degradation observed for larger systems.

The scalability analysis further indicates that current hardware limitations constitute the primary obstacle to large-scale implementation. Although the logical circuit depth grows relatively slowly, hardware compilation introduces significant overhead in circuit depth and two-qubit gate resources. Since two-qubit operations remain the dominant source of error on contemporary quantum processors, improvements in hardware connectivity, compilation efficiency, and error mitigation are expected to directly enhance the achievable performance of variational quantum receivers.

In this context, this work illustrates how concepts from classical MIMO communication can be reformulated within a quantum-information framework by combining coherent multi-qubit interactions with variational quantum learning. The proposed approach establishes a connection between communication theory, quantum information processing, and variational quantum computing, and may provide a useful foundation for future quantum networking and communication architectures. The results also highlight the importance of developing scalable, noise-resilient receiver designs capable of operating effectively on near-term quantum hardware.

\subsection{Current Limitations and Future Research}

While the proposed framework provides a systematic investigation of variational quantum receivers for quantum MIMO communication, several limitations should be acknowledged.

The results presented in this work are obtained using noise-model-augmented quantum simulations rather than physical quantum hardware. Although the adopted noise model incorporates depolarizing errors, thermal relaxation, and readout imperfections representative of noisy intermediate-scale quantum (NISQ) devices, real quantum processors exhibit additional hardware-specific effects, including crosstalk, calibration drift, correlated errors, and connectivity constraints. These factors may influence practical performance and are not fully captured within the present simulation framework. Experimental validation on contemporary quantum hardware therefore remains an important direction for future investigation.

The numerical analysis is also restricted to moderate system sizes, with detailed studies performed for systems containing up to eight qubits. While this range is sufficient to reveal the principal scaling trends of the proposed architecture, larger systems may introduce additional challenges associated with circuit depth, optimization complexity, cumulative noise accumulation, and barren-plateau effects. Extending the analysis to larger quantum MIMO configurations is therefore necessary to fully characterize the scalability of the approach.

The proposed receiver is evaluated within the NISQ paradigm and does not incorporate advanced error-mitigation or fault-tolerant techniques. Methods such as measurement-error mitigation, zero-noise extrapolation, probabilistic error cancellation, and quantum error correction have the potential to improve decoding performance and extend the operational regime of variational quantum receivers. Understanding how these techniques interact with the proposed communication framework remains an important topic for future research.

It is also important to emphasize that the present work does not claim a proven asymptotic quantum computational advantage over classical MIMO detection algorithms. Rather, the primary objective of this study is to demonstrate the feasibility of variational quantum equalization for quantum MIMO communication under realistic NISQ noise conditions. Establishing a rigorous quantum advantage would require formal complexity-theoretic analysis together with large-scale implementations on quantum hardware, which remain important directions for future research.

Despite these limitations, the results provide a proof-of-concept demonstration of quantum MIMO communication with variational decoding and provide a foundation for future studies involving larger-scale systems, hardware implementation, and fault-tolerant quantum communication architectures.

\section{Conclusion}\label{SecV}

This work presented a variational quantum receiver for quantum MIMO communication and investigated its performance under realistic NISQ noise conditions. The results demonstrate that adaptive variational decoding can recover transmitted information under realistic NISQ noise conditions while exposing the fundamental trade-offs among system size, circuit complexity, entanglement generation, and noise accumulation. Together, the numerical and theoretical analyses provide a foundation for understanding the scalability and practical feasibility of variational quantum communication architectures.

Beyond the specific receiver architecture studied here, the results highlight the potential of variational quantum algorithms as communication-theoretic primitives capable of performing adaptive signal recovery in multi-qubit quantum channels. Unlike conventional receiver designs that rely on explicit channel estimation and matrix inversion, the proposed approach learns an approximate decoding transformation through variational optimization using measurement-derived performance metrics, suggesting an alternative paradigm for communication processing in quantum-enabled networks. This capability may become increasingly valuable as communication architectures evolve toward environments where classical and quantum information processing coexist.

The proposed framework is relevant to the development of future hybrid classical-quantum communication networks in which quantum processors, quantum memories, and quantum networking components operate alongside conventional communication infrastructure. The channel model investigated in this work provides a physically motivated extension of classical MIMO communication concepts into the quantum domain by representing interference through coherent multi-qubit interactions rather than classical linear mixing. As a result, the proposed approach offers a useful platform for studying information transmission, signal processing, and receiver design in quantum communication environments. In such settings, variational quantum receivers may serve as adaptive interfaces between transmitted information and quantum processing resources, enabling signal recovery, channel adaptation, and information extraction within larger quantum-enabled network architectures.

The results additionally emphasize the relevance of NISQ-era communication architectures. Although current quantum hardware remains constrained by limited qubit counts and noise, the present study demonstrates that variational decoding remains operational under realistic noise models and can be systematically analyzed within quantum communication settings.

The broader significance of this work lies in the presentation of a framework for studying scalable quantum communication theory, variational quantum algorithms, and quantum information processing within a unified architecture. Future research may extend the present framework through hardware implementation, larger-scale quantum MIMO systems, advanced error-mitigation techniques, fault-tolerant receiver designs, and more realistic quantum channel models. Such developments could contribute to future quantum communication networks in which adaptive quantum receivers provide a flexible mechanism for information recovery and channel adaptation.The present study therefore provides a foundation for further investigation of variational quantum learning techniques in quantum communication systems involving multi-qubit interactions and noisy quantum channels.
\bibliographystyle{apsrev4-1}  
\bibliography{Qmimo}

\begin{thebibliography}{31}%
\makeatletter
\providecommand \@ifxundefined [1]{%
 \@ifx{#1\undefined}
}%
\providecommand \@ifnum [1]{%
 \ifnum #1\expandafter \@firstoftwo
 \else \expandafter \@secondoftwo
 \fi
}%
\providecommand \@ifx [1]{%
 \ifx #1\expandafter \@firstoftwo
 \else \expandafter \@secondoftwo
 \fi
}%
\providecommand \natexlab [1]{#1}%
\providecommand \enquote  [1]{``#1''}%
\providecommand \bibnamefont  [1]{#1}%
\providecommand \bibfnamefont [1]{#1}%
\providecommand \citenamefont [1]{#1}%
\providecommand \href@noop [0]{\@secondoftwo}%
\providecommand \href [0]{\begingroup \@sanitize@url \@href}%
\providecommand \@href[1]{\@@startlink{#1}\@@href}%
\providecommand \@@href[1]{\endgroup#1\@@endlink}%
\providecommand \@sanitize@url [0]{\catcode `\\12\catcode `\$12\catcode `\&12\catcode `\#12\catcode `\^12\catcode `\_12\catcode `\%12\relax}%
\providecommand \@@startlink[1]{}%
\providecommand \@@endlink[0]{}%
\providecommand \url  [0]{\begingroup\@sanitize@url \@url }%
\providecommand \@url [1]{\endgroup\@href {#1}{\urlprefix }}%
\providecommand \urlprefix  [0]{URL }%
\providecommand \Eprint [0]{\href }%
\providecommand \doibase [0]{http://dx.doi.org/}%
\providecommand \selectlanguage [0]{\@gobble}%
\providecommand \bibinfo  [0]{\@secondoftwo}%
\providecommand \bibfield  [0]{\@secondoftwo}%
\providecommand \translation [1]{[#1]}%
\providecommand \BibitemOpen [0]{}%
\providecommand \bibitemStop [0]{}%
\providecommand \bibitemNoStop [0]{.\EOS\space}%
\providecommand \EOS [0]{\spacefactor3000\relax}%
\providecommand \BibitemShut  [1]{\csname bibitem#1\endcsname}%
\let\auto@bib@innerbib\@empty
\bibitem [{\citenamefont {Tse}\ and\ \citenamefont {Viswanath}(2005)}]{tse2005fundamentals}%
  \BibitemOpen
  \bibfield  {author} {\bibinfo {author} {\bibfnamefont {D.}~\bibnamefont {Tse}}\ and\ \bibinfo {author} {\bibfnamefont {P.}~\bibnamefont {Viswanath}},\ }\href {\doibase 10.1017/CBO9780511807213} {\emph {\bibinfo {title} {Fundamentals of Wireless Communication}}}\ (\bibinfo  {publisher} {Cambridge University Press},\ \bibinfo {year} {2005})\BibitemShut {NoStop}%
\bibitem [{\citenamefont {Telatar}(1999)}]{telatar1999capacity}%
  \BibitemOpen
  \bibfield  {author} {\bibinfo {author} {\bibfnamefont {E.}~\bibnamefont {Telatar}},\ }\href {\doibase 10.1002/ett.4460100604} {\bibfield  {journal} {\bibinfo  {journal} {European Transactions on Telecommunications}\ }\textbf {\bibinfo {volume} {10}},\ \bibinfo {pages} {585} (\bibinfo {year} {1999})}\BibitemShut {NoStop}%
\bibitem [{\citenamefont {Poor}(1994)}]{poor1994introduction}%
  \BibitemOpen
  \bibfield  {author} {\bibinfo {author} {\bibfnamefont {H.~V.}\ \bibnamefont {Poor}},\ }\href {\doibase 10.1007/978-1-4757-2341-0} {\emph {\bibinfo {title} {An Introduction to Signal Detection and Estimation}}},\ \bibinfo {edition} {2nd}\ ed.\ (\bibinfo  {publisher} {Springer},\ \bibinfo {year} {1994})\BibitemShut {NoStop}%
\bibitem [{\citenamefont {Kay}(1993)}]{kay1993}%
  \BibitemOpen
  \bibfield  {author} {\bibinfo {author} {\bibfnamefont {S.~M.}\ \bibnamefont {Kay}},\ }\href@noop {} {\emph {\bibinfo {title} {Fundamentals of Statistical Signal Processing, Volume I: Estimation Theory}}}\ (\bibinfo  {publisher} {Prentice Hall},\ \bibinfo {year} {1993})\BibitemShut {NoStop}%
\bibitem [{\citenamefont {Nielsen}\ and\ \citenamefont {Chuang}(2010)}]{nielsen2010quantum}%
  \BibitemOpen
  \bibfield  {author} {\bibinfo {author} {\bibfnamefont {M.~A.}\ \bibnamefont {Nielsen}}\ and\ \bibinfo {author} {\bibfnamefont {I.~L.}\ \bibnamefont {Chuang}},\ }\href {\doibase 10.1017/CBO9780511976667} {\emph {\bibinfo {title} {Quantum Computation and Quantum Information}}}\ (\bibinfo  {publisher} {Cambridge University Press},\ \bibinfo {year} {2010})\BibitemShut {NoStop}%
\bibitem [{\citenamefont {Wilde}(2013)}]{wilde2013quantum}%
  \BibitemOpen
  \bibfield  {author} {\bibinfo {author} {\bibfnamefont {M.~M.}\ \bibnamefont {Wilde}},\ }\href {\doibase 10.1017/CBO9781139525343} {\emph {\bibinfo {title} {Quantum Information Theory}}}\ (\bibinfo  {publisher} {Cambridge University Press},\ \bibinfo {year} {2013})\BibitemShut {NoStop}%
\bibitem [{\citenamefont {Shannon}(1948)}]{shannon1948mathematical}%
  \BibitemOpen
  \bibfield  {author} {\bibinfo {author} {\bibfnamefont {C.~E.}\ \bibnamefont {Shannon}},\ }\href {\doibase 10.1002/j.1538-7305.1948.tb01338.x} {\bibfield  {journal} {\bibinfo  {journal} {Bell System Technical Journal}\ }\textbf {\bibinfo {volume} {27}},\ \bibinfo {pages} {379} (\bibinfo {year} {1948})}\BibitemShut {NoStop}%
\bibitem [{\citenamefont {Holevo}(1998)}]{holevo1998capacity}%
  \BibitemOpen
  \bibfield  {author} {\bibinfo {author} {\bibfnamefont {A.~S.}\ \bibnamefont {Holevo}},\ }\href {\doibase 10.1109/18.651037} {\bibfield  {journal} {\bibinfo  {journal} {IEEE Transactions on Information Theory}\ }\textbf {\bibinfo {volume} {44}},\ \bibinfo {pages} {269} (\bibinfo {year} {1998})}\BibitemShut {NoStop}%
\bibitem [{\citenamefont {Schumacher}\ and\ \citenamefont {Westmoreland}(1997)}]{schumacher1997sending}%
  \BibitemOpen
  \bibfield  {author} {\bibinfo {author} {\bibfnamefont {B.}~\bibnamefont {Schumacher}}\ and\ \bibinfo {author} {\bibfnamefont {M.~D.}\ \bibnamefont {Westmoreland}},\ }\href {\doibase 10.1103/PhysRevA.56.131} {\bibfield  {journal} {\bibinfo  {journal} {Physical Review A}\ }\textbf {\bibinfo {volume} {56}},\ \bibinfo {pages} {131} (\bibinfo {year} {1997})}\BibitemShut {NoStop}%
\bibitem [{\citenamefont {Mikki}(2019)}]{mikki2019quantum}%
  \BibitemOpen
  \bibfield  {author} {\bibinfo {author} {\bibfnamefont {S.}~\bibnamefont {Mikki}},\ }\href {\doibase 10.2528/PIERC19032404} {\bibfield  {journal} {\bibinfo  {journal} {Progress In Electromagnetics Research C}\ }\textbf {\bibinfo {volume} {93}},\ \bibinfo {pages} {143} (\bibinfo {year} {2019})}\BibitemShut {NoStop}%
\bibitem [{\citenamefont {Koudia}\ and\ \citenamefont {Chatzinotas}(2025)}]{koudia2025crosstalk}%
  \BibitemOpen
  \bibfield  {author} {\bibinfo {author} {\bibfnamefont {S.}~\bibnamefont {Koudia}}\ and\ \bibinfo {author} {\bibfnamefont {S.}~\bibnamefont {Chatzinotas}},\ }\href {\doibase 10.1038/s41534-025-01113-x} {\bibfield  {journal} {\bibinfo  {journal} {npj Quantum Information}\ }\textbf {\bibinfo {volume} {11}},\ \bibinfo {pages} {162} (\bibinfo {year} {2025})}\BibitemShut {NoStop}%
\bibitem [{\citenamefont {Rehman}\ \emph {et~al.}(2025{\natexlab{a}})\citenamefont {Rehman}, \citenamefont {Oleynik}, \citenamefont {Koudia}, \citenamefont {Bayraktar},\ and\ \citenamefont {Chatzinotas}}]{oleynik2025diversity}%
  \BibitemOpen
  \bibfield  {author} {\bibinfo {author} {\bibfnamefont {J.~u.}\ \bibnamefont {Rehman}}, \bibinfo {author} {\bibfnamefont {L.}~\bibnamefont {Oleynik}}, \bibinfo {author} {\bibfnamefont {S.}~\bibnamefont {Koudia}}, \bibinfo {author} {\bibfnamefont {M.}~\bibnamefont {Bayraktar}}, \ and\ \bibinfo {author} {\bibfnamefont {S.}~\bibnamefont {Chatzinotas}},\ }\href {\doibase 10.1140/epjqt/s40507-025-00324-7} {\bibfield  {journal} {\bibinfo  {journal} {EPJ Quantum Technology}\ }\textbf {\bibinfo {volume} {12}},\ \bibinfo {pages} {18} (\bibinfo {year} {2025}{\natexlab{a}})}\BibitemShut {NoStop}%
\bibitem [{\citenamefont {Rehman}\ \emph {et~al.}(2025{\natexlab{b}})\citenamefont {Rehman}, \citenamefont {Rizvi}, \citenamefont {Koudia}, \citenamefont {Chatzinotas},\ and\ \citenamefont {Shin}}]{rehman2025mimo}%
  \BibitemOpen
  \bibfield  {author} {\bibinfo {author} {\bibfnamefont {J.~u.}\ \bibnamefont {Rehman}}, \bibinfo {author} {\bibfnamefont {S.~M.~A.}\ \bibnamefont {Rizvi}}, \bibinfo {author} {\bibfnamefont {S.}~\bibnamefont {Koudia}}, \bibinfo {author} {\bibfnamefont {S.}~\bibnamefont {Chatzinotas}}, \ and\ \bibinfo {author} {\bibfnamefont {H.}~\bibnamefont {Shin}},\ }\href {\doibase 10.1109/LCOMM.2025.3567246} {\bibfield  {journal} {\bibinfo  {journal} {IEEE Communications Letters}\ } (\bibinfo {year} {2025}{\natexlab{b}}),\ 10.1109/LCOMM.2025.3567246}\BibitemShut {NoStop}%
\bibitem [{\citenamefont {Cerezo}\ \emph {et~al.}(2021)\citenamefont {Cerezo} \emph {et~al.}}]{cerezo2021variational}%
  \BibitemOpen
  \bibfield  {author} {\bibinfo {author} {\bibfnamefont {M.}~\bibnamefont {Cerezo}} \emph {et~al.},\ }\href {\doibase 10.1038/s42254-021-00348-9} {\bibfield  {journal} {\bibinfo  {journal} {Nature Reviews Physics}\ }\textbf {\bibinfo {volume} {3}},\ \bibinfo {pages} {625} (\bibinfo {year} {2021})}\BibitemShut {NoStop}%
\bibitem [{\citenamefont {Mitarai}\ \emph {et~al.}(2018)\citenamefont {Mitarai} \emph {et~al.}}]{mitarai2018quantum}%
  \BibitemOpen
  \bibfield  {author} {\bibinfo {author} {\bibfnamefont {K.}~\bibnamefont {Mitarai}} \emph {et~al.},\ }\href {\doibase 10.1103/PhysRevA.98.032309} {\bibfield  {journal} {\bibinfo  {journal} {Physical Review A}\ }\textbf {\bibinfo {volume} {98}},\ \bibinfo {pages} {032309} (\bibinfo {year} {2018})}\BibitemShut {NoStop}%
\bibitem [{\citenamefont {Kandala}\ \emph {et~al.}(2017)\citenamefont {Kandala}, \citenamefont {Mezzacapo}, \citenamefont {Temme}, \citenamefont {Takita}, \citenamefont {Brink}, \citenamefont {Chow},\ and\ \citenamefont {Gambetta}}]{kandala2017hardware}%
  \BibitemOpen
  \bibfield  {author} {\bibinfo {author} {\bibfnamefont {A.}~\bibnamefont {Kandala}}, \bibinfo {author} {\bibfnamefont {A.}~\bibnamefont {Mezzacapo}}, \bibinfo {author} {\bibfnamefont {K.}~\bibnamefont {Temme}}, \bibinfo {author} {\bibfnamefont {M.}~\bibnamefont {Takita}}, \bibinfo {author} {\bibfnamefont {M.}~\bibnamefont {Brink}}, \bibinfo {author} {\bibfnamefont {J.~M.}\ \bibnamefont {Chow}}, \ and\ \bibinfo {author} {\bibfnamefont {J.~M.}\ \bibnamefont {Gambetta}},\ }\href {\doibase 10.1038/nature23879} {\bibfield  {journal} {\bibinfo  {journal} {Nature}\ }\textbf {\bibinfo {volume} {549}},\ \bibinfo {pages} {242} (\bibinfo {year} {2017})}\BibitemShut {NoStop}%
\bibitem [{\citenamefont {Preskill}(2018)}]{preskill2018nisq}%
  \BibitemOpen
  \bibfield  {author} {\bibinfo {author} {\bibfnamefont {J.}~\bibnamefont {Preskill}},\ }\href {\doibase 10.22331/q-2018-08-06-79} {\bibfield  {journal} {\bibinfo  {journal} {Quantum}\ }\textbf {\bibinfo {volume} {2}},\ \bibinfo {pages} {79} (\bibinfo {year} {2018})}\BibitemShut {NoStop}%
\bibitem [{\citenamefont {McClean}\ \emph {et~al.}(2018)\citenamefont {McClean} \emph {et~al.}}]{mcclean2018barren}%
  \BibitemOpen
  \bibfield  {author} {\bibinfo {author} {\bibfnamefont {J.~R.}\ \bibnamefont {McClean}} \emph {et~al.},\ }\href {\doibase 10.1038/s41467-018-07090-4} {\bibfield  {journal} {\bibinfo  {journal} {Nature Communications}\ }\textbf {\bibinfo {volume} {9}},\ \bibinfo {pages} {4812} (\bibinfo {year} {2018})}\BibitemShut {NoStop}%
\bibitem [{\citenamefont {Temme}\ \emph {et~al.}(2017)\citenamefont {Temme} \emph {et~al.}}]{temme2017error}%
  \BibitemOpen
  \bibfield  {author} {\bibinfo {author} {\bibfnamefont {K.}~\bibnamefont {Temme}} \emph {et~al.},\ }\href {\doibase 10.1103/PhysRevLett.119.180509} {\bibfield  {journal} {\bibinfo  {journal} {Physical Review Letters}\ }\textbf {\bibinfo {volume} {119}},\ \bibinfo {pages} {180509} (\bibinfo {year} {2017})}\BibitemShut {NoStop}%
\bibitem [{\citenamefont {Li}\ and\ \citenamefont {Benjamin}(2017)}]{li2017efficient}%
  \BibitemOpen
  \bibfield  {author} {\bibinfo {author} {\bibfnamefont {Y.}~\bibnamefont {Li}}\ and\ \bibinfo {author} {\bibfnamefont {S.~C.}\ \bibnamefont {Benjamin}},\ }\href {\doibase 10.1103/PhysRevX.7.021050} {\bibfield  {journal} {\bibinfo  {journal} {Physical Review X}\ }\textbf {\bibinfo {volume} {7}},\ \bibinfo {pages} {021050} (\bibinfo {year} {2017})}\BibitemShut {NoStop}%
\bibitem [{\citenamefont {Cover}\ and\ \citenamefont {Thomas}(2006)}]{cover2006elements}%
  \BibitemOpen
  \bibfield  {author} {\bibinfo {author} {\bibfnamefont {T.~M.}\ \bibnamefont {Cover}}\ and\ \bibinfo {author} {\bibfnamefont {J.~A.}\ \bibnamefont {Thomas}},\ }\href {\doibase 10.1002/047174882X} {\emph {\bibinfo {title} {Elements of Information Theory}}},\ \bibinfo {edition} {2nd}\ ed.\ (\bibinfo  {publisher} {Wiley-Interscience},\ \bibinfo {address} {Hoboken, NJ},\ \bibinfo {year} {2006})\BibitemShut {NoStop}%
\bibitem [{\citenamefont {Schuld}\ \emph {et~al.}(2020)\citenamefont {Schuld} \emph {et~al.}}]{schuld2020circuit}%
  \BibitemOpen
  \bibfield  {author} {\bibinfo {author} {\bibfnamefont {M.}~\bibnamefont {Schuld}} \emph {et~al.},\ }\href {\doibase 10.1103/PhysRevA.101.032308} {\bibfield  {journal} {\bibinfo  {journal} {Physical Review A}\ }\textbf {\bibinfo {volume} {101}},\ \bibinfo {pages} {032308} (\bibinfo {year} {2020})}\BibitemShut {NoStop}%
\bibitem [{\citenamefont {Helstrom}(1976)}]{helstrom1976quantum}%
  \BibitemOpen
  \bibfield  {author} {\bibinfo {author} {\bibfnamefont {C.~W.}\ \bibnamefont {Helstrom}},\ }\href@noop {} {\emph {\bibinfo {title} {Quantum Detection and Estimation Theory}}}\ (\bibinfo  {publisher} {Academic Press},\ \bibinfo {address} {New York},\ \bibinfo {year} {1976})\BibitemShut {NoStop}%
\bibitem [{\citenamefont {Schuld}\ and\ \citenamefont {Petruccione}(2019)}]{schuld2019quantum}%
  \BibitemOpen
  \bibfield  {author} {\bibinfo {author} {\bibfnamefont {M.}~\bibnamefont {Schuld}}\ and\ \bibinfo {author} {\bibfnamefont {F.}~\bibnamefont {Petruccione}},\ }\href {\doibase 10.1007/978-3-319-96424-9} {\emph {\bibinfo {title} {Quantum Machine Learning}}}\ (\bibinfo  {publisher} {Springer},\ \bibinfo {year} {2019})\BibitemShut {NoStop}%
\bibitem [{\citenamefont {Spall}(1998{\natexlab{a}})}]{spall1998implementation}%
  \BibitemOpen
  \bibfield  {author} {\bibinfo {author} {\bibfnamefont {J.~C.}\ \bibnamefont {Spall}},\ }\href {\doibase 10.1109/7.705889} {\bibfield  {journal} {\bibinfo  {journal} {IEEE Transactions on Aerospace and Electronic Systems}\ } (\bibinfo {year} {1998}{\natexlab{a}}),\ 10.1109/7.705889}\BibitemShut {NoStop}%
\bibitem [{\citenamefont {Spall}(1998{\natexlab{b}})}]{spall1998overview}%
  \BibitemOpen
  \bibfield  {author} {\bibinfo {author} {\bibfnamefont {J.~C.}\ \bibnamefont {Spall}},\ }\href@noop {} {\bibfield  {journal} {\bibinfo  {journal} {Johns Hopkins APL Technical Digest}\ }\textbf {\bibinfo {volume} {19}},\ \bibinfo {pages} {482} (\bibinfo {year} {1998}{\natexlab{b}})}\BibitemShut {NoStop}%
\bibitem [{\citenamefont {Kingma}\ and\ \citenamefont {Ba}(2014)}]{kingma2014adam}%
  \BibitemOpen
  \bibfield  {author} {\bibinfo {author} {\bibfnamefont {D.~P.}\ \bibnamefont {Kingma}}\ and\ \bibinfo {author} {\bibfnamefont {J.}~\bibnamefont {Ba}},\ }\href {\doibase 10.48550/arXiv.1412.6980} {\bibfield  {journal} {\bibinfo  {journal} {arXiv preprint arXiv:1412.6980}\ } (\bibinfo {year} {2014}),\ 10.48550/arXiv.1412.6980}\BibitemShut {NoStop}%
\bibitem [{\citenamefont {Bottou}\ \emph {et~al.}(2018)\citenamefont {Bottou}, \citenamefont {Curtis},\ and\ \citenamefont {Nocedal}}]{bottou2018optimization}%
  \BibitemOpen
  \bibfield  {author} {\bibinfo {author} {\bibfnamefont {L.}~\bibnamefont {Bottou}}, \bibinfo {author} {\bibfnamefont {F.~E.}\ \bibnamefont {Curtis}}, \ and\ \bibinfo {author} {\bibfnamefont {J.}~\bibnamefont {Nocedal}},\ }\href {\doibase 10.1137/16M1080173} {\bibfield  {journal} {\bibinfo  {journal} {SIAM Review}\ }\textbf {\bibinfo {volume} {60}},\ \bibinfo {pages} {223} (\bibinfo {year} {2018})}\BibitemShut {NoStop}%
\bibitem [{\citenamefont {Powell}(1994)}]{powell1994direct}%
  \BibitemOpen
  \bibfield  {author} {\bibinfo {author} {\bibfnamefont {M.~J.~D.}\ \bibnamefont {Powell}},\ }\href {\doibase 10.1007/978-94-015-8330-5_4} {\bibfield  {journal} {\bibinfo  {journal} {Advances in Optimization and Numerical Analysis}\ }\textbf {\bibinfo {volume} {275}},\ \bibinfo {pages} {51} (\bibinfo {year} {1994})}\BibitemShut {NoStop}%
\bibitem [{\citenamefont {{Qiskit contributors}}(2025)}]{qiskit}%
  \BibitemOpen
  \bibfield  {author} {\bibinfo {author} {\bibnamefont {{Qiskit contributors}}},\ }\href {\doibase 10.5281/zenodo.2573505} {\enquote {\bibinfo {title} {Qiskit: An open-source framework for quantum computing},}\ } (\bibinfo {year} {2025})\BibitemShut {NoStop}%
\bibitem [{\citenamefont {Fuchs}\ and\ \citenamefont {van~de Graaf}(1999)}]{fuchs1999cryptographic}%
  \BibitemOpen
  \bibfield  {author} {\bibinfo {author} {\bibfnamefont {C.~A.}\ \bibnamefont {Fuchs}}\ and\ \bibinfo {author} {\bibfnamefont {J.}~\bibnamefont {van~de Graaf}},\ }\href {\doibase 10.1109/18.761271} {\bibfield  {journal} {\bibinfo  {journal} {IEEE Transactions on Information Theory}\ }\textbf {\bibinfo {volume} {45}},\ \bibinfo {pages} {1216} (\bibinfo {year} {1999})}\BibitemShut {NoStop}%
\end{thebibliography}%

\end{document}